\documentclass[12pt]{article}

\usepackage{amsmath,amssymb}
\usepackage{multirow}
\usepackage[pdftex]{graphicx}
\usepackage{natbib}
\usepackage{mathrsfs}
\usepackage{bm, xcolor}
\usepackage{mathrsfs}
\usepackage{verbatim}
\usepackage{longtable}
\usepackage{threeparttable}
\usepackage{color}
\usepackage{amsmath}
\usepackage{url}

\usepackage{enumerate}
\usepackage{amssymb}
\usepackage{amsbsy}
\usepackage{amsmath}
\usepackage{amsthm}
\usepackage{amsfonts}
\usepackage{latexsym}
\usepackage{xifthen} 
\usepackage{color}
\usepackage{manfnt}
\usepackage{ifthen}
\usepackage{bm}
\usepackage{caption}
\usepackage{subcaption}

\usepackage{algorithm}
\usepackage{algorithmic}

\usepackage{float}

\usepackage{cleveref}
\crefformat{equation}{(#2#1#3)}
\crefrangeformat{equation}{(#3#1#4) to~(#5#2#6)}
\crefname{equation}{}{}
\Crefname{equation}{}{}

\def\calC{\mathcal{C}}
\def\calF{\mathcal{F}}
\def\calI{\mathcal{I}}
\def\calP{\mathcal{P}}
\def\calM{\mathcal{M}}

\def\P{\mathbb{P}}

\newcommand{\wt}[1]{\widetilde{#1}} 

\newcommand{\mR}{\mathbb{R}}

\newcommand{\E}{\mathbb{E}} 
\renewcommand{\P}{\mathbb{P}} 
\newcommand{\var}{{\rm Var}} 
\newcommand{\cov}{\mathop{\rm Cov}} 

\newcommand{\indic}[1]{\mbf{1}\left\{#1\right\}} 

\makeatletter
\def\singlespace{\def\baselinestretch{1}\@normalsize}

\newtheoremstyle{mythmstyle}
{8 pt} 
{3 pt} 
{} 
{} 
{\bfseries} 
{.} 
{.5em} 
{} 

\theoremstyle{plain}

\makeatletter
\def\thm@space@setup{%
	\thm@preskip=6pt plus 1pt minus 1pt
	\thm@postskip=\thm@preskip 
}
\makeatother

\newtheorem{theorem}{Theorem}[section]
\newtheorem{lemma}[theorem]{Lemma}

\newtheorem{proposition}[theorem]{Proposition}
\newtheorem{remark}{Remark}

\newtheorem*{example*}{Example}
\newtheorem{definition}{Definition}
\newtheorem*{definition*}{Definition}

\newtheorem*{remark*}{Remark}
\crefname{definition}{\textbf{definition}}{definitions}
\Crefname{definition}{Definition}{Definitions}
\crefname{assumption}{\textbf{assumption}}{assumptions}
\Crefname{assumption}{Assumption}{Assumptions}

\renewcommand{\hat}{\widehat}

\def\singlespace{\def\baselinestretch{1}\@normalsize}

\makeatother

\def\newpage{\vfill\eject}
\def\wh{\widehat}

\def\wt{\widetilde}

\newdimen\biblioindent    \biblioindent=30pt

 at 7truept

\def\beqr{\begin{eqnarray}}
	\def\eeqr{\end{eqnarray}}
\def\beqrs{\begin{eqnarray*}}
	\def\eeqrs{\end{eqnarray*}}

\def\beq{\begin{equation}}
\def\eeq{\end{equation}}
\def\beqn{\begin{eqnarray}}
\def\eeqn{\end{eqnarray}}
\def\beqnn{\begin{eqnarray*}}
\def\eeqnn{\end{eqnarray*}}

\def\hDash{\bot\!\!\!\bot}
\def\wh{\widehat}
\def\wt{\widetilde}

\def\defby{\stackrel{\mbox{\textrm{\rm\tiny def}}}{=}}

\def\calI{\mathcal{I}}
\def\calF{\mathcal{F}}

\def\calP{\mathcal{P}}

\def\calM{\mathcal{M}}

\def\ep{\epsilon}

\def\indic{\mathbb I} 

\newcommand{\ind}{1{\hskip -2.5 pt}\hbox{I}}

\begin{document}
	\renewcommand{\baselinestretch}{1.3}

	\title {\bf  Asymptotic Distribution-Free Independence Test for High Dimension Data}
	\author{ Zhanrui Cai$^*$, Jing Lei$^\dagger$, Kathryn Roeder$^\dagger$\\      
	$^*$Faculty of Business and Economics, The University of Hong Kong\\
	$^\dagger$Department of Statistics and Data Science, Carnegie Mellon University\\   }
	\date{\empty}
	
	\maketitle
	\renewcommand{\baselinestretch}{1.5}
	\baselineskip=24pt
	\noindent{\bf Abstract:}

		Test of independence is of fundamental importance in modern data analysis, with broad applications in variable selection, graphical models, and causal inference. 
	When the data is high dimensional and the potential dependence signal is sparse, independence testing becomes very challenging without distributional or structural assumptions.
	In this paper, we propose a general framework for independence testing by first fitting a classifier that distinguishes the joint and product distributions, and then testing the significance of the fitted classifier.
	This framework allows us to borrow the strength of the most advanced classification algorithms developed from the modern machine learning community, making it applicable to high dimensional, complex data.
	By combining a sample split and a fixed permutation, our test statistic has a universal, fixed Gaussian null distribution that is independent of the underlying data distribution. 
	Extensive simulations demonstrate the advantages of the newly proposed test compared with existing methods. We further apply the new test to a single cell data set to test the independence between two types of single cell sequencing measurements, whose high dimensionality and sparsity make existing methods hard to apply.

	\par \vspace{9pt} \noindent {\it Key words and phrases: }Test of independence, sample splitting, neural network.

	\pagestyle{plain}
	
	\newpage
	
\section{Introduction}
Test of independence is a fundamental question in data analysis and statistical inference. Considering two multivariate random vectors $X $ and $Y$, we are interested in testing whether the two random vectors are independent, namely, $H_0: X \hDash Y$.   Such testing problems are relevant in many statistical learning problems, including variable selection in regression, Gaussian graphical models, Markov random fields, and causal inference \citep{fan2020statistical, maathuis2018handbook, imbens2015causal}.  In traditional statistical literature, one may choose the Pearson correlation to measure the independence between $X$ and $Y$ when the data has a jointly normal distribution, or opt for the rank correlation when both $X$ and $Y$ are univariate. With the development of information technology, researchers are now able to collect complex and potentially high dimensional data with potentially highly nonlinear dependence. How to perform tests of independence for modern data is a challenging and important problem in the contemporary statistical community.

In the past two decades, there have been a series of substantial developments in the testing of independence for general $X$ and $Y$ without assuming their parametric distributions. A natural starting point is to study the difference between $\mathcal P_{X,Y}$, the joint measure of $(X,Y)$, and $\mathcal P_X\times\mathcal P_Y$, the product measure of $X$ and $Y$.  In one of the most well-known papers on this topic,  \cite{szekely2007measuring} proposed the distance correlation by measuring the weighted integrated squared difference between the characteristic functions of $\mathcal P_{X,Y}$ and $\mathcal P_X\times \mathcal P_Y$, which is later shown to be equivalent to the maximum mean discrepancies in the machine learning community \citep{sejdinovic2013equivalence}, and closely related to the Hilbert-Schmidt independence criterion \citep{gretton2005measuring}. Extensions of distance correlation have been widely discussed \citep{szekely2013distance, huo2016fast, yao2018testing}. \cite{zhu2017projection} relaxed the moment constraint in distance correlation by combining the Hoeffding coefficient with projection pursuit. Other than comparing characteristic functions, there are also novel methods that compare the density functions \citep{berrett2019nonparametric}, and the cumulative distribution functions \citep{heller2012consistent,cui2019distribution, moon2020interpoint}.  \cite{kong2019composite} and \cite{chatterjee2021new} used the appealing idea of conditional mean variance to evaluate the dependence between two random variables. More recently, \cite{shi2020distribution} and \cite{deb2021multivariate}  developed the first distribution-free  independence test for multivariate random vectors. They define multiple ranks using the theory of measure transportation and propose (multivariate) rank versions of distance covariance and energy statistic for independence testing. But in practice, the computation for measure transportation will grow quickly with the sample size and dimension, which restricts the application of those two tests to large-scale datasets. High dimensional independence test has recently been studied by \cite{zhu2020distance} and \cite{gao2021asymptotic}. In comparison, our work is more generally applicable as we allow the dependence signal in high dimensional vectors to be very sparse, which is a benefit of implementing the advanced machine learning algorithms.

Our work is motivated by challenges arising in single-cell multimodal omics, a research area labeled `Method of the Year 2019' by Nature Methods.  This technological advance builds on the recent breakthroughs in sequencing the RNA of single cells and promises greater insights into gene regulatory networks, cell lineages, and trajectories by permitting the measurement of multiple omics on the same cell \citep{zhu2020single, schier2020single}. Of particular interest are simultaneous measurements of gene expression (RNA-seq) and chromatin accessibility (ATAC-seq).  ATAC-seq identifies active regulatory sequences in the genome by finding open chromatin, which determines whether a gene will be actively transcribed.  For this reason, it is widely assumed that RNA-seq and ATAC-seq will co-vary.  But both data sources tend to be high dimensional and extremely sparse, positing great challenges to performing statistical independence tests for the two random vectors. For example, the data we analyze consists of 11,188 blood cells, each with RNA-seq and ATAC-seq read counts. The dimension of RNA-seq is 29,717  and the dimension of ATAC-seq is 143,887. Only 6.35\% entries in the RNA-seq and 5.66\% entries in the ATAC-seq are non-zero, making all current independence testing methods practically infeasible.

The purpose of this paper is to build a distribution-free test of independence that is powerful even under high dimensional, complex data. 
Existing methods use U-statistics to directly estimate the integrated squared difference between the joint distribution and the product distribution, in the forms of characteristic functions, density functions, or cumulative distributions. Such U-statistics often fail to pick up the hidden signal when there are many noise dimensions in the data, and often require cumbersome resampling procedures to calibrate the null distribution.  Our proposal deviates from these methods by aiming at a different and adaptive quantity: Instead of the integrated squared difference between distribution functions, our method seeks to find \emph{any} potential difference between the joint and product distributions by constructing a classification problem between these two distributions.  
By leveraging recent developments in two sample testing and sample splitting \citep{kim2019global, hu2020distribution, kim2021classification}, we develop a test that is more flexible and can borrow strength from the most powerful classification tools, such as deep neural networks, from the machine learning community.  It is particularly powerful for high dimensional data when proper regularizations (such as sparsity) are enforced on the classifier. 

The proposed method consists of three steps: sample splitting, classification, and rank-sum comparison. We fist split the index set $\calI = \{1, \dots, n\}$ into two subsets $\calI_1 = \{1, 2, \dots, n_1\}$ and $\calI_2 = \{n_1+1, \dots, n\}$.  Let $D_{1A} = \{(X_i, Y_i), i\in\calI_1\}$ and $D_{2A} = \{(X_i, Y_i), i\in\calI_2\}$ be the two subsets of the data. Then we generate two correspondingly permuted datasets by cyclically permuting $Y$ in each of the two subsets. Let $D_{1B} = \{(X_i', Y_{i}'), i\in\calI_1\}$ and $D_{2B} = \{(X_i', Y_i'), i\in\calI_2\}$, where $X_i'=X_i$ for all $i$, $Y_i'=Y_{i+1}$ for $i\notin \{n_1,n\}$, and $Y_{n_1}'=Y_1$, $Y_n'=Y_{n_1+1}$. In the classification step, we train a classifier that aims to distinguish $D_{1A}$ from $D_{1B}$, because the sample points in $D_{1A}$ are generated from $\mathcal P_{X,Y}$ while those in $D_{1B}$ have marginal distribution $\mathcal P_{X}\times\mathcal P_Y$ and weak dependency between sample points. Next, in the rank-sum comparison step we compare the predicted class probabilities in $D_{2A}$ and $D_{2B}$. Under $H_0$, the predicted class probabilities of $D_{2A}$ and $D_{2B}$ should have the same distribution, while under $H_1$, those predicted probabilities of $D_{2A}$ and $D_{2B}$ should be different if the classifier is able to pick up the difference between $\mathcal P_{X,Y}$ and $\mathcal P_X\times\mathcal P_Y$. This intuition motivates a rank-sum test to compare the predicted class probabilities of the two samples. The main technical challenge is that the sample points in $D_{2A}$ and $D_{2B}$ are dependent, thus classical U-statistics theory can not be directly applied. Our theoretical development uses Hoeffding's projection to decompose the test statistic into sums of sparsely dependent random variables, and uses a version of Stein's method for sparsely dependent data to establish the normal approximation of the test statistic.

To sum up, the proposed method has the following advantages.

(i) {\it Completely nonparametric.} We require very few assumptions on the data to ensure the test's validity. Under $H_0$, the type I error control is automatically guaranteed by sample splitting and the single permutation. Under $H_1$, the test will have good power as long as the classifier is better than a random guess, which is practically feasible given the powerful neural networks.

(ii) {\it Asymptotic distribution-free and computationally efficient.} Our test statistic has a standard normal asymptotic null distribution. This is in critical contrast to other current independence tests that have non-explicit distributions and require the computationally expensive bootstraps to obtain $p$-values \citep{szekely2007measuring, heller2012consistent, berrett2019nonparametric}. For the most recent distribution-free independence tests \citep{shi2020distribution, deb2021multivariate}, the limiting null distributions are still weighted  $\chi^2(1)$, without an analytic form. Although \cite{shi2020distribution} listed the thresholds for some combinations of dimensions of $X$ and $Y$, it still needs at least one round of numerical approximation when the dimensions exceed those in \cite{shi2020distribution}. Such improved computational efficiency makes our method particularly appealing for the aforementioned single cell sequencing data.

(iii) {\it Applicability to high dimensional data.} The test is suitable for high dimensional data. Existing tests based on degenerate U-statistics are hard to apply and have limited power when the data dimension is high and the dependence signal is very sparse. By taking the classification perspective, we can take advantage of adaptive and structured classifiers to pick up weak signals from high dimensional data.  Moreover, our framework allows $X$, $Y$ to take value in infinite-dimensional spaces, as long as the likelihood ratio is well defined.

(iv) {\it Flexibility and generality.} The method described in this paper is just one example from a general framework. All three steps (permutation, classification, and calibration) can be carried out with other variants that are more suitable to the problem at hand.  For example, one can use other dimension reduction or variable selection methods when distinguishing the two distributions, and/or use different two-sample testing methods, such as two-sample $t$-test, to calibrate the significance of classification.  When the original sample $(X_i, Y_i)$ has a time-series or random field structure as the index $i$ changes from $1$ to $n$, one can also consider other types of permutations that are more suitable for the particular dependence structure across sample points.


\section{Test of Independence by Sample Splitting and Classification}
\subsection{Preliminaries and basic ideas}

Consider independent observations $\{(X_i, Y_i):1\le i\le n\}$ of a pair of random variables $X$ and $Y$ with joint distribution  $\calP_{X,Y}$ in a space $\mathcal X\times\mathcal Y$. Let $\calP_X$ and $\calP_Y$ be the marginal distributions of $X$ and $Y$ respectively. We are interested in testing
\beqrs
H_0: \calP_{X,Y} = \calP_X \times\calP_Y\quad\mbox{versus} \quad H_1: \calP_{X,Y} \neq \calP_X \times\calP_Y,
\eeqrs
where $\calP_X \times\calP_Y$  denotes the product distribution.

Most existing methods for independence testing focus on a quantity of the form $$\int w(x,y)\phi(G(x,y),G_1(x)G_2(y))dxdy,$$
where $G(\cdot)$, $G_1(\cdot)$, $G_2(\cdot)$ are joint and marginal distribution functions, $w$ is a weight function, and $\phi$ is a discrepancy measure.  This framework covers nearly all the popularly studied independence testing methods, including distance correlation \citep{szekely2007measuring}, Hilbert-Schimidt independence criterion \citep{gretton2005measuring,gretton2007kernel}, rank-correlation based methods \citep{heller2012consistent,moon2020interpoint}, and mutual information based methods \citep{berrett2019nonparametric}.  While enjoying elegant theoretical properties, these methods rely on specific choices of $w$, $\phi$, and $G$ functions, making them hard to apply for high-dimensional, complex data.  Moreover, the null distributions of the corresponding test statistic usually depend on the unknown underlying distribution $\mathcal P_{X,Y}$ and must be approximated using resampling methods. 

The key feature of our method is that it does not rely on a pre-chosen set of functions $(w,\phi, G)$.  Instead, our method begins with fitting a flexible classifier to distinguish $\mathcal P_{X, Y}$ and $\mathcal P_X\times\mathcal P_Y$, and then tests whether the fitted classifier does anything different from random guessing.  
Suppose we have two equal-sized samples, one from $\mathcal P_{X,Y}$ and one from $\mathcal P_X\times\mathcal P_Y$, and we associate a label $K=1$ ($K=0$) for each sample point from $\mathcal P_{X,Y}$ ($\mathcal P_X\times \mathcal P_Y$).  We will discuss how to obtain these samples in the next subsection.
Under $H_0$, the two samples have the same distribution $\mathcal P_X\times\mathcal P_Y$, so any classifier trying to distinguish these two samples would behave like a random guess. On the other hand, under $H_1$, any classifier that can detect the difference between these two distributions should do better than random guess, which can be tested on a holdout pair of samples from the two distributions.

More specifically,
the conditional label probability
$$
\theta(x,y)=\mathbb P(K=1|x,y)
$$
is related to the likelihood ratio
\beqr\label{likelihood_ratio}
L(x,y)\defby \frac{\theta(x,y)}{1-\theta(x,y)} = \frac{d\calP_{X, Y}}{d(\calP_X\times \calP_Y)}.
\eeqr
Therefore, $\theta(x,y)$ reduces the data dimension to $1$, while largely capturing the difference between $\calP_{X,Y}$ and $\calP_X\times\calP_Y$ as guranteed by the following result.
Under the null hypothesis, $\theta(x,y)\equiv 1/2$ and the likelihood ratio $L(x,y)\equiv 1$, which corresponds to a degenerate case.

\begin{proposition}\label{pro:separation}
	Let $\mathcal P$, $\mathcal Q$  be two probability distributions on a common measurable space such that $\mathcal P\ll \mathcal Q$ and
	the Radon-Nikodym derivative $d \mathcal P / d\mathcal Q$ has a continuous distribution under $\mathcal Q$. Let $V\sim \mathcal P$ and $W\sim\mathcal Q$ be independent and $d_{tv}(\cdot,\cdot)$ be the total variation distance between two probability measures, then
	$$
	\frac{1}{4}d_{\rm tv}(\mathcal P,\mathcal Q)\le\frac{1}{2}- \mathbb P\left\{\frac{d\mathcal P}{d\mathcal Q}(V)<\frac{d\mathcal P}{d\mathcal Q}(W)\right\}\le \frac{1}{2}d_{\rm tv}(\mathcal P,\mathcal Q)\,.
	$$ 
\end{proposition}

\begin{remark}[Dropping the continuity assumption] If $(d\mathcal P/d\mathcal Q)(W)$ has point mass, then it is possible to have $(d\mathcal P/d\mathcal Q)(V)=(d\mathcal P/d\mathcal Q)(W)$.  In this case one can associate each of $V$ and $W$ with an independent $U(0, 1)$ random variable, $\zeta$ and $\eta$, and rank them with randomized tie-breaking 
	\beqrs
	\indic\left\{\frac{d\mathcal P}{d\mathcal Q}(V)<\frac{d\mathcal P}{d\mathcal Q}(W)\right\}
	+\indic(\zeta<\eta)\indic\left\{\frac{d\mathcal P}{d\mathcal Q}(V)=\frac{d\mathcal P}{d\mathcal Q}(W)\right\}\,.
	\eeqrs
	All the theory, including \Cref{pro:separation}, goes through the same for such a random tie-breaking ranking scheme with more careful bookkeeping.  Therefore, in the rest of this paper, we will proceed under the assumption that $\theta(X,Y)$ and its estimate $\widehat\theta(X,Y)$ are continuous under $\mathcal P_X\times\mathcal P_Y$ for notational simplicity.
\end{remark}

Such a classification-testing procedure consists of a fitting part and testing part, which need to be carried out on separate subsamples.
Splitting the sample reduces the sample size used for both classification and testing. But the benefits are quite substantial: First, in high-dimensional data, the signal is often quite weak and concentrates on a low-dimensional subspace or submanifold hidden in the high-dimensional ambient space. It is often more efficient to find out the direction of the signal and then conduct hypothesis tests targeted specifically in that signal direction.  The reduced sample sizes can be viewed as our investment in finding the most promising direction of the signal.  Second, sample splitting provides great flexibility in the choice of classification algorithms, such as black-box methods and deep neural networks, which are particularly powerful in handling complex data.

Even if we split the sample to carry out the classification and test, another challenge remains: How do we obtain samples from the two distributions $\mathcal P_{X,Y}$ and $\mathcal P_X\times\mathcal P_Y$, as required by both the classification and the testing steps?  We provide a sample-size efficient answer to this question in the next subsection.

\subsection{Sample Splitting and Cyclic Permutation}

As discusssed in the previous subsection, the classification and testing procedures need to be carried out on separate subsamples to ensure the validity.  Suppose  we split the index set $\calI = \{1, \dots, n\}$ into two subsets $\calI_1 = \{1, 2, \dots, n_1\}$ and $\calI_2 = \{n_1+1, \dots, n\}$, $n_2 = n - n_1$, so that the subsample  $\mathcal I_1$ is used for classification and  $\mathcal I_2$ is used for testing.
However, after such a sample split we still do not have a sample from $\mathcal P_X\times\mathcal P_Y$ for classification or testing.  A simple idea is to further split $\mathcal I_1$ into $\mathcal I_{11}$ and $\mathcal I_{12}$, and permute the sample pairs in $\mathcal I_{12}$ to form a sample from $\mathcal P_X\times \mathcal P_Y$.  A similar second split and permutation can be applied to $\mathcal I_2$ for the testing purpose.  
Although this approach is simple and straightforward to implement, it further splits an already reduced sample size. A natural question is whether one can avoid such a second split and use the sample more efficiently.  We provide a positive answer below.

To avoid the second split,
denote $D_{1A} = \{(X_i, Y_i), i\in\calI_1\}$ the subsample in $\mathcal I_1$, and its cyclicly permuted version $D_{1B} = \{(X_i', Y_{i}'), i\in\calI_1\}$, where $X_i'=X_i$ for all $i$, $Y_i'=Y_{i+1}$ for $1\le i\le n_1-1$, and $Y_{n_1}'=Y_1$.  Similarly $D_{2A}=\{(X_i,Y_i):i\in\mathcal I_2\}$ denotes the subsample in $\mathcal I_2$, and its cyclicly permuted version $D_{2B} = \{(X_i', Y_i'), i\in\calI_2\}$, with $X_i'=X_i$ for all $i$, $Y_i'=Y_{i+1}$ for $n_1+1\le i\le n-1$, and $Y_n'=Y_{n_1+1}$.  Our plan is to treat $D_{jA}$, $D_{jB}$ as approximately independent samples from $\mathcal P_{X,Y}$ and $\mathcal P_X\times\mathcal P_Y$ for classification ($j=1$) and two-sample testing ($j=2$), because the dependence between the original and cyclicly permuted samples are very sparse.


Suppose we apply a classification algorithm on $D_{1A}$, $D_{1B}$ with labels $K=1$ for sample points in $D_{1A}$ and labels $K=0$ for those in $D_{1B}$, resulting in a function estimate $\wh\theta(x,y)$ of $\theta(x,y)=\mathbb P(K=1|x,y)$ as defined in \eqref{likelihood_ratio}.  To test the significance of the classifier, we use the rank-sum statistic
\begin{equation}\label{eq:R}
	R = \frac{1}{n_2^2}\sum_{i,j\in\mathcal I_2}\indic\left\{\wh\theta(X_i, Y_i)<\wh\theta(X_j', Y_j') \right\}\,.
\end{equation}

If $\wh\theta$ is close to $\theta$ under $H_1$ then \Cref{pro:separation} suggests we should reject $H_0$ if $R$ is too small.  As detailed in the next subsection, combining the two-sample $U$-statistic theory and Stein's method for sparsely dependent random variables, we have the following asymptotic scaling of $R$ under $H_0$:
$$
\mathrm{Var}(\sqrt{n_2}R) \approx \hat\sigma^2
$$
with
\beqr\label{eq:sigmahat}
\wh\sigma^2 = \frac{1}{6} - \frac{2}{n_2}\sum_{i=n_1+1}^n \wh h_1(X_i, Y_i)\wh h_1(X_i', Y_{i}')- \frac{2}{n_2}\sum_{i=n_1+1}^n \wh h_1(X_{i+1}, Y_{i+1})\wh h_1(X_i', Y_i')\,,
\eeqr
where $\wh h_1(x,y)=1/2-\wh F_{2*}(\wh\theta(x,y))$, with $\wh F_{2*}$ the empirical distribution function of $\{\wh\theta(X_i',Y_i'):i\in \mathcal I_2\}$, and using the convention $(X_{n+1},Y_{n+1})=(X_{n_1+1},Y_{n_1+1})$.
Thus we arrive at the following split-permute-classification-test procedure.

\begin{algorithm}[h]
	\caption{Test of independence via classification significance}
	\label{alg1}
	\begin{algorithmic}
		\STATE 1. Input data $D = \{(X_1, Y_1), \dots, (X_n, Y_n)\}$, classificaion algorithm $\mathcal A$.
		\STATE 2.  Split $\{1,\dots,n\}$ into subsets $\calI_1$ and $\calI_2$ to form subsamples $D_{1A}$, $D_{2A}$, and cyclicly permuted subsamples $D_{1B}$, $D_{2B}$ as described above.
		\STATE 3. 
		Apply $\mathcal A$ to $D_{1A}$ and $D_{1B}$ to obtain the estimated class probability function $\wh\theta(\cdot, \cdot)$;
		\STATE 4.  Calculate the $p$-value $\defby\Phi\{\sqrt{n_2}(R-1/2)/\hat\sigma\}$ with $R$, $\hat\sigma$ given by \eqref{eq:R} and \eqref{eq:sigmahat}, where $\Phi(\cdot)$ is the standard normal distribution function.
	\end{algorithmic}
\end{algorithm}

\noindent
{\bf Remark 1: split ratio.} To implement Algorithm \ref{alg1}, one needs to choose the sizes of $\calI_1$ and $\calI_2$. While a large $\calI_1$ will train a more accurate classifier, it also leads to a smaller testing data set $\calI_2$. Thus it is important to balance the trade-off between classification and testing data. In our simulations, we found an equal-split performs very well. Without further notations, we assume $|\calI_1| = |\calI_2|$ throughout the paper.

\noindent
{\bf Remark 2: choice of the classifier.} In principle, our method can work with any classification algorithm $\mathcal A$. However, the classification problem in our method is quite challenging. By construction, each coordinate in the two populations $D_{1A}$, $D_{1B}$ have the same mean value, and the only difference is the dependence structure among the columns.  Therefore, linear methods such as logistic regression cannot perform very well, and nonlinear methods such as support vector machine \ would require a good choice of kernel. In practice, we choose neural networks due to their great flexibility and adaptivity to complex structures in the data.

\section{Theoretical Justifications}\label{sec:theory}
In the split-permute testing procedure described in \Cref{alg1}, both the classifier and two-sample test are obtained using an originally paired subsample together with its cyclicly permuted version.  Therefore the samples are not completely independent and the theoretical properties of the resulting test statistic deserve careful analysis.  We first establish the asymptotic conditional distribution of the test statistic conditioning on a given fitted label probability function $\wh\theta$.  It turns out that the null asymptotic conditional distribution is independent of $\wh\theta$ and asymptotically distribution-free, while the estimated likelihood ratio needs to be better than random guess under the alternative.  We will discuss the performance of classification using the cyclic permuted data in \Cref{sec:classification_accuray}.

\subsection{Asymptotic distribution of test statistic}\label{sec:asym_dist}
Before presenting the theoretical results, we describe some necessary notations. 
Let $F_{1*}(\cdot)$, $F_{2*}(\cdot)$ be the cumulative distribution functions of $\wh\theta(X, Y)$ under $\mathcal P_{X,Y}$ and $\mathcal P_X\times\mathcal P_Y$, respectively.  
Let $\E_*(\cdot)$, $\P_*(\cdot)$, $\cov_*(\cdot)$ and $\var_*(\cdot)$ denote the conditional expectation, probability, covariance and variance given $\wh\theta$ (or equivalently, given the first subsample).  
For $k = 3,4$, define
\beqr\label{a34}
A_k = \frac{6}{n_2^k\var_*(R)^{k/2}}\Big( \sum_{i\in\calI_2}  \E_*\left|F_{2*}\{\wh\theta(X_i, Y_i)\} - \frac{1}{2}\right|^k + \sum_{i\in\calI_2}\E_*\left|F_{1*}\{\wh\theta(X_i', Y_i')\} - \frac{1}{2}\right|^k\Big).
\eeqr

We first derive the asymptotic behavior of the test statistic under $H_0$. The proof begins with decomposing the U-statistic $R$ into its projection $\tilde{R}$ and the remaining term, as detailed in Lemma 1. Specifically, let
\beqrs
\tilde{R} =  \frac{1}{n_2}\sum_{i\in\calI_2}\left[\frac{1}{2} - F_{2*}\{\wh\theta(X_i, Y_i)\} \right] + \frac{1}{n_2}\sum_{i\in\calI_2}\left[ F_{2*}\{\wh\theta(X_i', Y_i')\}  - \frac{1}{2}\right]
\eeqrs
Lemma 1 shows that $R  - \frac{1}{2} = \tilde{R} +O_p(n_2^{-1})$. Then we prove the conditional Berry-Essen bound of $\tilde{R}$ and the unconditional asymptotic normality of $R$. The theoretical results under $H_0$ are summarized in Theorem \ref{thm: null}.

\begin{theorem}\label{thm: null}
	Under $H_0$, assume $(X_i, Y_i)$, $i\in\calI_2$ are i.i.d samples from $\calP_{X,Y}$,  and  $\wh\theta(x, y)$ is a function such that $\wh\theta(X_1,Y_1)$ is continuous, and $F_{2*}\{\wh \theta(X, Y)\} \neq g_1(X)+g_2(Y)$ for any $g_1(\cdot)$ and $g_2(\cdot)$.
	Then
	\beqrs
	\sup_{s\in\mR}\left| \P_*\left(\frac{\sqrt{n_2} \tilde R  }{\sigma_*}\leq s\right) - \Phi(s) \right| \leq c(\sqrt{A_3} +\sqrt{A_4})
	\eeqrs
	where $0\leq c<8$ is a constant, $A_3$ and $A_4$ are defined in (\ref{a34}), and
	\begin{align*}
		\sigma_*^2 :=&  \frac{1}{6} - 2\cov_*\left[F_{2*}\{\wh\theta(X_2, Y_2)\},F_{1*}\{\wh\theta(X_{1}, Y_{2})\} \right]-
		2\cov_*\left[F_{2*}\{\wh\theta(X_1, Y_1)\},F_{1*}\{\wh\theta(X_{1}, Y_{2})\} \right]\,.
	\end{align*}
	Under the additional assumption of $n_2^{1/3}\sigma_*^2\rightarrow\infty$, we have $\hat\sigma/\sigma_*-1=o_{P}(1)$ and the test statistic $\sqrt{n_2}(R - 1/2)/\wh \sigma$ converges in distribution to $N(0, 1)$ as $n_1$ and $n_2\rightarrow\infty$.
\end{theorem}

We discuss the convergence rate and conditions for Theorem \ref{thm: null} in the following remarks.
\begin{remark}
	The right hand side of the Berry-Essen bound in Theorem \ref{thm: null} consists of two terms: $\sqrt{A_3}$ and $\sqrt{A_4}$. Here $\sqrt{A_3}$ is the dominating term, and is of order $n_2^{-1/4}$ when $\sigma_*^2$ is of constant order. We can further improve the bound rate to the classical $n_2^{-1/2}$ and relax the condition on $\sigma_*^2$ to $n_2^{1/2}\sigma_*^2\rightarrow\infty$ by applying Theorem 2.2 of \cite{jirak2016berry}. The cost is a slightly more complicated condition on the constant term in the Berry-Essen bound.
\end{remark}

\begin{remark}
	Conditioning on the estimated probability function $\hat\theta$, our test statistic $R$ is a two-sample $U$-statistic. Its asymptotic normality requires its kernel to be non-degenerate, such that the asymptotic variance $\sigma_*^2>0$.
	This non-degeneracy condition is further equivalent to $F_{2*}\{\wh \theta(X, Y)\}$ cannot be written in the form of $g_1(X)+g_2(Y)$ for any functions $g_1,~g_2$, which is mild because $F_{2*}\{\wh \theta(X, Y)\} = g_1(X)+g_2(Y)$ is equivalent to 1) $g_1(X)+g_2(Y)$ follows $U(0,1)$ and 2) $\wh \theta(X, Y)  = W\{ g_1(X)+g_2(Y)\}$, for some strictly monotone increasing $W: [0,1]\rightarrow[0, 1]$. Common classifiers (logistic regression, random forest, SVM, neural network) can be easily verified to satisfy this non-degeneracy condition.
\end{remark}

\begin{theorem}\label{thm: alt}
	Under $H_1$, assume $(X_i, Y_i)$, $i\in\calI_2$ are i.i.d samples from $\calP_{X,Y}$, 
	and there exists a strictly monotone function $g$ such that 
	\beqr\label{eq:alt_thm_condition}
	\E_*\left|\frac{g(\wh\theta(X', Y'))}{1-{g(\wh\theta(X', Y'))}} - \frac{\theta(X', Y')}{1-\theta(X', Y')}\right| < 1/4-\mu/2 - c
	\eeqr
	holds with probability tending to 1 for some positive constant $c$. Here $\mu = \P \{\theta(X, Y)<\theta(X', Y')\}$, with $(X,Y)$, $(X',Y')$ independently generated from $\mathcal P_{X,Y}$ and $\mathcal P_X\times \mathcal P_Y$ respectively.   
	Then, as $n_1$ and $n_2\rightarrow\infty$,  the test statistic $\sqrt{n_2}(R - 1/2)/\wh \sigma \stackrel{p}{\rightarrow}\infty$.
\end{theorem}

The condition required for the power guaranteee under the alternative is substantially weaker than the asymptotic normality under the null.  This is because we no longer need to lower bound the variance term. 

It is remarkable that we do not need to assume the classifier to be consistent to have valid type-I and type-II error control. The type I error control is automatically guaranteed by the cyclic permutation and holds for arbitrary classifiers, because under $H_0$, $(X_1,Y_1)$ and $(X_1,Y_2)$ have the same distribution and any classifier will not be able to distinguish the two samples.
For the type II error control, equation \eqref{eq:alt_thm_condition} is much weaker than consistency, as it only requires $\hat\theta$ to be close to $\theta$ up to a monotone transform and within some constant error bound.
These properties are especially appealing in practice. For example, many nonparametric tests that rely on kernel density estimations need to carefully choose the kernel bandwidth to guarantee the correct type-I error rate. In our case,  even though the classifier (such as a neural network) may have many tunning parameters to choose from, the test is always valid, and the power is non-trivial whenever the classifier can pick up even only a part of the difference between the joint and product distributions.  

Next, we present a local alternative analysis where the dependence signal changes with the sample size. 	To quantity the signal, we use the likelihood ratio defined in (\ref{likelihood_ratio}). Specifically, consider $(X', Y')$ and $(X'', Y'')$ independently drawn from $\calP_X\times \calP_Y$. We define
\beqr\label{eq: local_quantity}
\delta = \E\left\{ \left|L(X', Y') - L(X'', Y'')\right| \right\}.
\eeqr
By Proposition \ref{pro:separation}, we know that $\delta\asymp d_{tv}(\calP_{X, Y}, \calP_X\times \calP_Y)$. 
Thus $\delta$ measures the distance between the null hypothesis and the local alternative.  	$\E\{L(X', Y') \} = 1$ and $\delta=0$ if and only if $\P\{L(X', Y') =1\}=1$, which is equivalent to $H_0$.    Our local alternative analysis focuses on the case $\delta\rightarrow 0$ as $(n_1,n_2)\rightarrow\infty$.

We introduce extra notation to analyze the local alternative. Let $\mu = \P \{\theta(X, Y)<\theta(X', Y')\}$,  $\mu_* = \P_* \{\wh\theta(X, Y)<\wh\theta(X', Y')\}$ and $F_1(\cdot)$, $F_2(\cdot)$ be the cumulative distribution functions of $\theta(X, Y)$ under $\mathcal P_{X,Y}$ and $\mathcal P_X\times\mathcal P_Y$, respectively. And define
\begin{equation}\label{eq:Rp}
	R' = \frac{1}{n_2^2}\sum_{i,j\in\mathcal I_2}\indic\left\{\theta(X_i, Y_i)<\theta(X_j', Y_j') \right\}\,.
\end{equation}
Based on equation (S1) in  Lemma 1, one can easily calculate the variance for the projection of $\sqrt{n_2}R'$ to be $\sigma_0^2:=\cov(V_1+V_2+V_3,V_2)$, where $V_i = F_{1}\{\theta(X_i', Y_i')\}  - F_{2}\{\theta(X_i, Y_i)\}$.  While $\sigma_0^2$ is complicated and hard to understand, we also define $\sigma^2$ and show that $\sigma_0^2$ is actually sufficient close to $\sigma^2$ under the local alternative hypothesis. Specifically, $\sigma_0^2$ can be approximated by
\beqrs
\sigma^2 = \frac{1}{6} - 2\cov\left[F_2\{\theta(X_2, Y_2)\},F_1\{\theta(X_{1}, Y_{2})\} \right]-
2\cov\left[F_2\{\theta(X_1, Y_1)\},F_1\{\theta(X_{1}, Y_{2})\} \right]\,,
\eeqrs 
because the joint distribution $\mathcal P_{X,Y}$ gets increasingly closer to the product distribution $\mathcal P_X\times\mathcal P_Y$.  For the same reason, $\sigma^2$ further converges to a quantity depending only on the product distribution $\mathcal P_X\times\mathcal P_Y$.  Thus it is reasonable to assume the variance term $\sigma_0^2$ is bounded away from zero in the local asyptotic population sequence:
$$
\sigma_0^2\ge c >0
$$
for some constant $c$ not depending on the sample size.

\begin{theorem}\label{thm: local}
	Under the local alternative with \eqref{eq: local_quantity} for a sequence $\delta=o(1)$, assume $(X_i, Y_i)$, $i\in\calI_2$ are i.i.d samples from $\calP_{X,Y}$, $\theta(\cdot)$ has a continuous distribution under both $\mathcal P_{X,Y}$ and $\mathcal P_X\times\mathcal P_Y$,  $\mu_* - \mu = o_p(n_2^{-1/2})$, $\sigma_0^2\ge c$ for some constant $c>0$, and
	\beqr\label{condition: local}
	\E_*\left[ \left|\indic\{ \wh\theta(X, Y) < \wh\theta(X', Y')\} -  \indic\{\theta(X, Y)<\theta(X', Y')\}\right| \right] = o_p(1).
	\eeqr
	Then
	\beqrs
	\frac{\sqrt{n_2}(R - 1/2)}{\wh \sigma} = Z  -\frac{\sqrt n_2 \delta}{4\sigma} + o_p(1).
	\eeqrs
	where $Z\overset{d}{\rightarrow} N(0, 1)$ as $n_1$ and $n_2$ goes to infinity.
\end{theorem}

As a consequence, when the distance between the local alternative and the null vanishes at the same or a slower rate as $n_2^{-1/2}$,  the limiting distribution of the test statistic under the local alternative becomes a location-shited normal distribution with unit variance.

\begin{remark}
	The conditions in Theorem \ref{thm: local} are stronger than those required in the fixed population versions in Theorems \ref{thm: null} and \ref{thm: alt}. This is because the local alternative hypothesis can be close to the null as fast as $n_2^{-1/2}$ and a more delicate treatment of the estimation error is needed to establish the asymptotic distribution.
	In particular, equation (\ref{condition: local}) typically holds when $\hat\theta$ is a consistent estimate of $\theta$ up to a strictly monotone transform, whereas equation (\ref{eq:alt_thm_condition}) only requires a constant error accracy.  The most stringent condition is $\mu_* - \mu = o_p(n_2^{-1/2})$.
	Let $\Delta_i=\indic\{\wh\theta(X_i, Y_i)<\wh\theta(X_i', Y_i')\} - \indic\{\theta(X_i, Y_i)<\theta(X_i', Y_i')\}$. Then $\mu^*-\mu=\mathbb E_*\Delta_i$.
	If a parametric estimate $\hat\theta$ is used, then typically $\Delta_i= O_P(n_1^{-1/2})$.  So the required condition holds if $n_1\gg n_2$.  In the pratically preferred case of $n_1\asymp n_2$, we  have $\Delta_i\asymp n_1^{-1/2}$, but 
	$\mu_* - \mu=\mathbb E_*\Delta_i$ can still be much smaller than $\Delta_i$ if the random variable $\Delta_i$ is centered around zero and not highly skewed. We also provide a simple numerical example that verifies the condition $\mu_* - \mu = o_p(n_2^{-1/2})$ in section C of the supplement. 
\end{remark}

\subsection{Classification accuracy under cyclic permutation}\label{sec:classification_accuray}
A remaining question regarding the procedure is whether we have any formal guarantees on the estimator $\wh\theta$ because it is not obtained from a standard independent two sample data, but from only a single sample, with the second sample obtained from cyclically permuting the original sample.  The quality of such $\wh\theta$ would depend on the particular form of the estimator and the data distribution.  Intuitively, the weak dependence caused by the cyclic permutation among the sample points should be negligible, and the resulting estimator would behave similarly to those obtained from genuine independent two-sample data. Here for an illustrative purpose, we prove the consistency of the classifier obtained  under (1) a classical low-dimensional M-estimation and (2) a high-dimensional lasso-based sparse regression. Note that both the low dimensional and high dimensional models are trained on the first subset of data $\calI_1$ with $n_1=|\calI_1|$. For notation simplicity of the consistency analysis, we drop the subscript and use $n$ instead of $n_1$ only in section \ref{sec:classification_accuray} and its proofs.

\subsubsection{Low-dimensional M-estimation}
Define the objective function as
\beqrs
M(X, Y, X', Y'; \beta) \defby M_1(X, Y;\beta) + M_2(X', Y'; \beta),
\eeqrs
where $\beta\in\mR^p$ is the unknown parameter in the classifier. Here $(X,Y)$ and $(X',Y')$ are independent realizations from $\mathcal P_{X,Y}$ and $\mathcal P_X\times \mathcal P_Y$, respectively.  We use $\mathcal P$ to denote the joint distribution of $(X,Y,X',Y')$.  Then the objective function is $\E\{M(X, Y, X', Y'; \beta) \}$, where the expectation is taken with respect to $\calP$. For example, we can choose $M_1(x,y;\beta)=-\ell_1(x,y;\beta)$ and  $M_2(x',y';\beta)=-\ell_2(x',y';\beta)$, with some class-specific binary classification loss functions $\ell_1(\cdot)$, $\ell_2(\cdot)$, such as the hinge loss or the logistic loss function. 
Let $\beta_0$ be the true parameter that maximizes the objective function. Using the cyclicly permuted data, the classifier is trained by maximizing the empirical criterion function
\beqrs
M_{n}(\beta)\defby\frac{1}{n}\sum_{i=1}^n\left\{  M_1(X_i, Y_i; \beta) + M_2(X_i', Y_i'; \beta) \right\},
\eeqrs
Denote $\wh\beta_n$ as the maximizer of  $M_{n}(\beta)$. The consistency of  $\wh\beta_n$ is established in Theorem \ref{Mestimation}.

\begin{theorem}\label{Mestimation}
	Suppose  $(X_i, Y_i)$, $i\in\calI$, are independent observations drawn from $\mathcal P_{X,Y}$. Let $\calM = \{ M(x, y, x', y'; \beta) :  \beta\in \mathcal{B}\}$ be a class of measurable functions such that $N_{[\mkern5mu ]} (\ep, \calM, \calP)<\infty$ for every $\ep>0$, and $\E\left[\{  M(X, Y, X', Y'; \beta) \}^4\right]<\infty$. Suppose the true parameter $\beta_0$ is identifiable, i.e.,
	\beqrs
	\sup_{\beta: d(\beta,\beta_0)\geq\ep} \E\{M(X, Y, X', Y'; \beta) \} < \E\{M(X, Y, X', Y'; \beta_0) \},
	\eeqrs		
	where $d(\cdot)$ is a distance measure.	Then any sequence of estimators $\wh\beta_n$ with $M_n(\wh\beta_n)\geq M_n(\beta_0) - o_p(1)$ converges in probability to $\beta_0$.
\end{theorem}

The condition for the class of objective functions $N_{[\mkern5mu ]} (\ep, \calM, L_1)<\infty$ is relatively standard for classical M-esimators. See \cite{van2000asymptotic} for several examples.  

A detailed proof of Theorem \ref{Mestimation} is given in Appendix E.11. The key of our proof is a strong law of large numbers resulting in the dependent data, proved by carefully decomposing the variance of the sum of dependent variables and applying the Borel-Cantelli lemma. Then we are able to show the uniform consistency of $M_n(\beta)$ in Lemma 8, which further implies consistency of $\wh\beta_n$ when combined with standard empirical processes and M-estimation results \citep{van2000asymptotic}.

\subsubsection{High dimensional regression}

We consider a scenario where the dimension can be large, compared to the sample size. Denote the dimension of $X$ as $d_1$ and the dimension of $Y$ as $d_2$.  Let $d = d_1+d_2$. Denote $Z = (X, Y)\sim\calP_{X,Y}$ and $Z' = (X', Y')\sim\calP_X\times\calP_Y$. We define 
\beqrs
g(z) = \frac{d \calP_{X,Y} }{d\calP_{X,Y} + d\calP_X\times\calP_Y}(z)
\eeqrs
Our goal is to estimate $g(z)$ while keeping in mind that $d_1$ and $d_2$ may be comparable or lager than the sample size $n$.  
In order to cope with high dimensionality, we assume that $g(z)$ has a sparse representation in a certain basis.  This would be particularly reasonable, for example, when only a few coordinates of $X$ and $Y$ are dependent.  Assume that $s_1$ out of $d$ coordinates of $Z=(X,Y)$ are dependent. Then $g(z)$ is essentially a function of $s_1$ variables instead of $d$ variables.  Consider all the $s_1$-way combinations of coordinates of $Z$, and use $K_n$ basis for each combination.
Specifically, let $\xi_1,\xi_2,...$ be a basis function of the $L_2$ space $\mathbb R^{s_1}\mapsto \mathbb R$. Let $K_n$ be a slowly growing number. We consider the basis
$\xi(Z) = \{\xi_k(Z_{j_1},Z_{j_2},...,Z_{j_{s_1}}):1\le k\le K_n, 1\le j_1<j_2<...<j_{s_1}\}$  with dimensionality $m\propto d^{s_1} K_n$, and assume that the function $g(z)=\xi(z)^T\beta^*$ with $\|\beta^*\|_0\le s_2\ll n$.  Such a hard sparsity assumption makes the presentation simpler and can be relaxed using a standard oracle-inequality argument.

Our starting point is that the function $g$ is the minimizer of the following problem
\begin{equation*}\label{eq:g_minimize}
	\min_h\E\left[(1-h(Z))^2 + (0-h(Z'))^2  \right]\,,
\end{equation*}
since we associated a label $K=1$ ($K=0$) for each sample point from $\mathcal P_{X,Y}$ ($\mathcal P_X\times \mathcal P_Y$). 
As a result, under the assumed basis expansion and sparse representation of $g$, $\beta^*$ is the minimizer of the problem
\beqr\label{hd_objective}
\min_{\beta} \beta^T\Gamma\beta - 2\gamma^T\beta,
\eeqr
where
\beqrs
\Gamma = \E\xi(Z)\xi(Z)^T +\E\xi(Z')\xi(Z')^T, \quad \gamma = \E\left[\xi(Z)\right],
\eeqrs

Now consider the empirical version with cyclic permuted $Z_1',\dots,Z_n'$, we estimate $\wh\beta$ by optimizing the regularized quadratic form
\beqr\label{eq: obj}
\min_{\beta} \beta^T\wh\Gamma\beta - 2\wh\gamma^T\beta + \lambda \|\beta\|_1,
\eeqr
Denote $\Xi_Z= (\xi(Z_1)^T,\dots, \xi(Z_n)^T)^T$ and $\Xi_{Z'}= (\xi(Z_1')^T,\dots, \xi(Z_n')^T)^T$, $\Xi = (\Xi_Z, \Xi_{Z'})$, then
\beqrs
\wh\Gamma = n^{-1}\Xi^T\Xi,\quad  \wh\gamma = n^{-1}\Xi_Z^T\mathbf{1}_{n\times1}. 
\eeqrs
Let $\mathcal{G}=  (g(Z_1)^T,\dots, g(Z_n)^T, g(Z_1')^T,\dots, g(Z_n')^T)^T\in\mR^{2n\times 1}$. 
Define the set $\calC_\alpha(S) = \{\Delta\in\mR^m: \|\Delta_{S^c}\|_1\leq\alpha\|\Delta_S\|_1 \}$. We assume the matrix $\Xi$ satisfies the {\it restricted eigenvalue} (RE) condition over $S$ with parameters $(\kappa, \alpha)$ if
\beqr\label{RE_condition}
\frac{1}{n}\|\Xi\Delta\|_2^2\geq \kappa \|\Delta\|_2^2, \quad \mbox{for all}\quad \Delta\in\calC_\alpha(S). 
\eeqr
We also define the residual with respect to the minimization problem (\ref{hd_objective}).  Note that the response vector is $(1,\dots, 1, 0\dots, 0)^T\in\mR^{2n}$ and the design matrix is $\Xi$, with parameter $\beta^*$. Thus we let $w_1(z) = 1 - \xi(z)^T\beta^*$ with $z = Z_1,\dots, Z_n$, and let $w_0(z) = -\xi(z)^T\beta^*$ with $z = Z_1', \dots, Z_n'$. Denote $w = (w_1(Z_1),\dots, w_n(Z_n), w_0(Z_1'),\dots, w_0(Z_n'))^T$.  

\begin{theorem}\label{thm: hd_theory}
	Assume that $\beta^*$ is supported on a subset $S\subset\{1,2,\dots, m\}$ with $|S| = s_2$, and each basis function is bounded on $[-B, B]$. Further assume the matrix $\Xi$ satisfies the restricted eigenvalue condition (\ref{RE_condition}) with parameters $(\kappa, 3)$, and $\lambda$ satisfies that $\lambda\geq4\|\Xi w/n \|_{\infty}$. Then the solution to the optimization problem (\ref{eq: obj}) satisfies
	\beqrs
	\|\wh\beta - \beta^*\|_2\leq  \frac{3}{2\kappa} \sqrt{s_2}\lambda.
	\eeqrs
	In particular, when taking $\lambda = C\sqrt{\log m /n}$, we have 
	$\|\wh\beta - \beta^*\|_2\leq C\sqrt{s_2  \log m /n}$
	with probability no less than $1-m^{-1}$ for some constant $C$ depending only on $\kappa$  and $B$.
\end{theorem}

The proof of Theorem \ref{thm: hd_theory} is given in the supplement. We can also relax the hard sparsity assumption on $\beta$ and use the oracle inequality version of the proof (Theorem 7.19 in \cite{wainwright2019high}) to prove the finite bound on $\wh\beta$.  The restricted eigenvalue condition is a standard one in the lasso literature. Here we directly assume the random design matrix $\Xi$ satisfies a restricted eigenvalue condition, which can hold with high probability if the population version $\Gamma$ satisfies the same condition with slightly different constants.  Recall that $m\propto d^{s_1} K_n$. Thus the error bound for $\beta$ is of order $\sqrt{s_1 s_2} \sqrt{\log d/n}$. When assuming $s_1$ and $s_2$ are constants, the dimension of the data is allowed to grow exponentially with the sample size.

\section{Numerical Validation}

In this section, we conduct numerical simulations to illustrate the performance of our method. For brevity, we will focus on the more challenging and interesting cases where both $X$ and $Y$ are high dimensional, and the dependence signal is sparse.  Specifically, we assume only the first element in $X$ and $Y$ are related: $Y_1 = a\times g(X_1) + \ep$, where the signal $a$ varies from $0$ to $1$. $Y_2, \dots, Y_{d_2}$, $X_1, X_2,\dots, X_{d_1}$, and $\ep$ all follow $N(0, 1)$ and are independent. The following models are considered:

\begin{itemize}
	\item M1:  $Y_1 = a\times X_1 + \ep$;
	\item M2:  $Y_1 = a\times \sin(X_1) + \ep$;
	\item M3:  $Y_1 = a\times \exp(X_1) + \ep$;
	\item M4:  $Y_1 = a\times\left\{ \indic{(X_1<0)} N(1, 1)  + \indic{(X_1>0)} N(-1, 1)   \right\}+ \ep$;
	\item M5: $Y_1 = a\times \log(4X_1^2)+\ep$;
	\item M6: $Y_1 = a\times 5\sqrt{|X_1|}+\ep$;
\end{itemize}

Our simulation models are similar to a variety of models that have been considered in the literature, though mostly in a less challenging case where $X$ and $Y$ are both low dimensional. For example, 
(M1) is one of the most popular models and have been considered in \cite{szekely2007measuring, huo2016fast, shi2020distribution, deb2021multivariate}, etc. Functional transformations similar as (M2) and (M3) have been considered in \cite{zhu2017projection} and \cite{zhu2020distance}. (M4) is the mixture  model and was used in \cite{heller2012consistent, biswas2016some,  deb2021multivariate}.
(M5) was previously used in \cite{szekely2007measuring} and \cite{deb2021multivariate}. (M6) has also been considered in \cite{huo2016fast} and \cite{zhu2020distance}.

As mentioned in the previous section, we choose the neural network to train the classifier and implement it by \texttt{TensorFlow} \citep{tensorflow2015whitepaper}. 
We use three layers of nodes (one input layer, one hidden layer, and one output layer). The number of nodes for the input layer is the dimension of the training data, and the number of nodes in the hidden layer is proportional to the data dimension. The output layer only contains one node since the task is binary classification. We further enforce the hidden layer with $L_1$ kernel regularization with regularization parameter varying from $10^{-4}$ to $10^{-3}$. The dropout rate \citep{srivastava2014dropout} for the hidden nodes also varies from 0.1 to 0.3. Details about the algorithm can be found in the supplemental code written in \texttt{python}. 

We compare the proposed method with other popular statistical independence tests, including the distance correlation (\cite{szekely2007measuring}, denoted by ``DC"), ranks of distance test  (\cite{heller2012consistent}, denoted by ``HHG"), and mutual information (\cite{berrett2019nonparametric}, denoted by ``MI"). Those competing tests are implemented with popular R packages: energy, HHG, and IndepTest, respectively. Because the proposed method is a {\bf C}ircularly {\bf P}ermuted {\bf C}lassification based independence test, we name it the {\it CPC} test.

\begin{figure}
	\centering
	\includegraphics[scale=0.6]{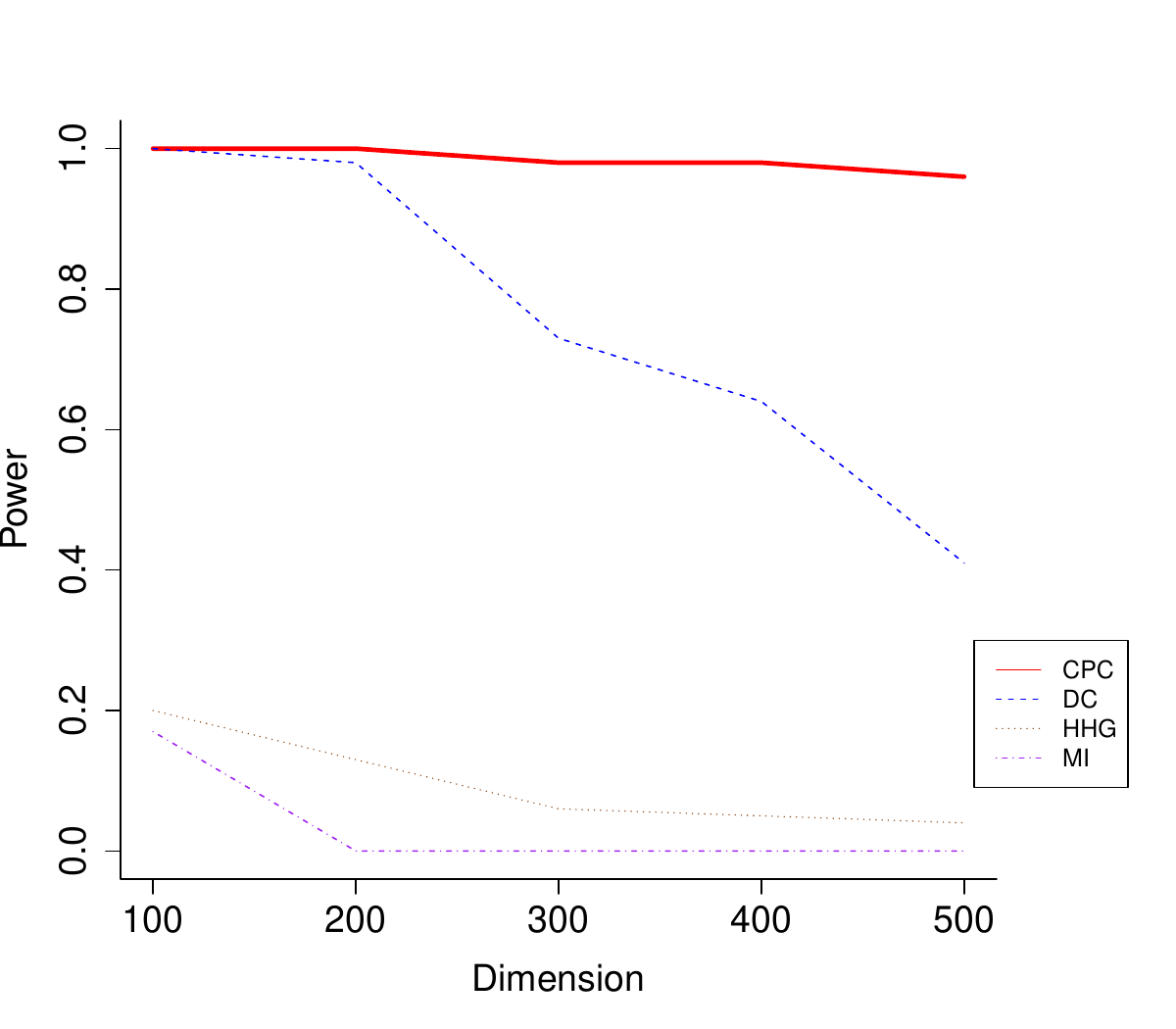}
	\caption{ The  power versus dimension of the proposed test (``CPC") compared with distance correlation (``DC"), ranks of distance test (``HHG"), and mutual information (``MI") when $n=1000$, $\alpha=0.05$.}\label{hd_power}
\end{figure}

We first look into the high dimensional effect on the independence tests by considering the linear model (M1), where $a$ is set to be 1. The performance of the tests when the dimension increases are summarized in Figure \ref{hd_power}. For the proposed method, it can detect the sparse dependence even when the dimension increases up to 500. The main reason is that we implement the $L_1$ penalization for the hidden layer, which greatly eliminates the noise in the data and preserves the desired sparse dependence signal. For comparison, the HHG and MI method suffer significantly from high dimensionality, while the distance correlation has surprisingly high power when the dimension $d_1$ and $d_2$ are less than 200, but its power still decreases dramatically when the dimension further increases.

\begin{figure}
	\centering
	\begin{subfigure}[b]{0.45\textwidth}
		\centering
		\includegraphics[scale=0.37]{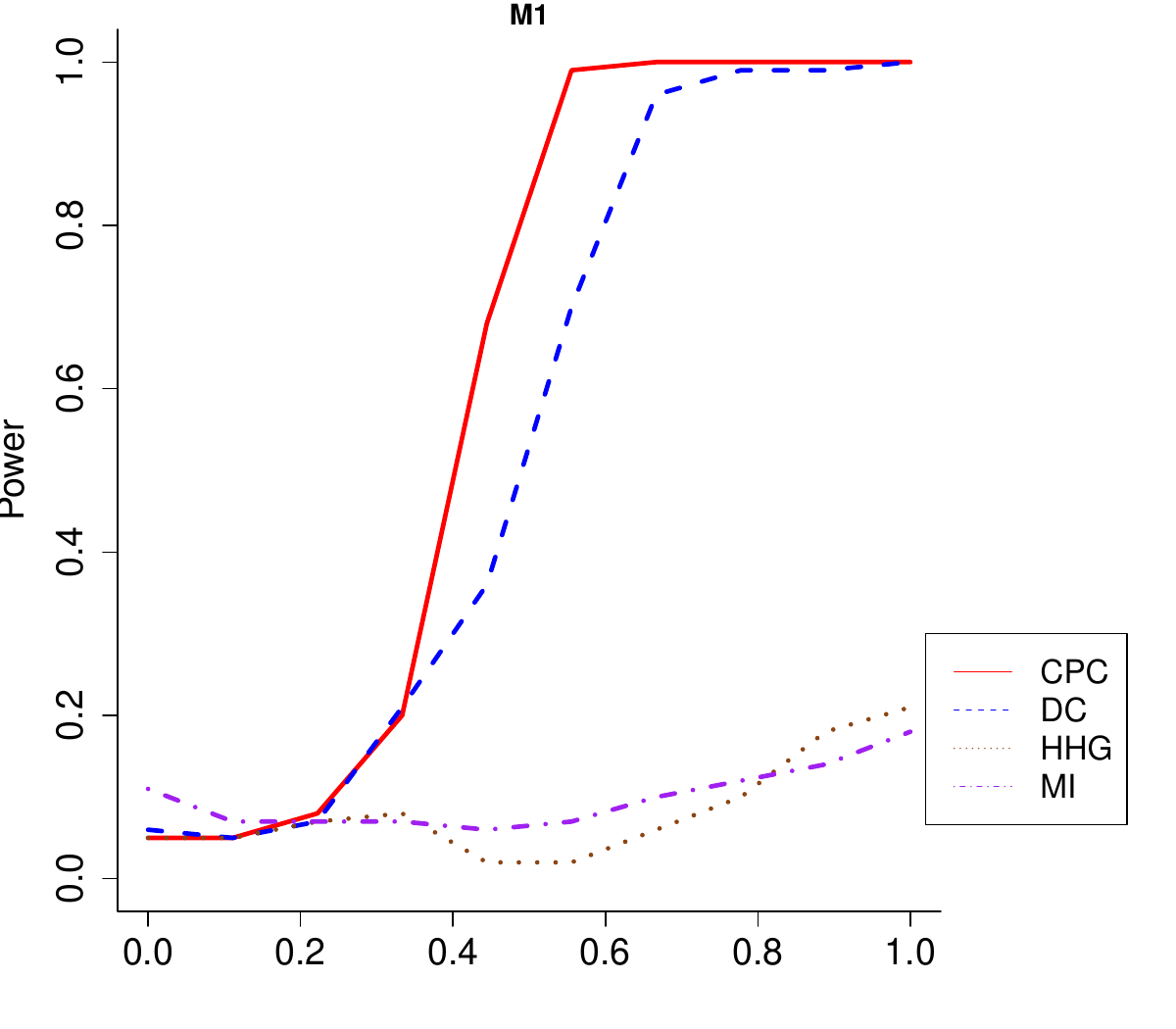}
	\end{subfigure}
	\hfill
	\begin{subfigure}[b]{0.45\textwidth}
		\centering
		\includegraphics[scale=0.37]{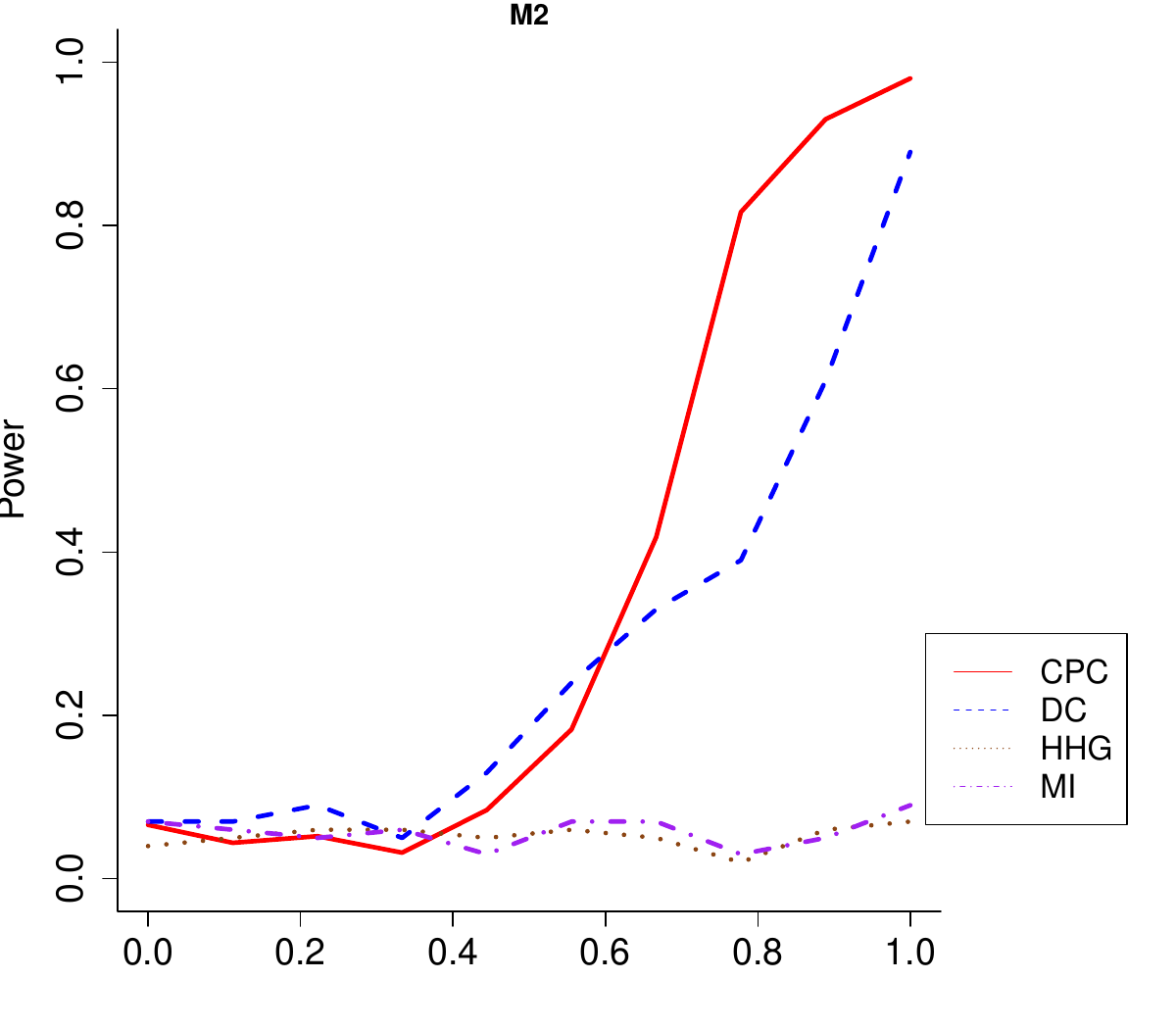}
	\end{subfigure}
	\centering
	\begin{subfigure}[b]{0.45\textwidth}
		\centering
		\includegraphics[scale=0.37]{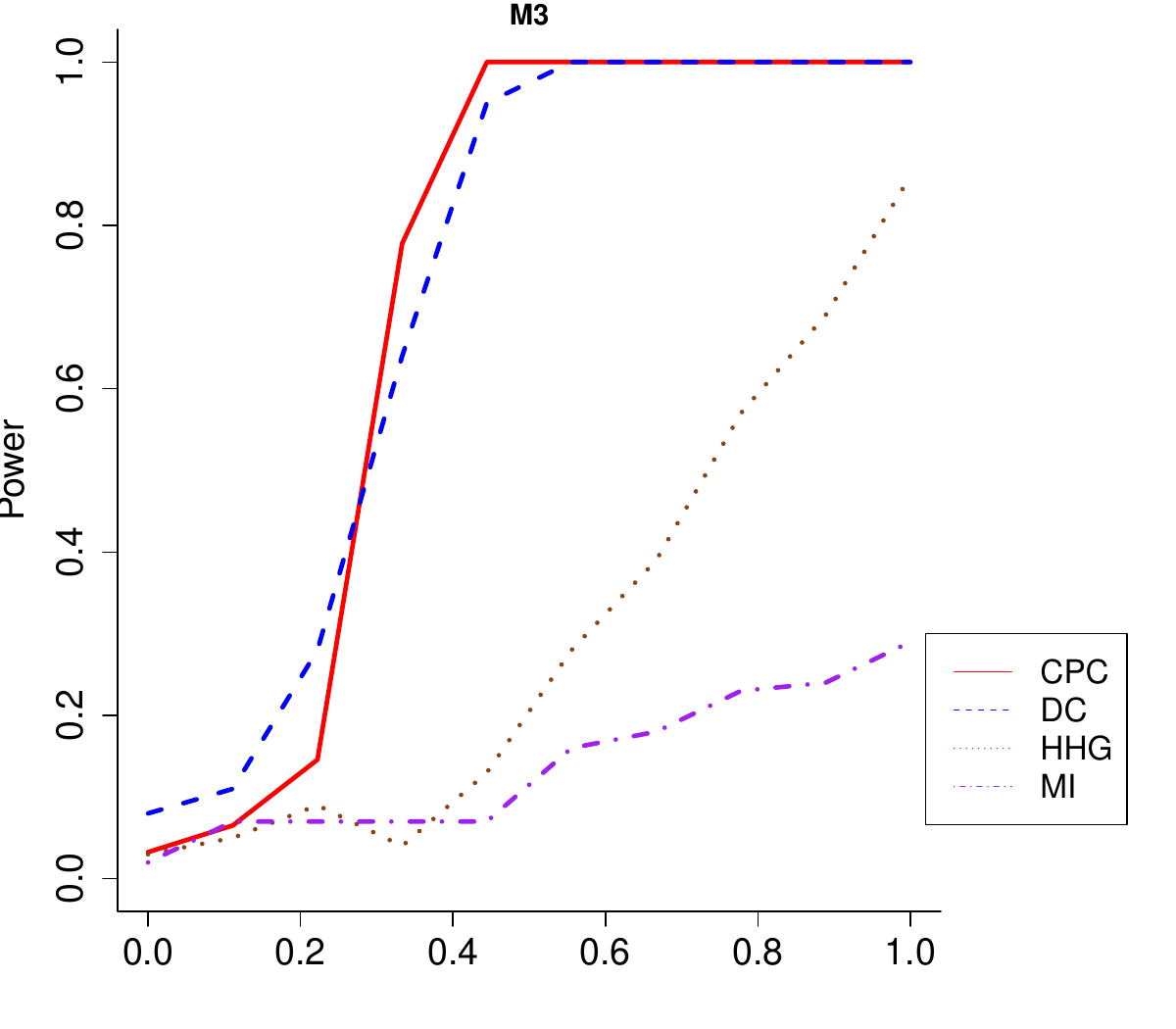}
	\end{subfigure}
	\hfill
	\begin{subfigure}[b]{0.45\textwidth}
		\centering
		\includegraphics[scale=0.37]{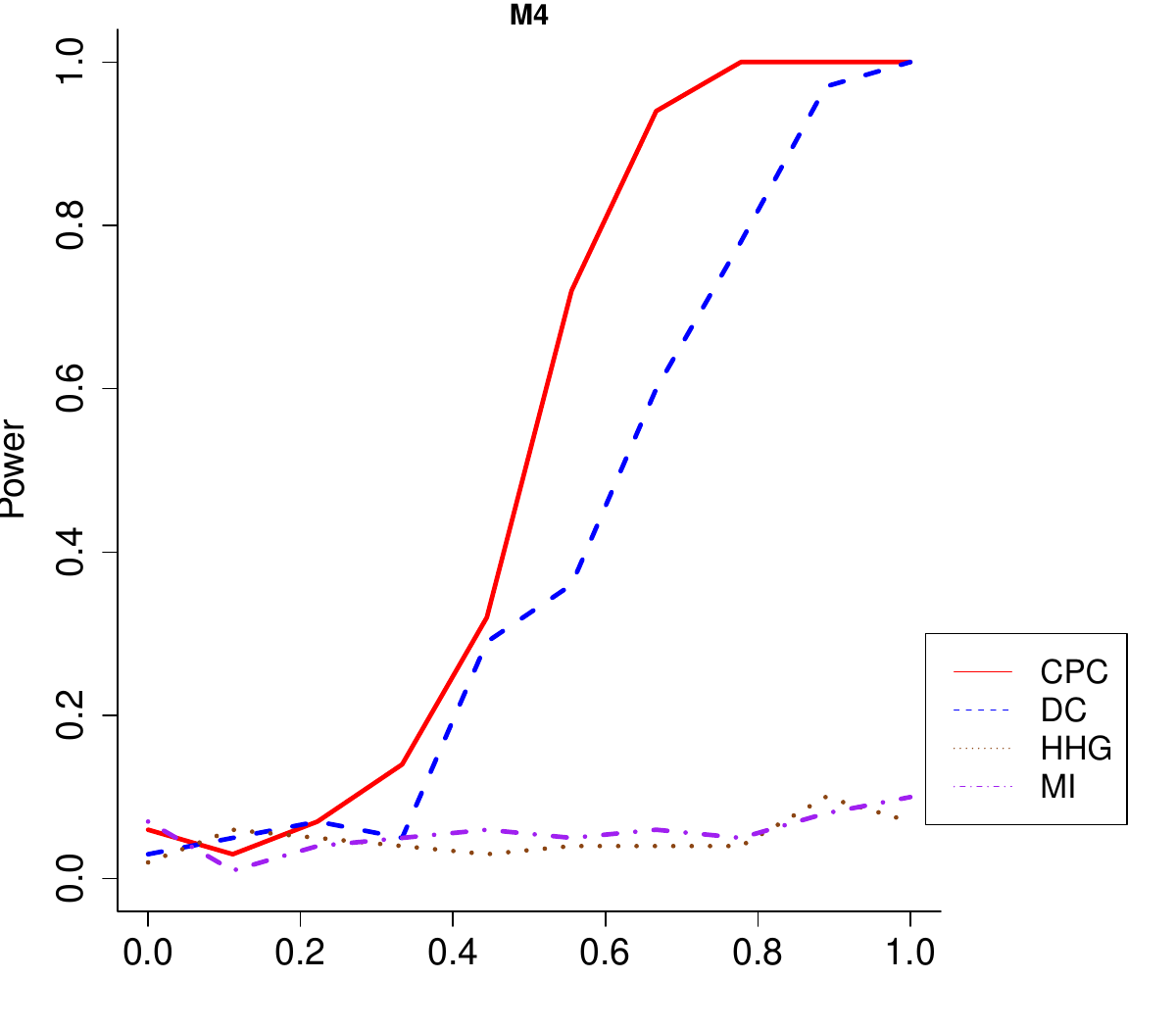}
	\end{subfigure}
	\begin{subfigure}[b]{0.45\textwidth}
		\centering
		\includegraphics[scale=0.37]{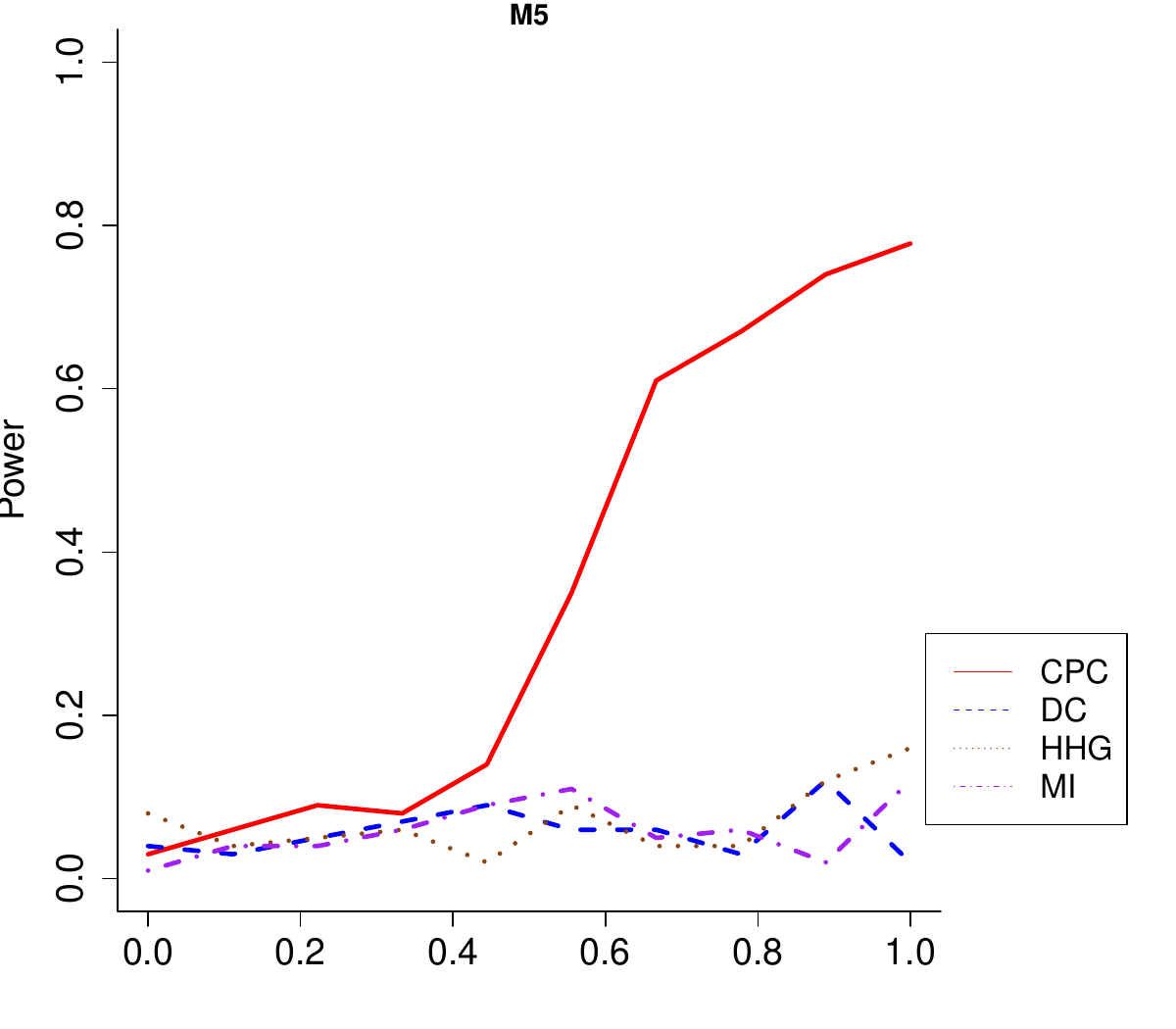}
	\end{subfigure}
	\hfill
	\begin{subfigure}[b]{0.45\textwidth}
		\centering
		\includegraphics[scale=0.37]{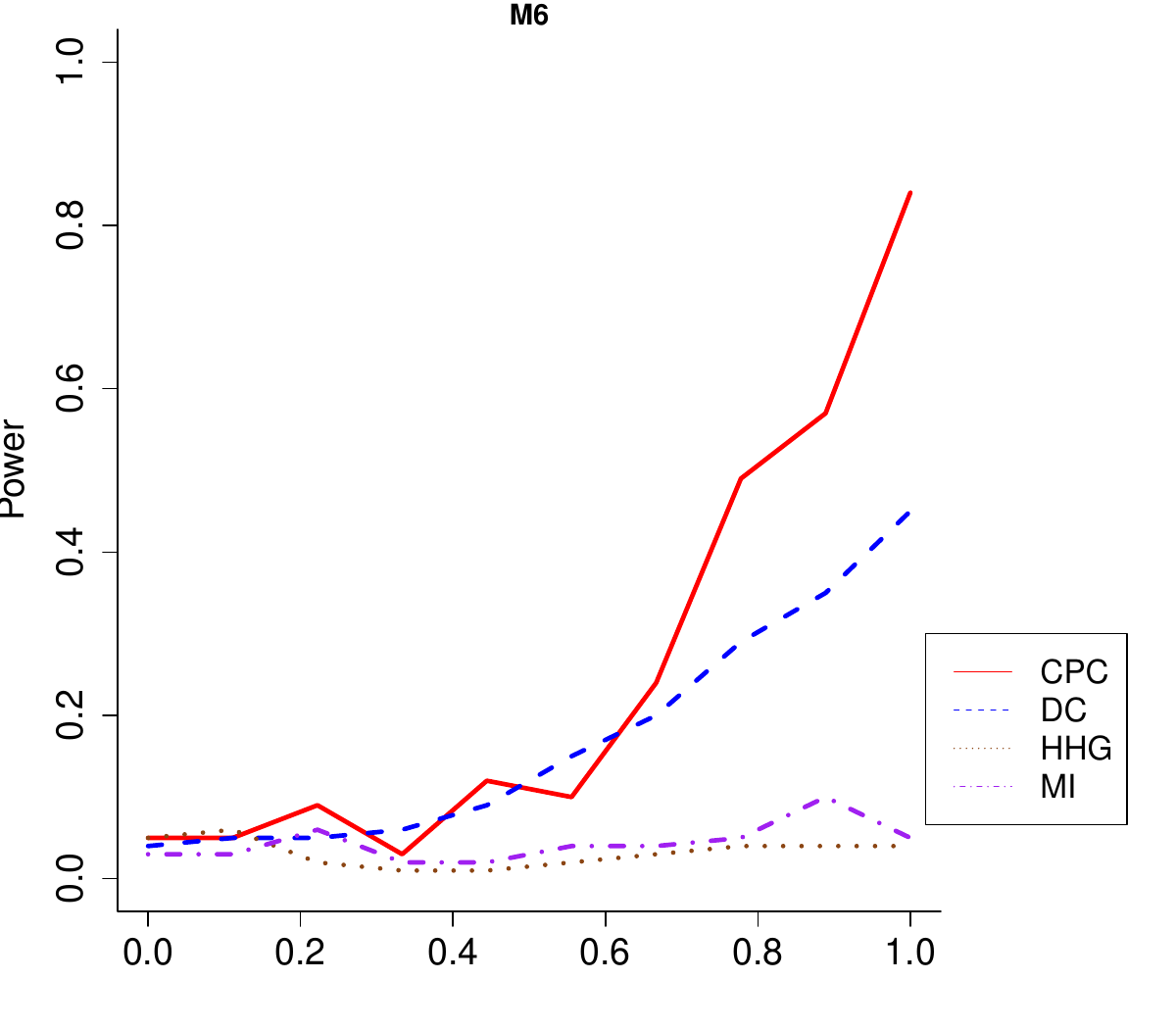}
	\end{subfigure}
	\caption{ The increasing power versus the signal $a$ of the proposed test (``CPC") compared with distance correlation (``DC"), ranks of distance test (``HHG"), and mutual information (``MI") when $n=1000$, $d_1=d_2=100$, $\alpha=0.05$. }\label{power_plot1}
\end{figure}

Next, we focus on fixed dimension $d_1 = d_2 = 100$ to ensure a relatively fair comparison, because otherwise, all current methods tend to have inferior power. We report the performance of all six models  (M1) - (M6)  in Figure \ref{power_plot1}, with sample size $n = 1000$ and significant level $\alpha=0.05$. Additional simulation when $\alpha = 0.1$ and $0.01$ are given in the supplementary material. All results are averaged over 1000 repetitions.   As expected, the proposed test has increasing power as the signal becomes stronger, with correct type-I error under the null hypothesis and high power when the signal exceeds a certain threshold. It performs particularly well for (M5), where all other tests have very low power even when the signal increases. The distance correlation also has considerable power, especially when the signal is strong and the dependence relationship is linear. The ranks of distance test and mutual information do not suit the high dimensional setting and have very low powers for almost all settings. 

While existing tests based on sample splitting tend to cause nonignorable power loss in practice\citep{wasserman2020universal, kim2020dimension}, this phenomenon is weakening in our test. In the simulations, the newly proposed test outperforms other tests that use the whole dataset. This is because half of the data is ``invested" to find the most promising dimension reduction directions, and improves power performance under $H_1$.  

Lastly, we compare the computing time of the tests. We still use the linear model in (M1) for simplicity. We restrict the computation memory to be 16 GB and compute the average computing time for one run of the test based on 1000 repetitions. Two settings are considered: 1) the sample size $n$ is fixed to be 1000, and dimension $d_1=d_2$ linearly increase from 100 to 500. 2) the dimension $d_1=d_2$ are fixed to be 100, and the sample size $n$ increases from 1000 to 5000. The time costs measured in minutes are reported in Tables \ref{time1} and \ref{time2}, respectively. We used permutation tests for distance correlation, HHG and mutual information to obtain $p$-values and the permutation replicate is set to be 200.  We observe that the computation time of the proposed test almost grows linearly with the dimension and sample size. For distance correlation, HHG and mutual information, the computation costs grow linearly with the dimension but grow at least quadratically with the sample size. The HHG method exceeds the memory constraint (16GB) when the sample size $n$ is larger than 2000, and we are unable to obtain its corresponding computation times in Table \ref{time2}. In general, the proposed test is much faster compared with other methods for large-scale data sets. Lastly, we only used regular CPU cores for the entire simulation. The computing time for our test can be further reduced when using advanced GPU cores.

\begin{table}[h]
	\centering
	\caption{The average computing time measured in minutes of the proposed test and distance correlation, rank of distance test, and mutual information when $d_1$ increases from 100 to 500. $d_2=d_1$, $n=1000$. }
	\begin{tabular}{ccccc}
		\hline
		$d_1, d_2$ & CPC & DC & HHG & MI \\ \hline
		100 & 0.025 & 0.009 & 0.105 & 0.425 \\  
		200 & 0.052 & 0.011 & 0.107 & 0.765 \\   
		300 & 0.086 & 0.015 & 0.108 & 1.104 \\   
		400 & 0.138 & 0.017 & 0.113 & 1.448 \\  
		500 & 0.144 & 0.020 & 0.122 & 1.837 \\  \hline
	\end{tabular}\label{time1}
\end{table}

\begin{table}[h]
	\centering
	\caption{The average computing time measured in minutes of the proposed test and distance correlation, rank of distance test, and mutual information when $n$ increases from 1000 to 5000. $d_1=d_2 = 100$. }
	\begin{tabular}{ccccc}
		\hline
		$n$ & CPC & DC & HHG & MI \\ \hline
		1000 & 0.023 & 0.009 & 0.105 & 0.417 \\  
		2000 & 0.040 & 0.046 & 0.470 & 1.785 \\  
		3000 & 0.055 & 0.099 & --  & 3.676 \\   
		4000 & 0.086 & 0.150 & --  & 6.499 \\   
		5000 & 0.086 & 0.201 & --  & 9.849 \\   \hline
	\end{tabular}\label{time2}
\end{table}

\section{Application to Single Cell Data}

The analysis of single cell sequencing data has fueled much discovery and innovation over recent years \citep{kulkarni2019beyond}, and recent advances in multimodal omics promise further progress. 
In this section, we apply the proposed test to a single cell dataset consisting of measurements of Peripheral blood mononuclear cells (PBMCs), publicly available on the 10X Genomics website \citep{pbmc}. The data contain measurements of ATAC-seq and RNA-seq in 11,898 cells, and we are interested in testing whether the two modes of measurement are independent. It has been widely assumed that ATAC-seq and RNA-seq are dependent because ATAC-seq identifies open chromatin sites that are available for transcription. For example, \cite{eltager2021scmoc} proposed to identify cell clusters using the co-measurements of RNA-seq and ATAC-seq from the same cell. In this section, we aim to provide solid statistical evidence for the dependence relationship among the two random vectors.

Each record in the dataset corresponds to a single cell. 	We perform quality control on these data before analysis.  The RNA-seq data initially consists of a vector of counts that we pre-process following the Seurat 3.0 pipeline \citep{stuart2019integrative}.  We retain cells that have counts from 50 - 10,000 genes to exclude almost empty and noisy cells. We set {\it minimum cells per gene} to be 1 to remove genes that are detected in cells less than this threshold. RNA-seq counts are then normalized by dividing each count by the total count for each cell and then scaling up to counts per million.  The ATAC-seq data is also derived from counts, however, because these fragments are distributed across the entire genome, the data were pre-processed to identify peaks, which are clusters of fragments that were inferred to indicate a single region of open chromatin; all of the fragments in the locality of a peak are counted and attributed to the peak location \citep{yan2020reads}. We retain cells whose peaks include from 50 to 15,000 counts. The {\it minimum cells per peak} is set as 1.  Peak counts are normalized by dividing each count by the total count for each cell and then scaling up to counts per million.

Overall 11,188 cells passed the quality control for both RNA-seq and ATAC-seq. The dimension of the RNA-seq data is 29,717  genes, for which only 6.35\% of the entries in the data matrix has non-zero values. For the ATAC-seq data, the dimension is 143,887 peaks and only 5.66\% entries have non-zero values. To achieve fast computation, we store the data in a sparse matrix and run the proposed algorithm and other competing algorithms in python and R, respectively. However, the distance correlation, HHG, and mutual information all reported errors in the algorithm because of exceeding the memory constraint of 16GB. It suggests that some substantial adaptations may be necessary to apply these existing tests of independence that are unsuitable for such high dimensional sparse datasets. For the proposed method, we use the neural network with 3 layers, where the hidden layer contains 2000 nodes. We only used CPU cores to train the algorithm, and it takes about 13.89 minutes to run the test. The test statistic is $-80.95$ and the corresponding $p$-value is practically 0. This strongly confirms that the RNA-seq and ATAC-seq are indeed dependent on each other.

\section{Discussion}

In this paper, we proposed a general framework for independence testing that is powerful in detecting sparse dependence signals in high dimensional data. We borrow the strength from the most powerful classification tools, such as neural networks, to boost power when the dependence between $X$ and $Y$ is sparse and weak. The proposed test statistic has a standard normal asymptotic distribution when the sample size is large.  In addition to such a distribution-free asymptotic null distribution, the new test has several advantages over existing works in both power performance and computing efficiency. We apply the new test to a single cell data set and confirmed a widely believed important hypothesis in the multimodal omics literature.

There are several potential directions to follow up. The idea in this paper can be readily applied in other related testing problems, including the test of mutual independence and the test of conditional independence \citep{cai2022distribution}, as well as related extensions in causal discovery \citep{cai2022causal} or high dimensional modelling \citep{tong2022model, cai2022asset}. By constructing two samples that have the same distribution under $H_0$ but different distributions under $H_1$, one can always transform those tests into a classification problem. Another interesting and unsolved problem is how to avoid the power loss caused by data splitting. One may switch the role of $\calI_1$ and $\calI_2$ and obtain another test statistic and $p$-value, which is dependent on the original one.   Another choice is to perform multiple sample splitting and obtain a sequence of test statistics and $p$-values, which are statistically dependent. Existing methods such as Cauchy combination test \citep{liu2020cauchy, cai2022modelfree} and averaging $p$-values \citep{vovk2020combining} could be applied to combine the results under cerntain restricting conditions. It will be very rewarding to study how to efficiently combine those dependent statistics and $p$-values in high dimensional independence testing problems.

\newpage
	\appendix

\section{Additional Simulations}

\begin{figure}
	\centering
	\begin{subfigure}[b]{0.45\textwidth}
		\centering
		\includegraphics[scale=0.37]{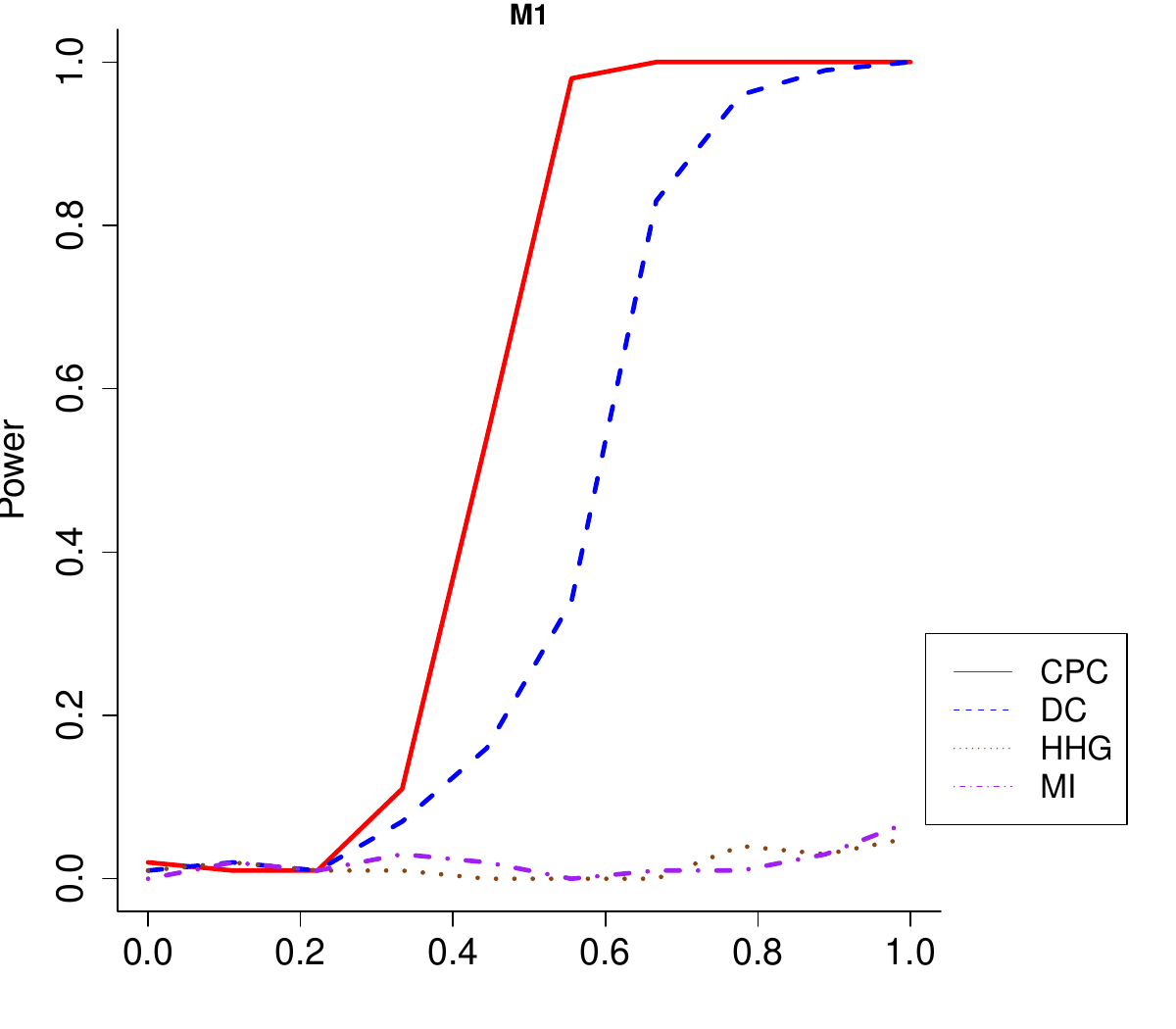}
	\end{subfigure}
	\hfill
	\begin{subfigure}[b]{0.45\textwidth}
		\centering
		\includegraphics[scale=0.37]{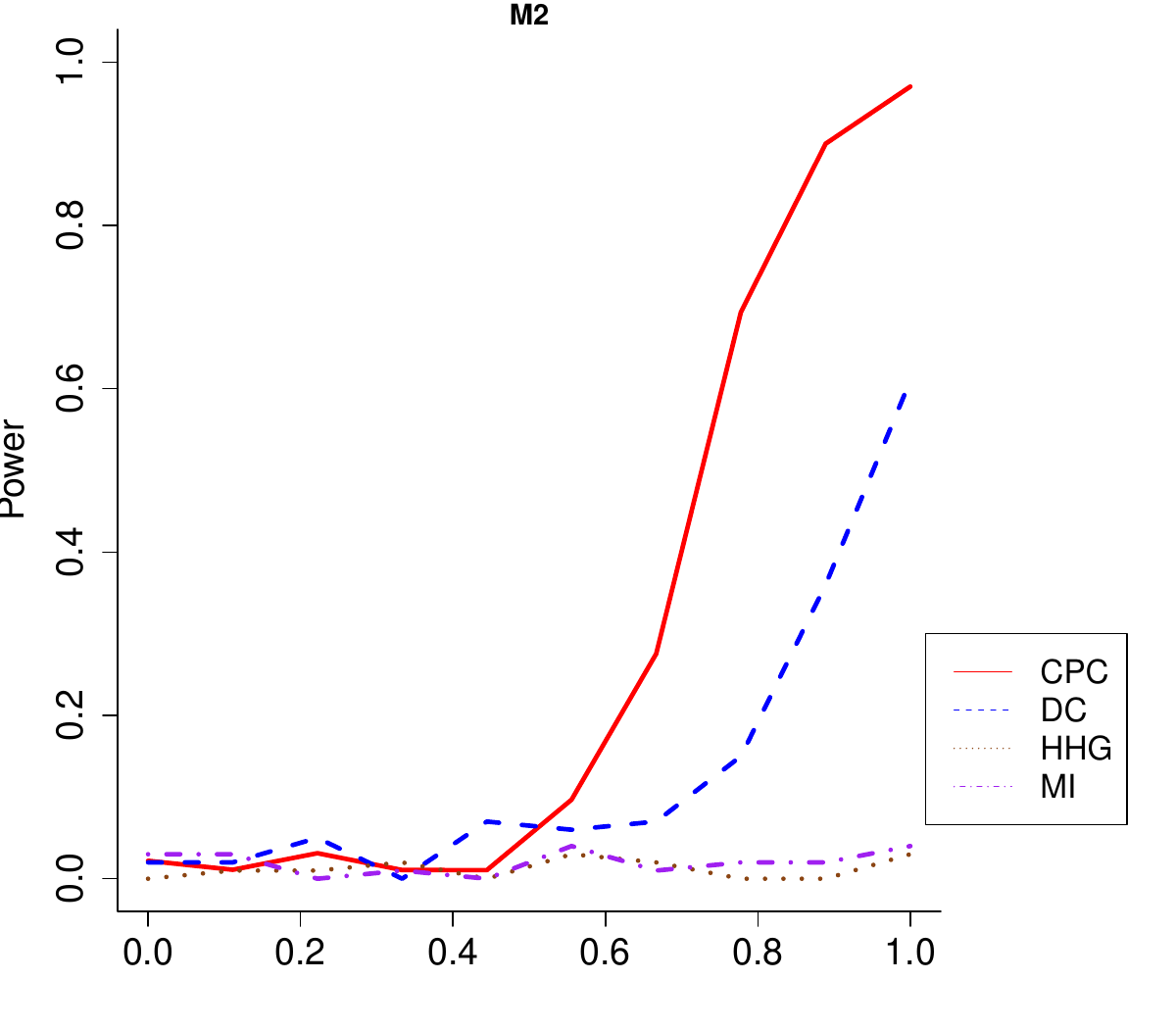}
	\end{subfigure}
	\centering
	\begin{subfigure}[b]{0.45\textwidth}
		\centering
		\includegraphics[scale=0.37]{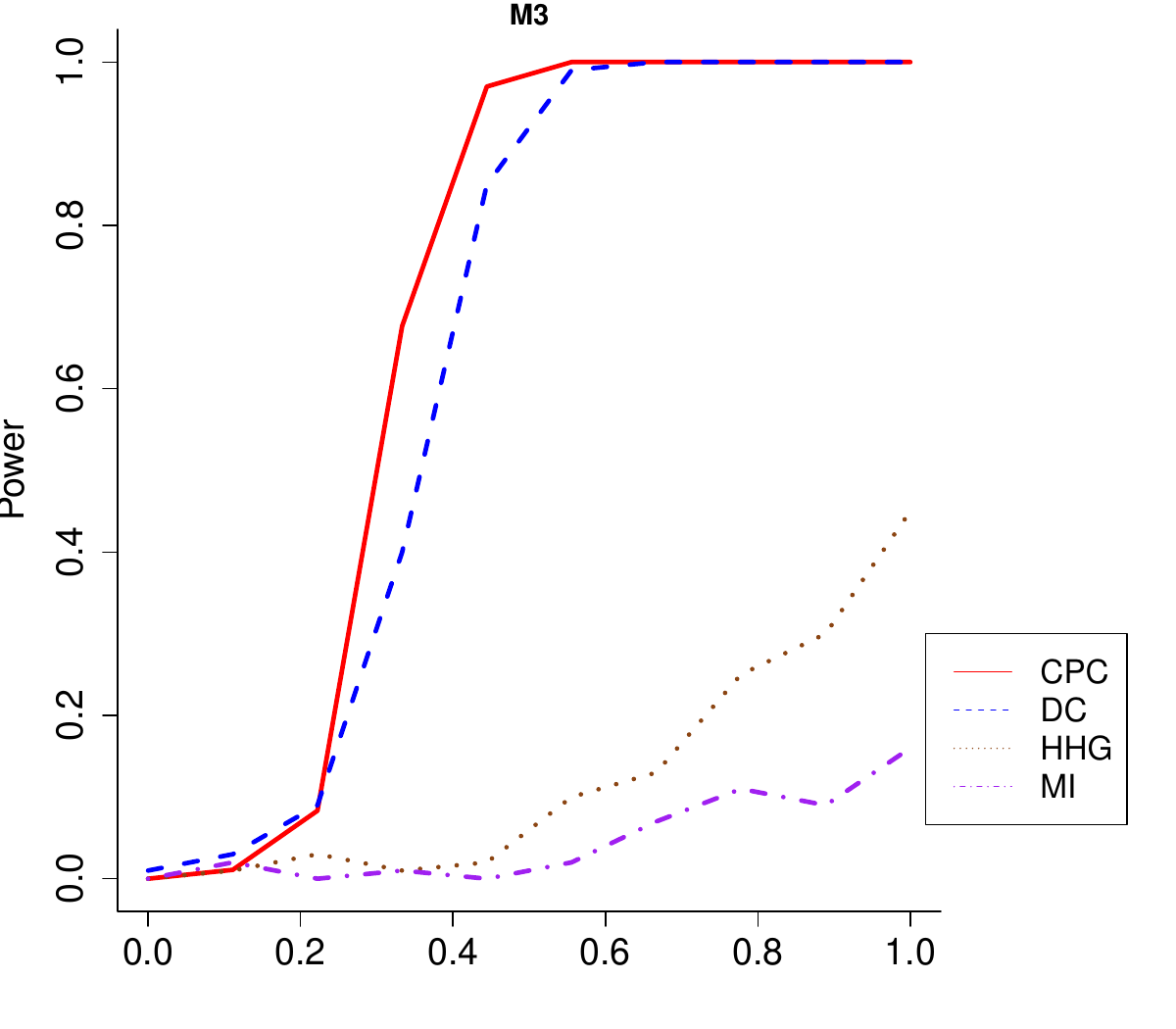}
	\end{subfigure}
	\hfill
	\begin{subfigure}[b]{0.45\textwidth}
		\centering
		\includegraphics[scale=0.37]{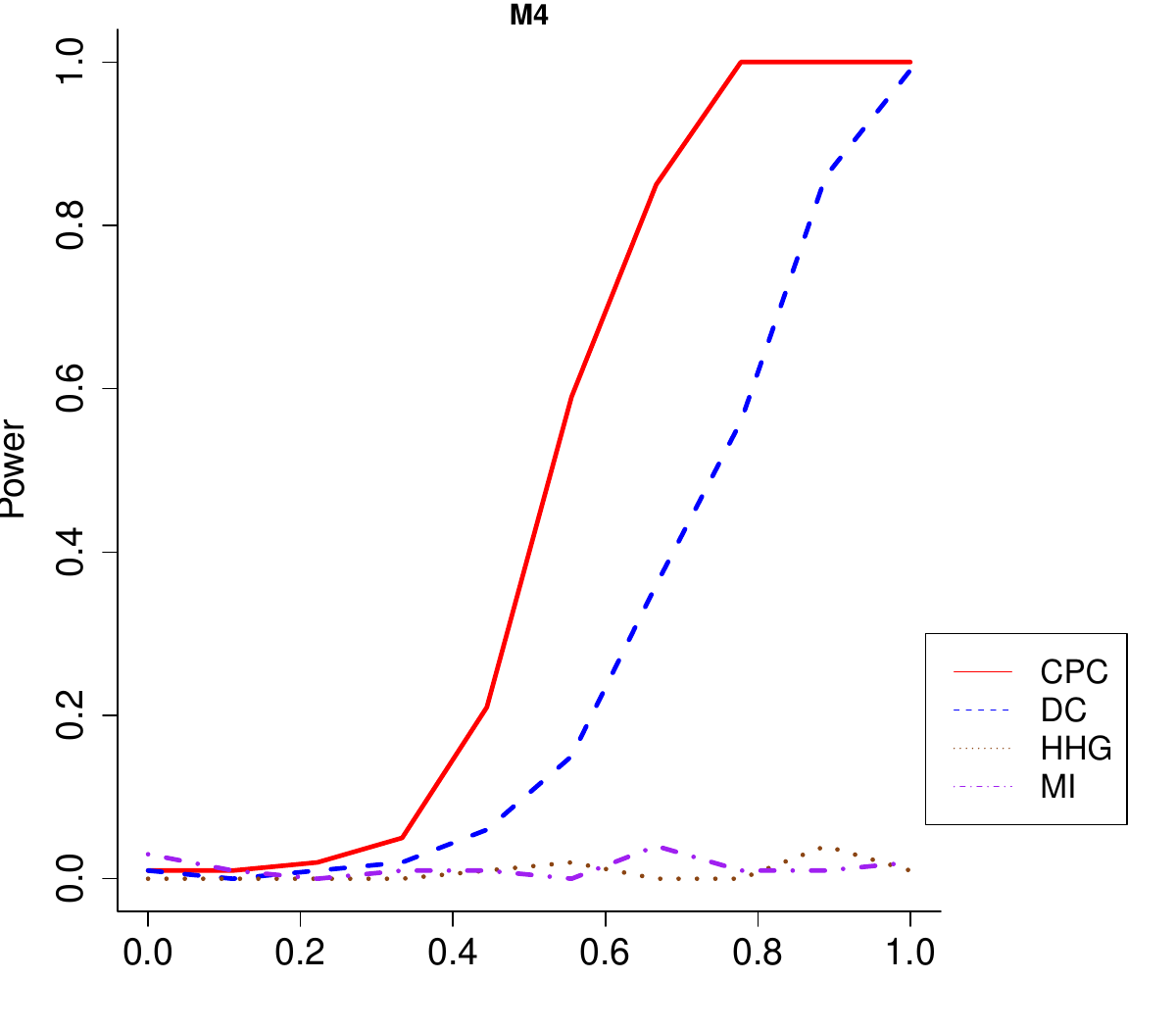}
	\end{subfigure}
	\begin{subfigure}[b]{0.45\textwidth}
		\centering
		\includegraphics[scale=0.37]{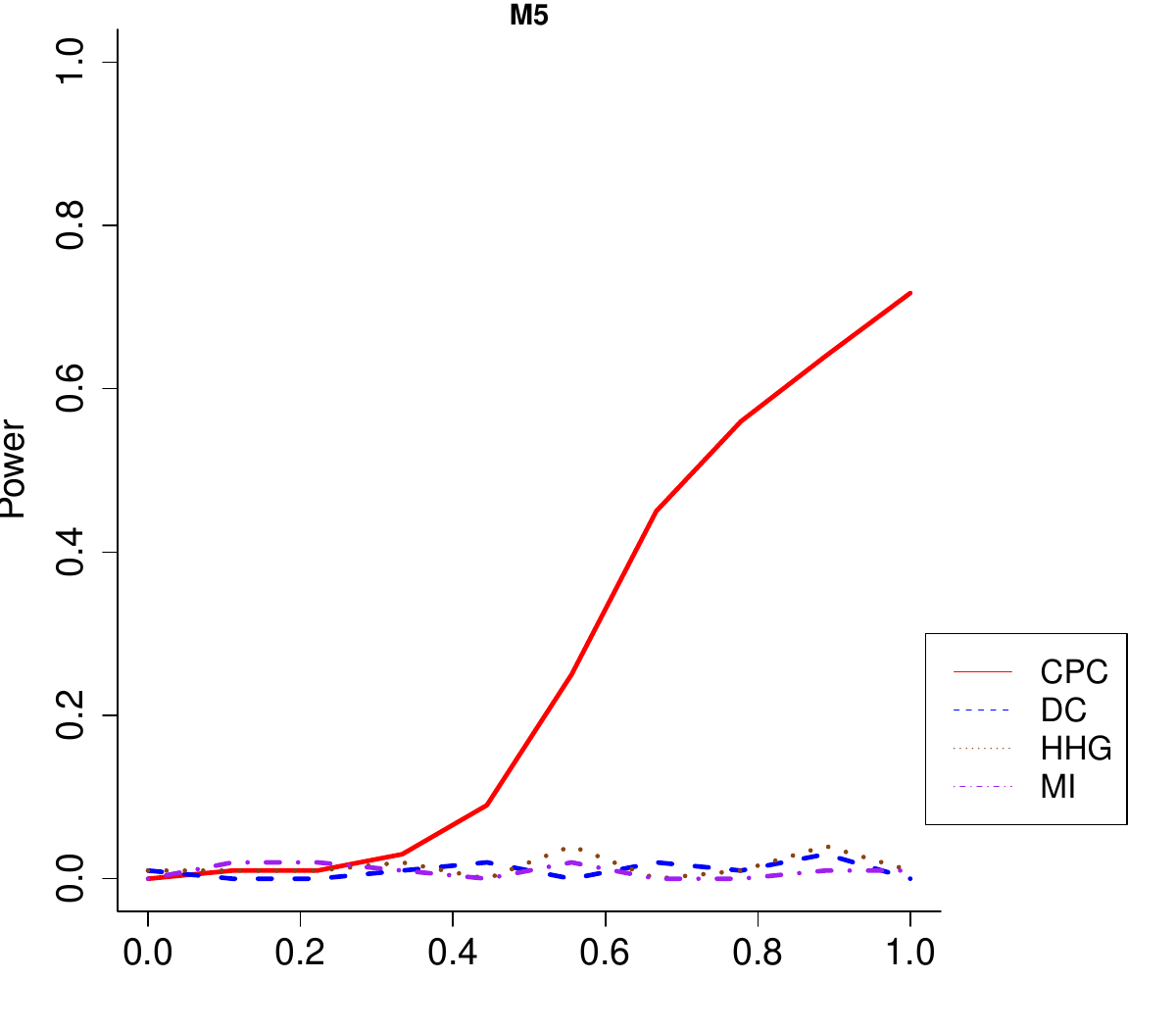}
	\end{subfigure}
	\hfill
	\begin{subfigure}[b]{0.45\textwidth}
		\centering
		\includegraphics[scale=0.37]{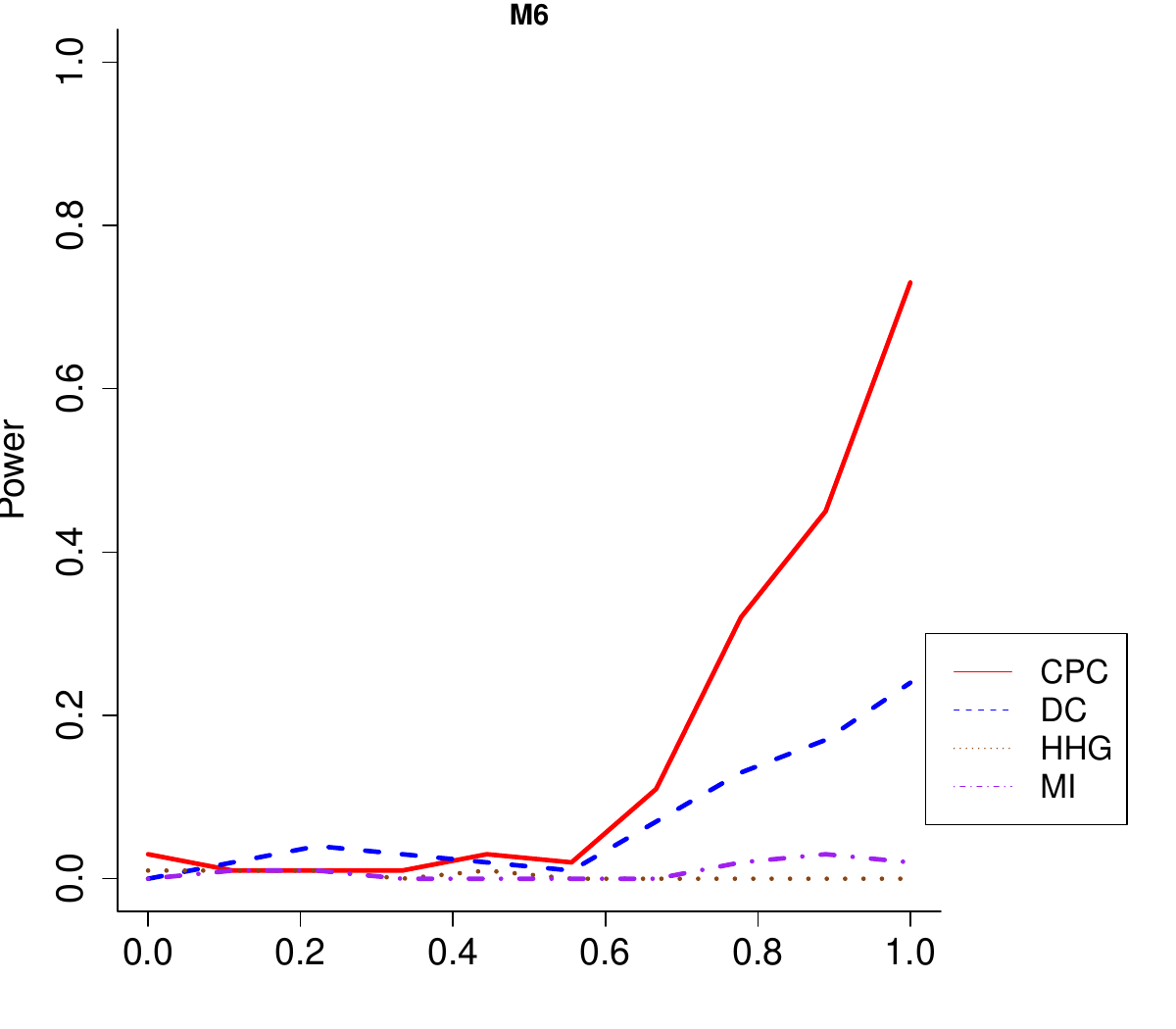}
	\end{subfigure}
	\caption{ The power versus the signal of the competing tests when $n=1000$, $d_1=d_2=100$, $\alpha=0.01$.}\label{power_case_plot2}
\end{figure}

\begin{figure}
	\centering
	\begin{subfigure}[b]{0.45\textwidth}
		\centering
		\includegraphics[scale=0.37]{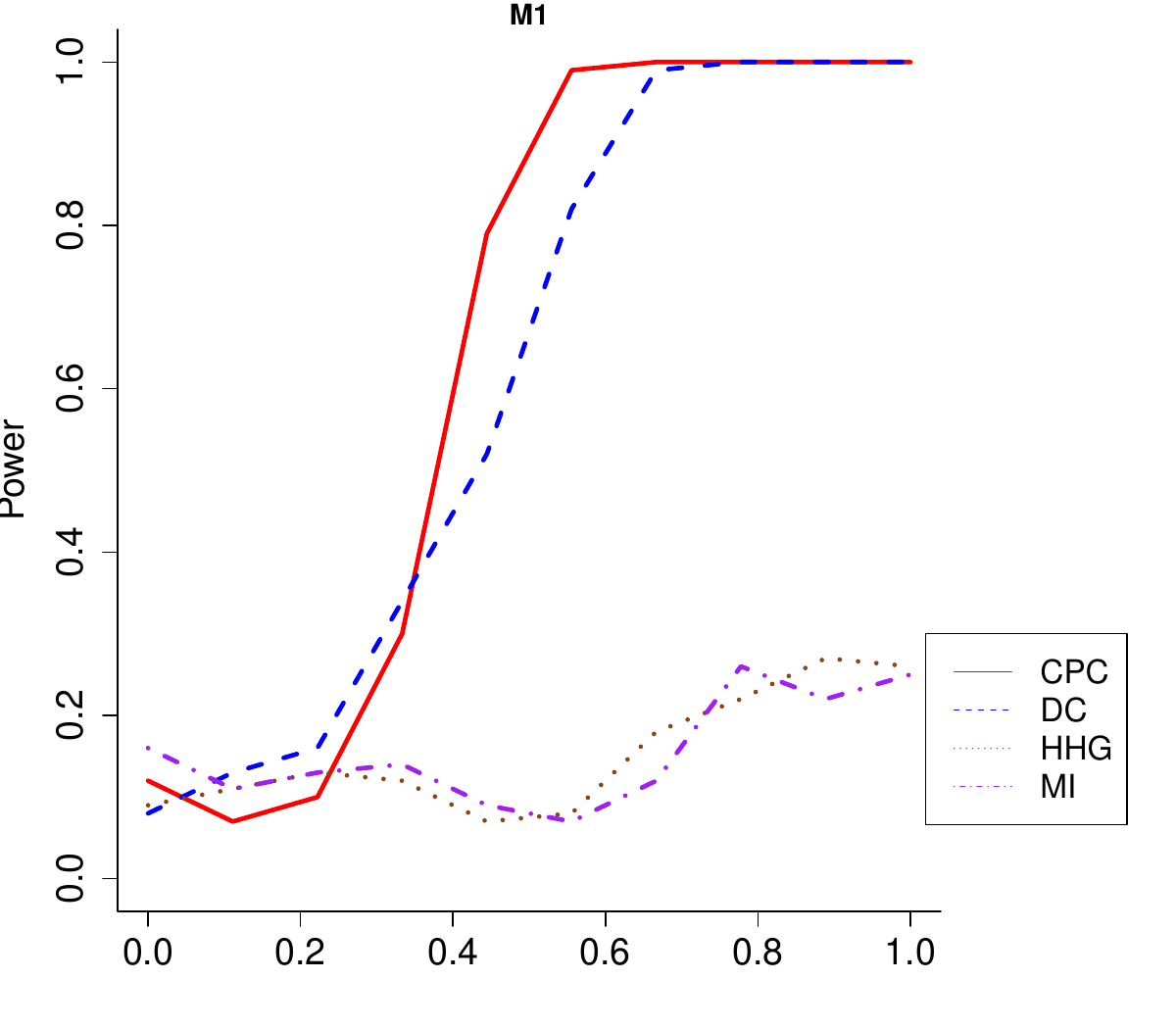}
	\end{subfigure}
	\hfill
	\begin{subfigure}[b]{0.45\textwidth}
		\centering
		\includegraphics[scale=0.37]{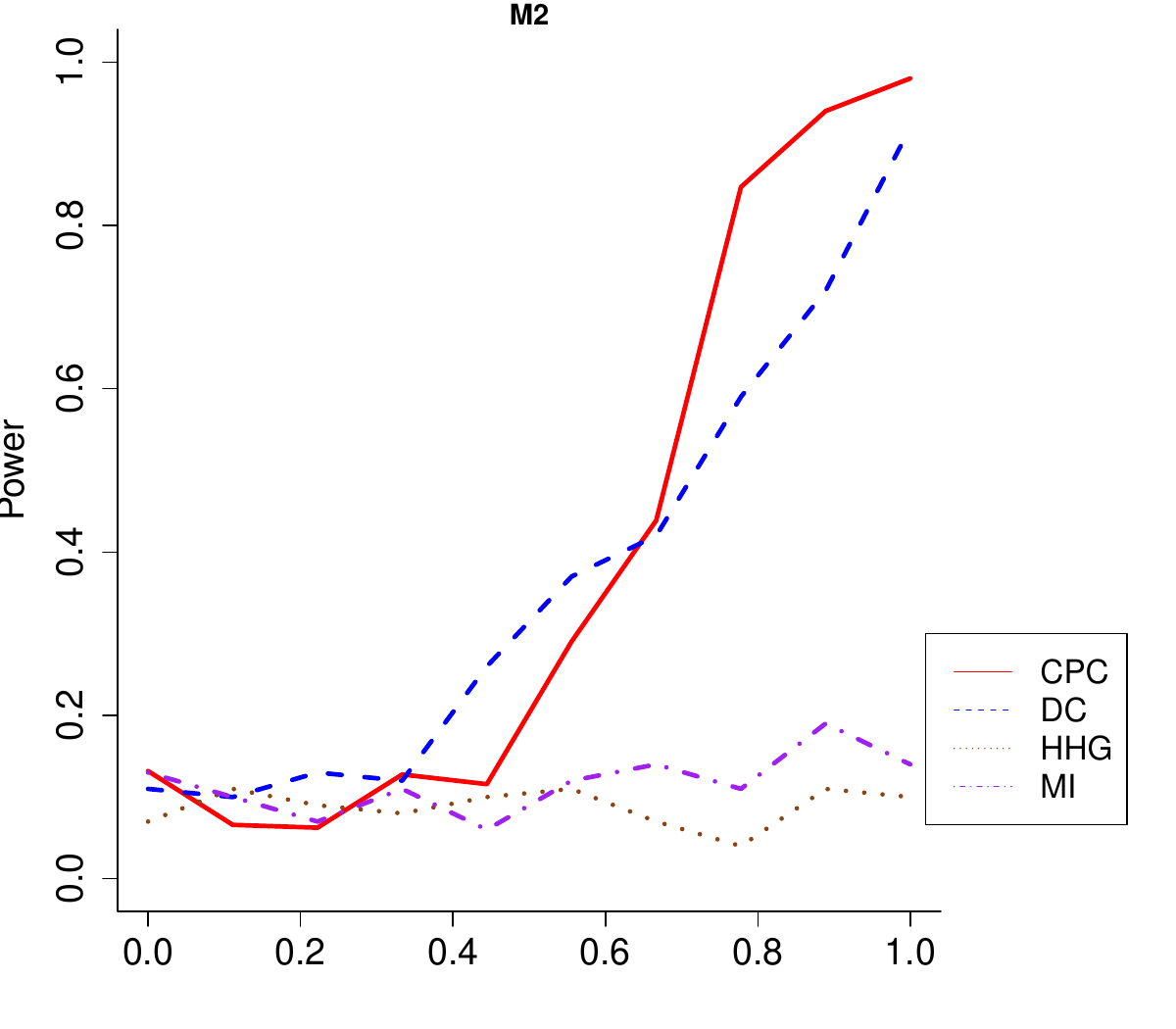}
	\end{subfigure}
	\centering
	\begin{subfigure}[b]{0.45\textwidth}
		\centering
		\includegraphics[scale=0.37]{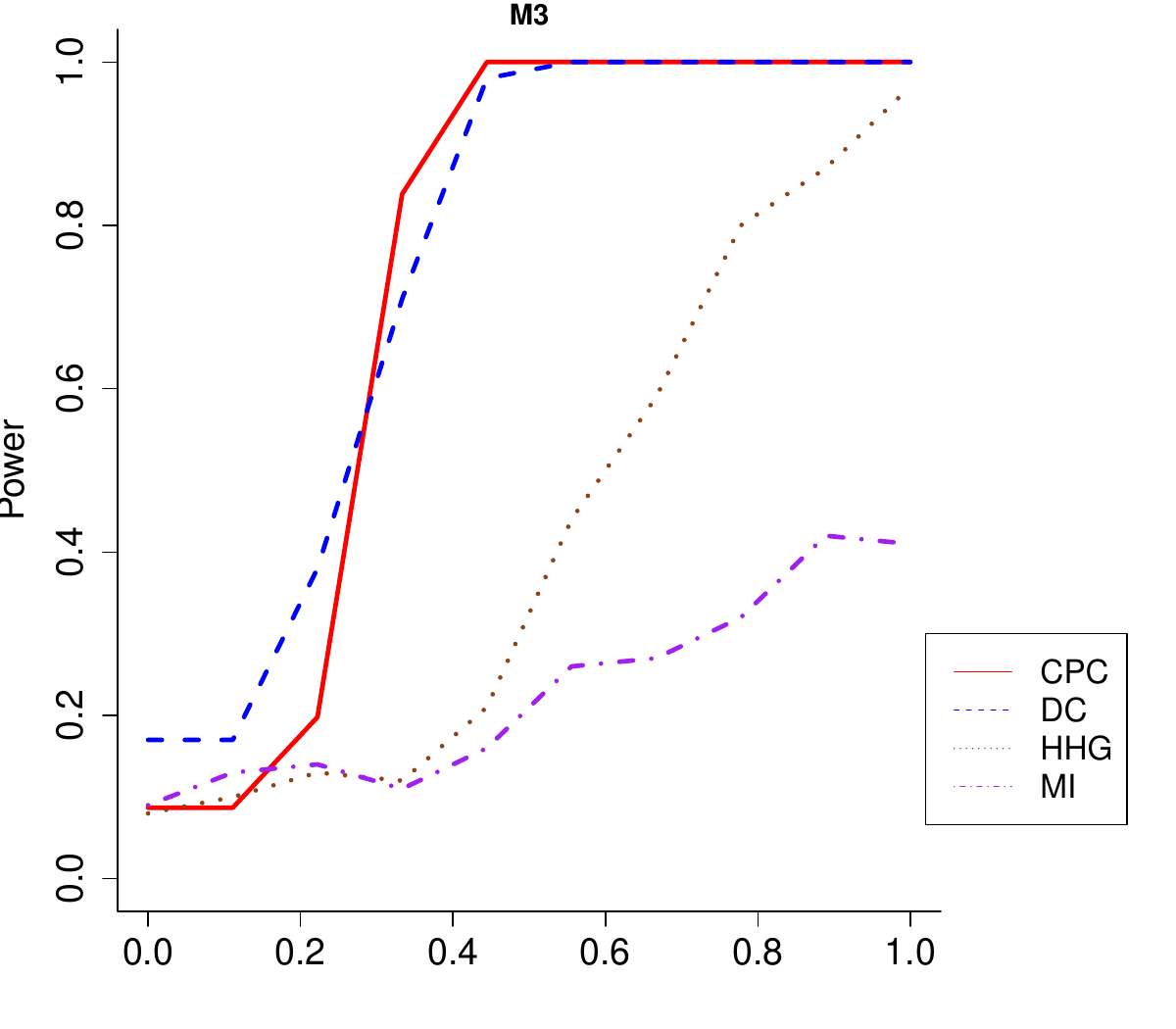}
	\end{subfigure}
	\hfill
	\begin{subfigure}[b]{0.45\textwidth}
		\centering
		\includegraphics[scale=0.37]{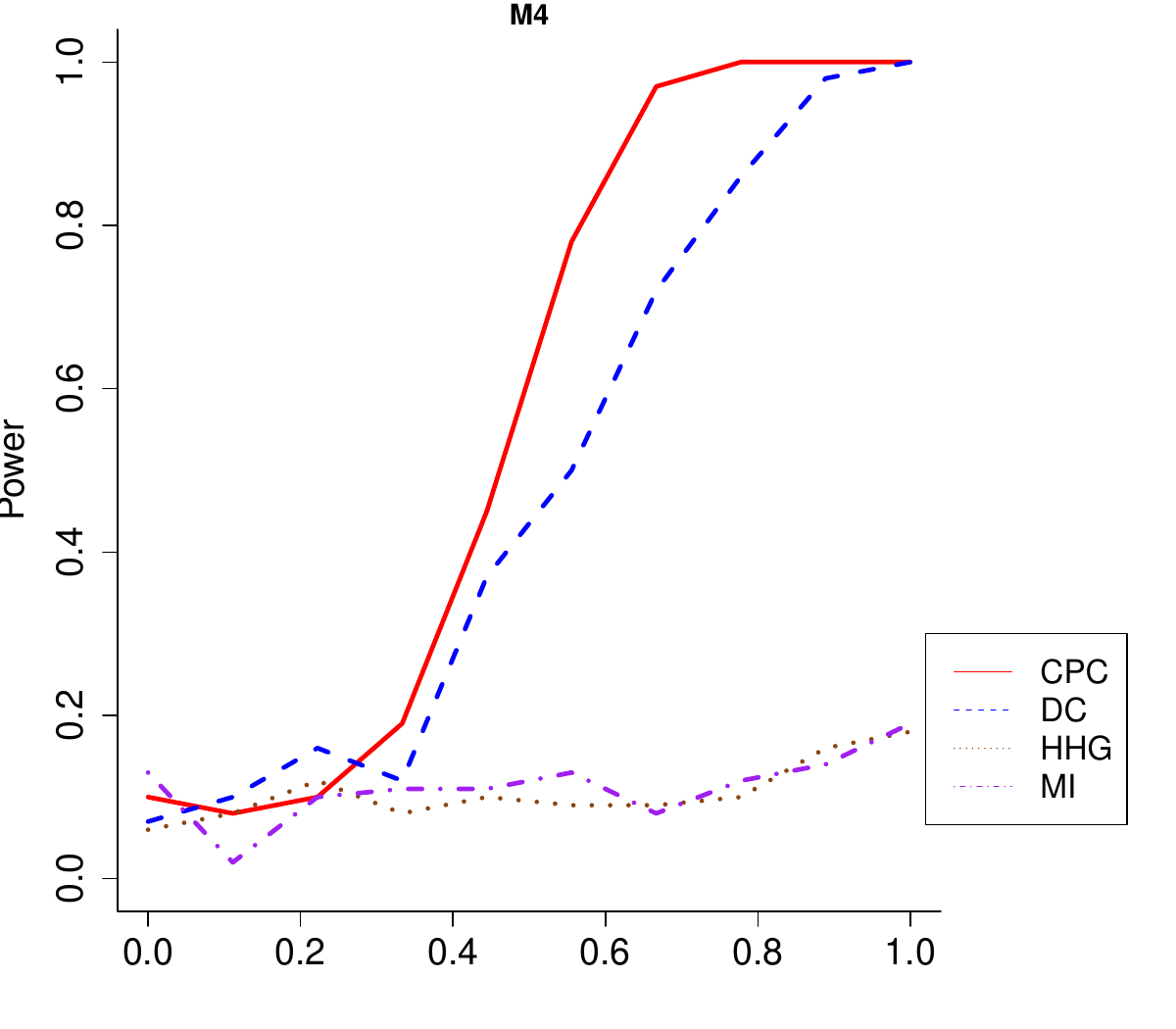}
	\end{subfigure}
	\begin{subfigure}[b]{0.45\textwidth}
		\centering
		\includegraphics[scale=0.37]{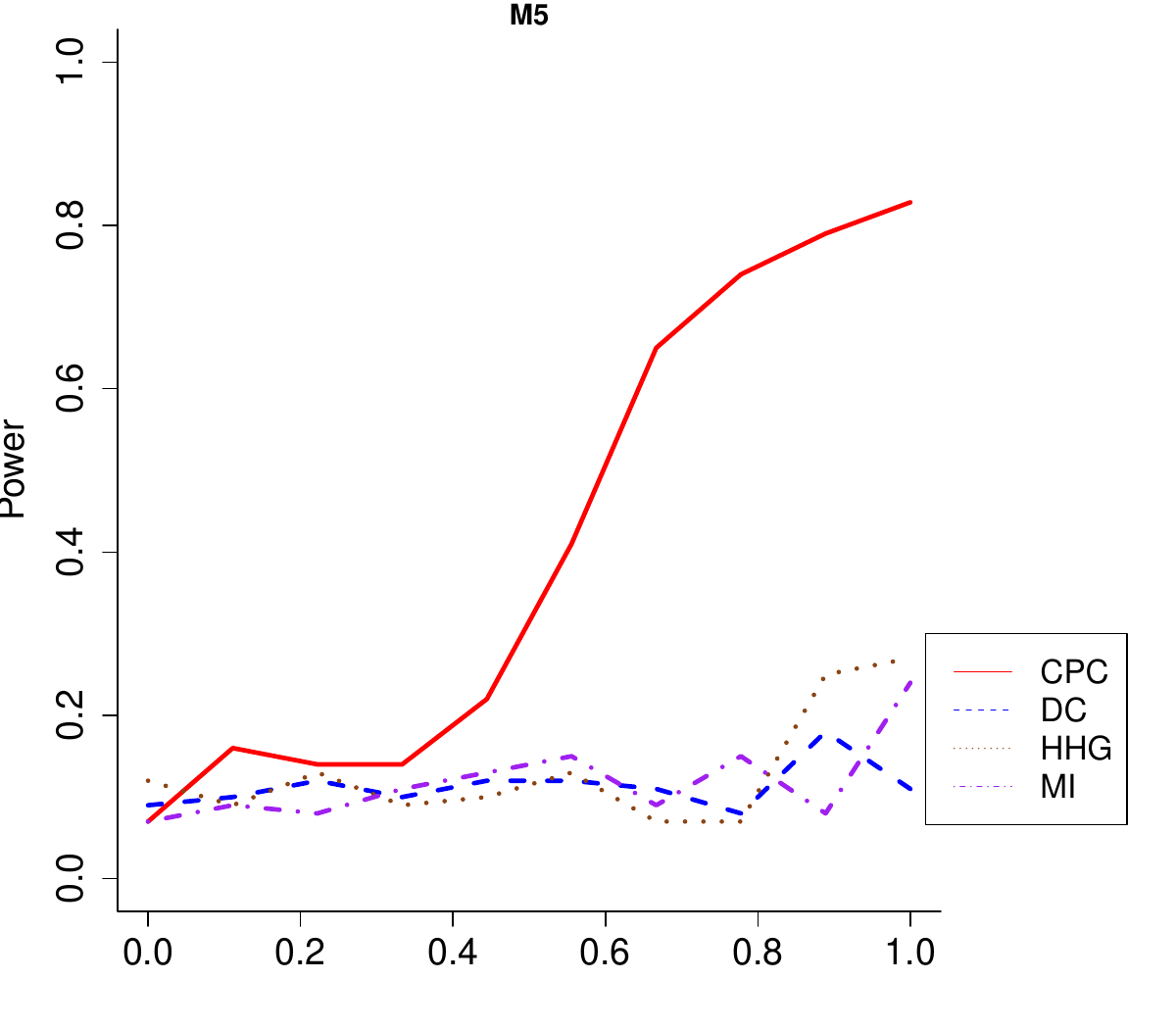}
	\end{subfigure}
	\hfill
	\begin{subfigure}[b]{0.45\textwidth}
		\centering
		\includegraphics[scale=0.37]{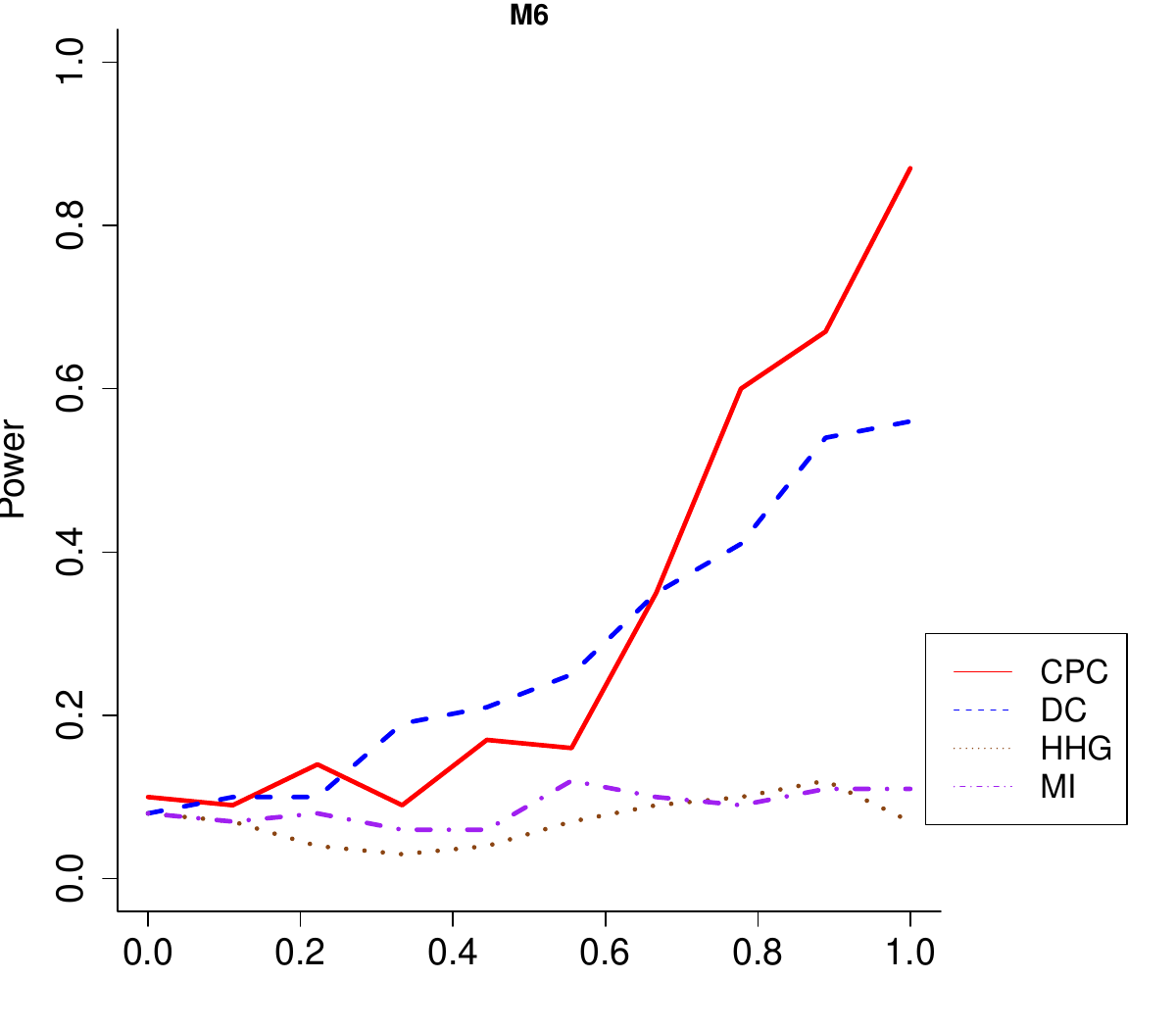}
	\end{subfigure}
	\caption{ The power versus the signal of the competing tests when $n=1000$, $d_1=d_2=100$, $\alpha=0.1$.}\label{power_case_plot3}
\end{figure}

Additional simulations for M1 - M6 when $\alpha=0.01$ and $\alpha=0.1$ are given in Figure \ref{power_case_plot2} and \ref{power_case_plot3}.

\begin{figure}
	\centering
	\begin{subfigure}[b]{0.45\textwidth}
		\centering
		\includegraphics[scale=0.37]{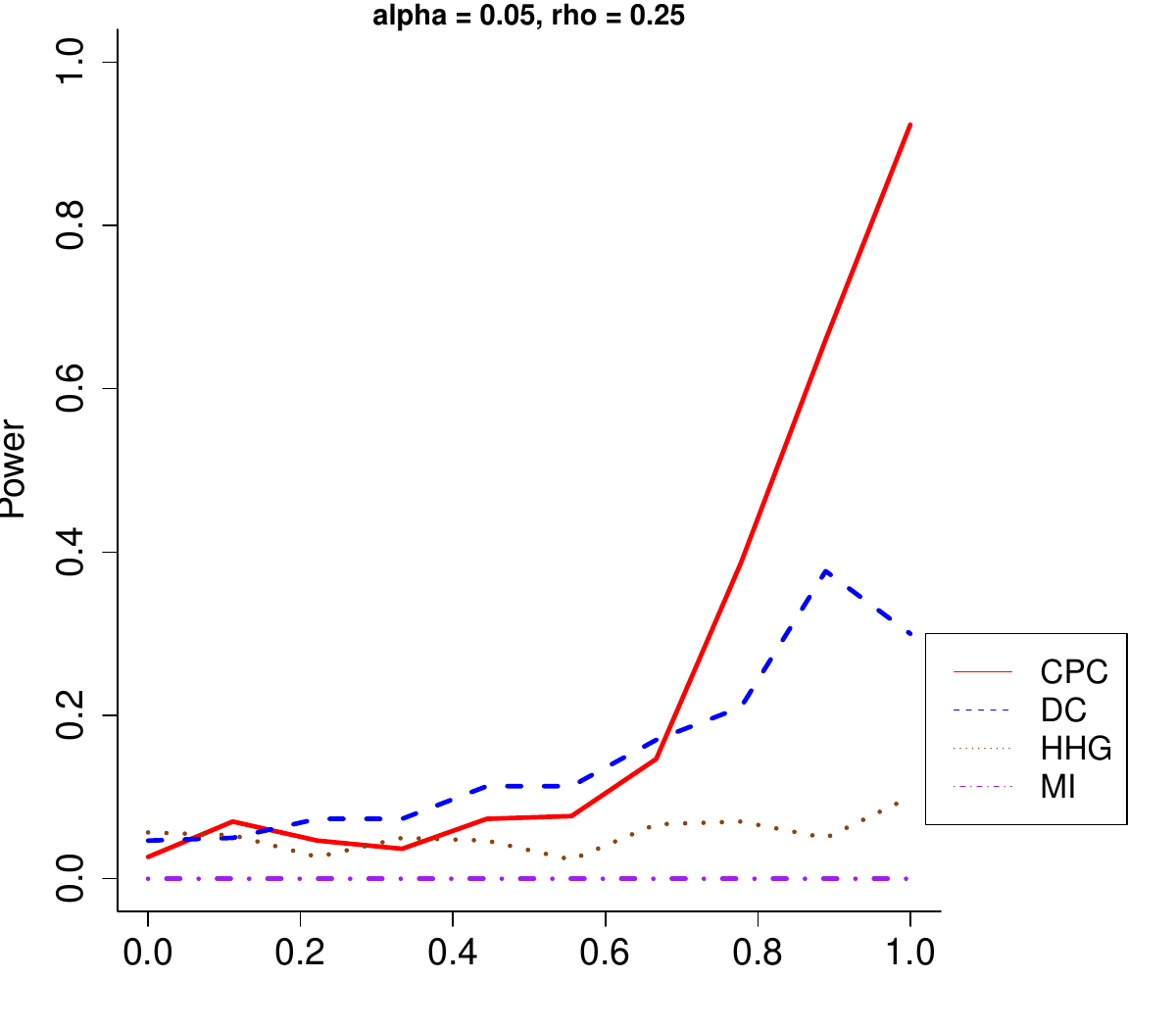}
	\end{subfigure}
	\hfill
	\begin{subfigure}[b]{0.45\textwidth}
		\centering
		\includegraphics[scale=0.37]{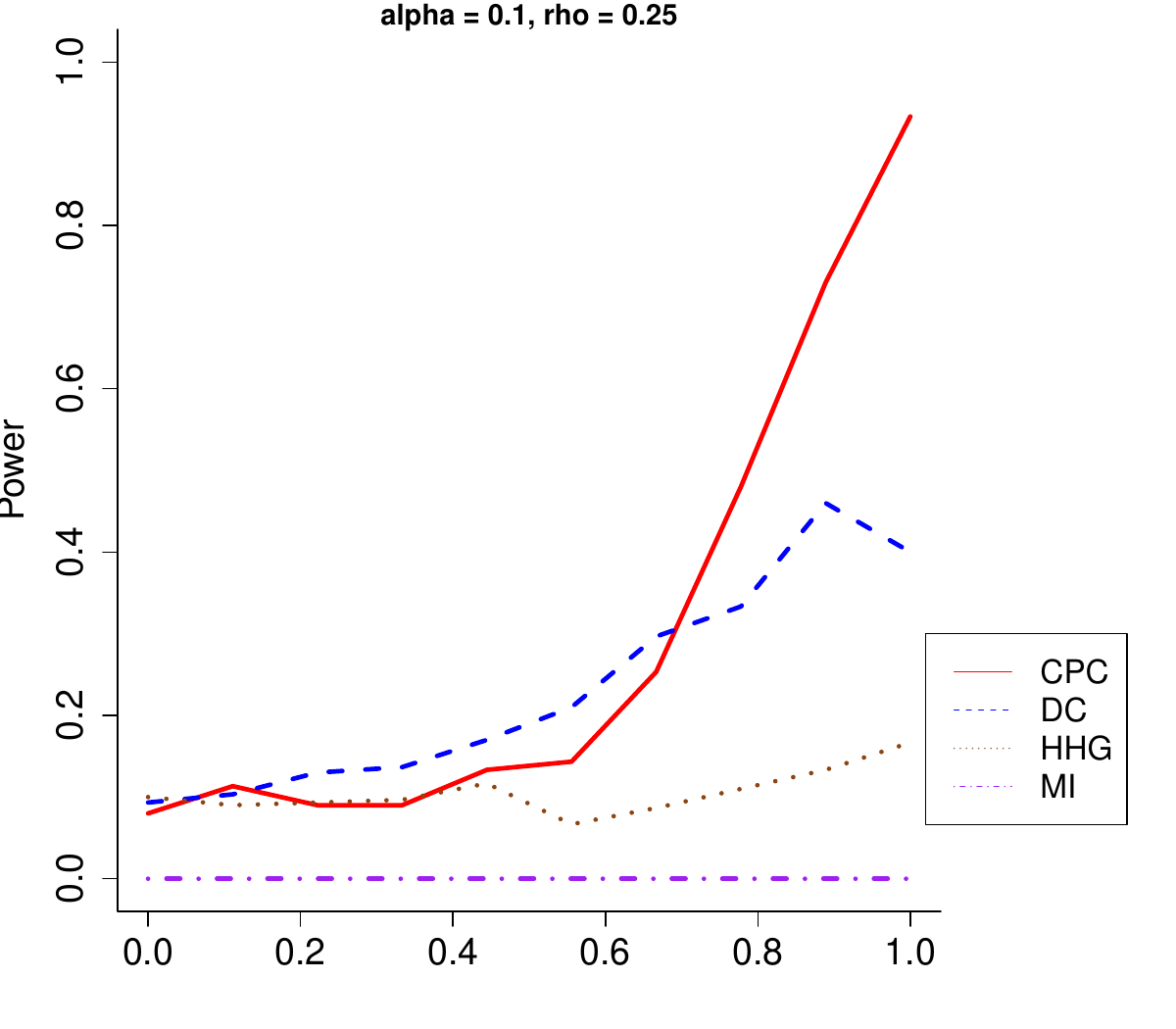}
	\end{subfigure}
	\centering
	\begin{subfigure}[b]{0.45\textwidth}
		\centering
		\includegraphics[scale=0.37]{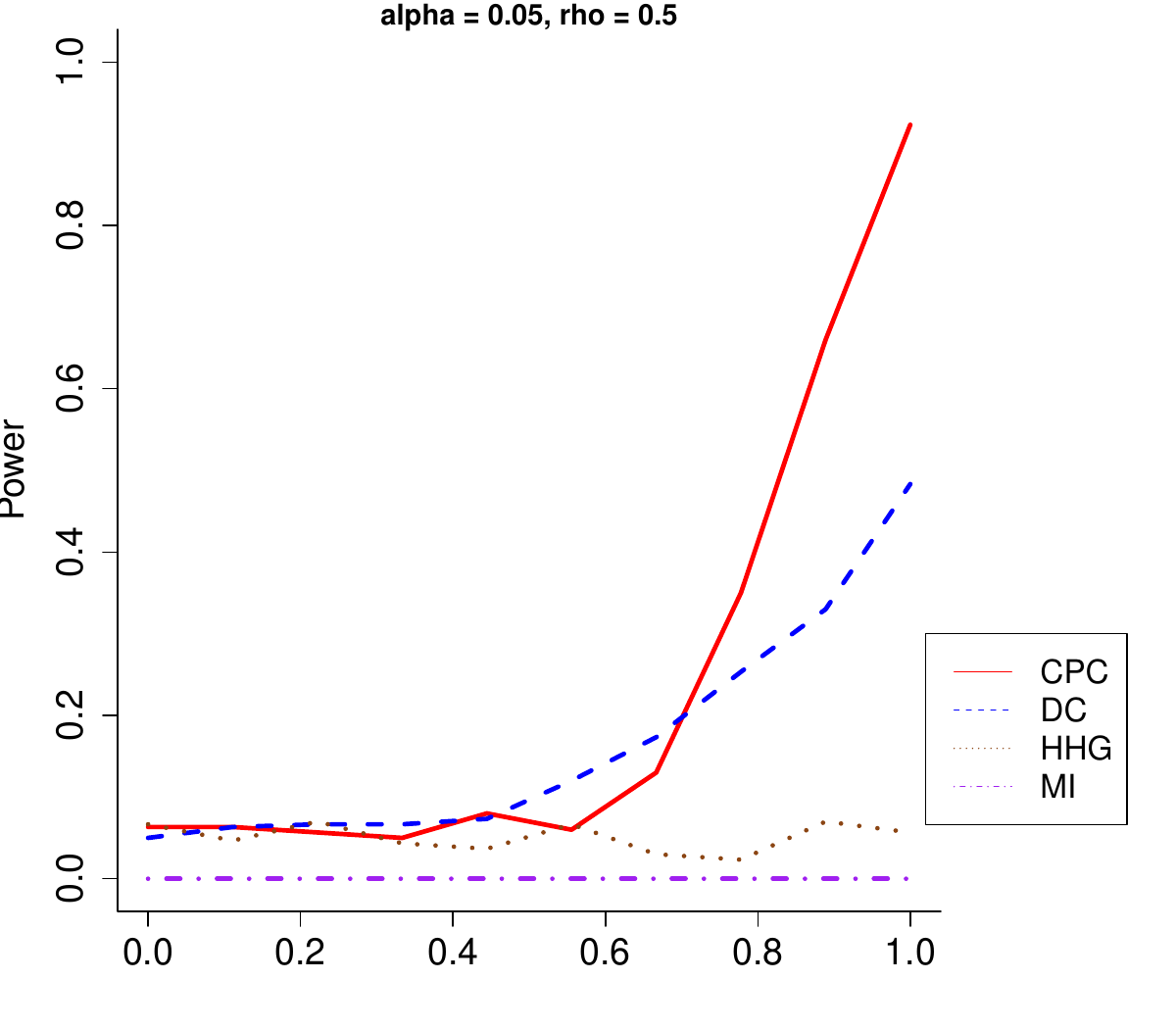}
	\end{subfigure}
	\hfill
	\begin{subfigure}[b]{0.45\textwidth}
		\centering
		\includegraphics[scale=0.37]{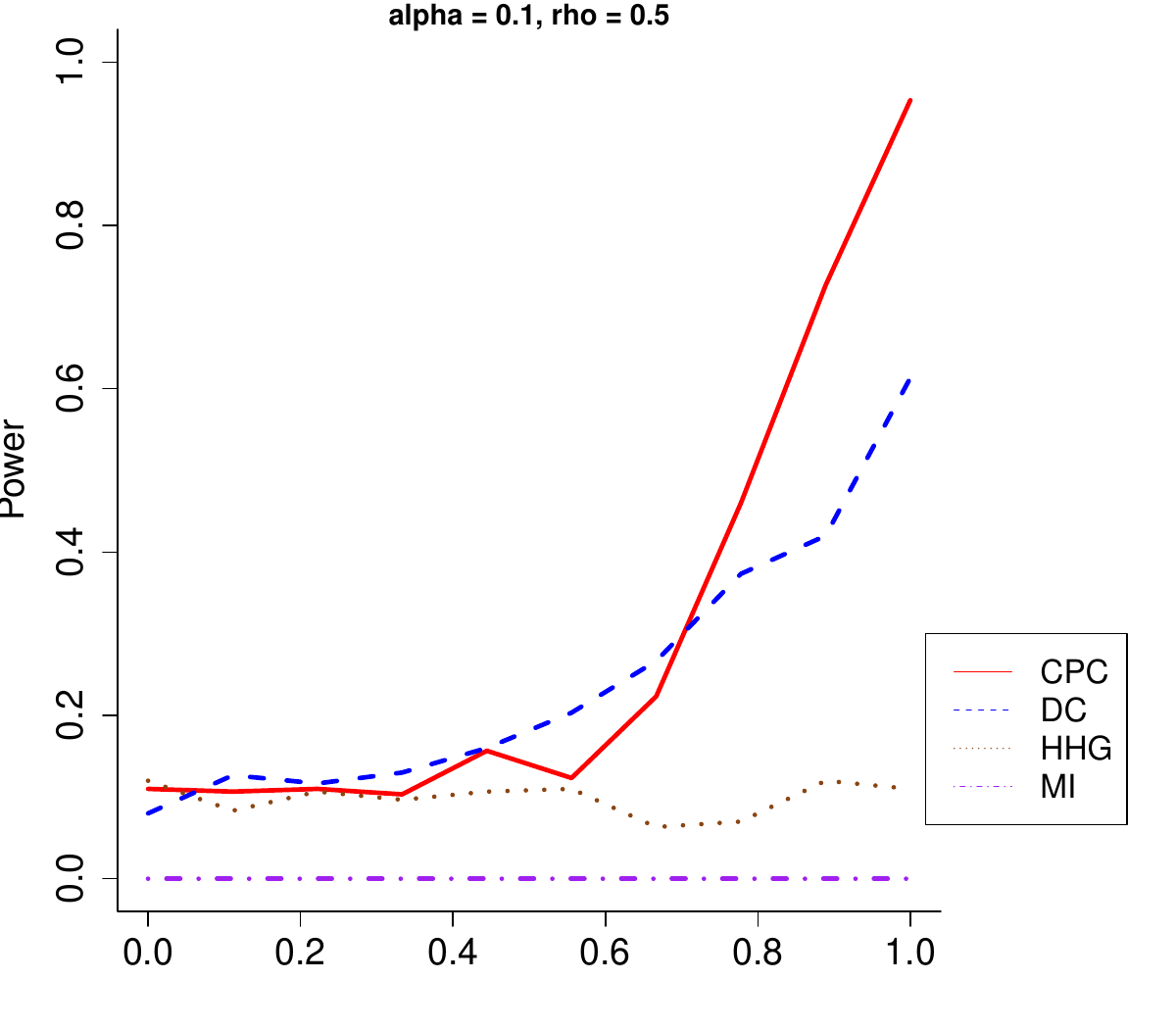}
	\end{subfigure}
	\begin{subfigure}[b]{0.45\textwidth}
		\centering
		\includegraphics[scale=0.37]{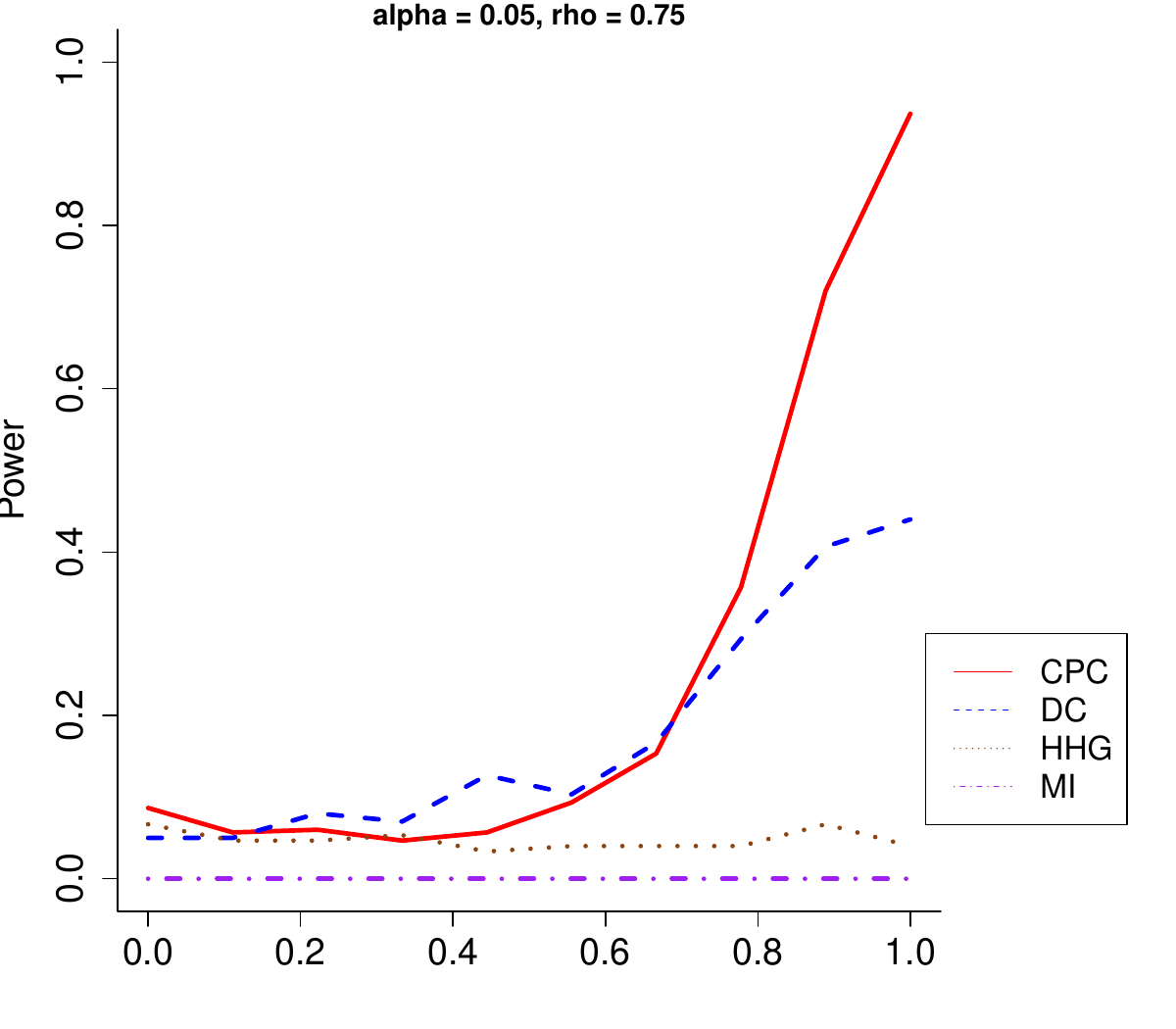}
	\end{subfigure}
	\hfill
	\begin{subfigure}[b]{0.45\textwidth}
		\centering
		\includegraphics[scale=0.37]{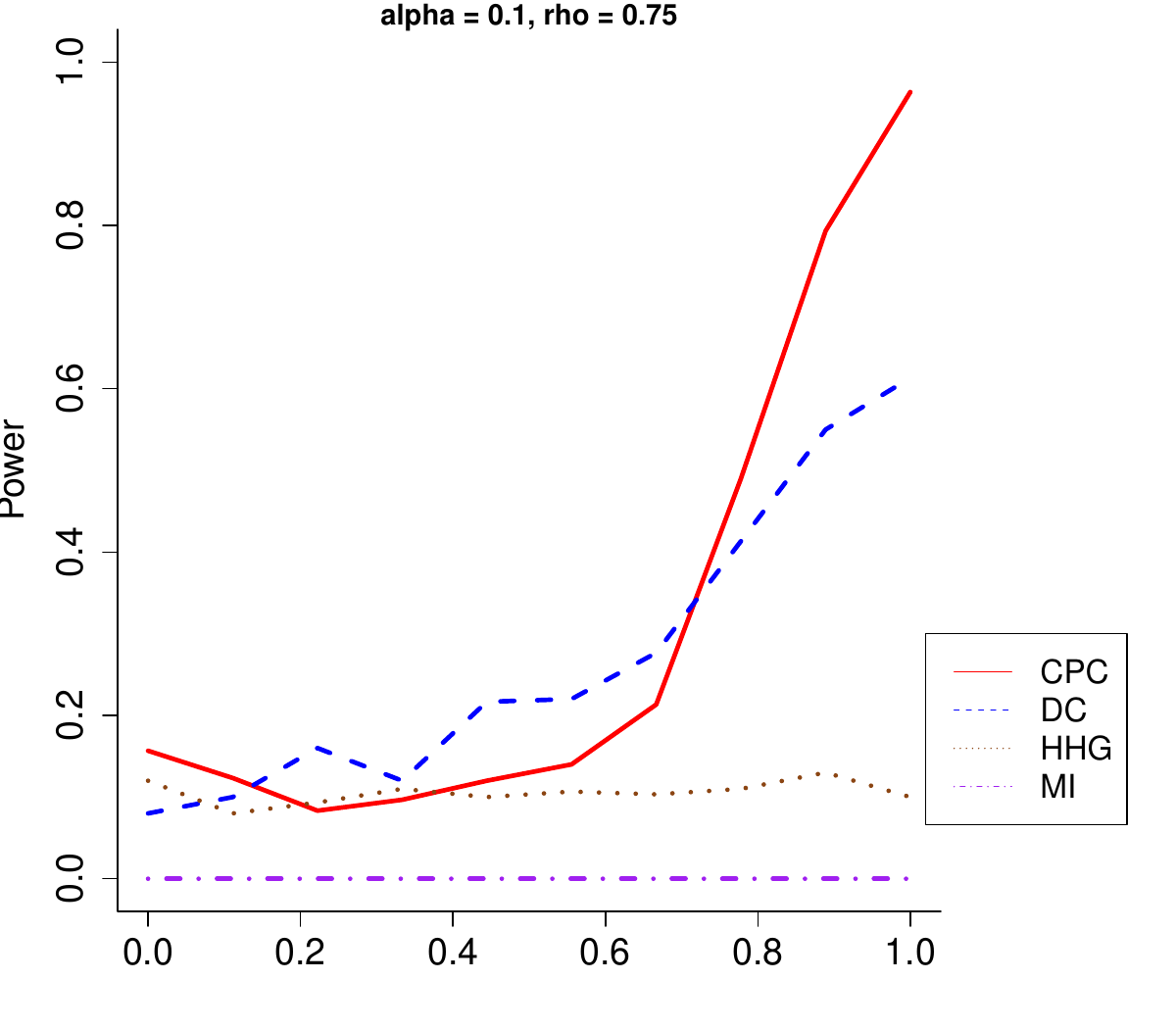}
	\end{subfigure}
	\caption{ The power versus the signal of the competing tests when the data is correlated. $n=1000$, $d_1=d_2=500$.}\label{power_case_cor}
\end{figure}

We also report the simulations when the data is correlated in Figure \ref{power_case_cor}. Consider the model $M1$, where we still assume only the fist element in $X$ and $Y$ are related, and let the signal $a$ vary from 0 to 1. We assume that $X$ follows a multivariate normal distribution with mean vector $\mathbf{0}$ and covariance matrix $\Sigma$, where $\Sigma_{i,j} = \rho^{|i-j|}$. Let $\rho\in\{0.25, 0.5, 0.75\}$. We report the power curves for all the methods when $\alpha = 0.05$ and $0.1$.

\begin{figure}
	\centering
	\begin{subfigure}[b]{0.45\textwidth}
		\centering
		\includegraphics[scale=0.37]{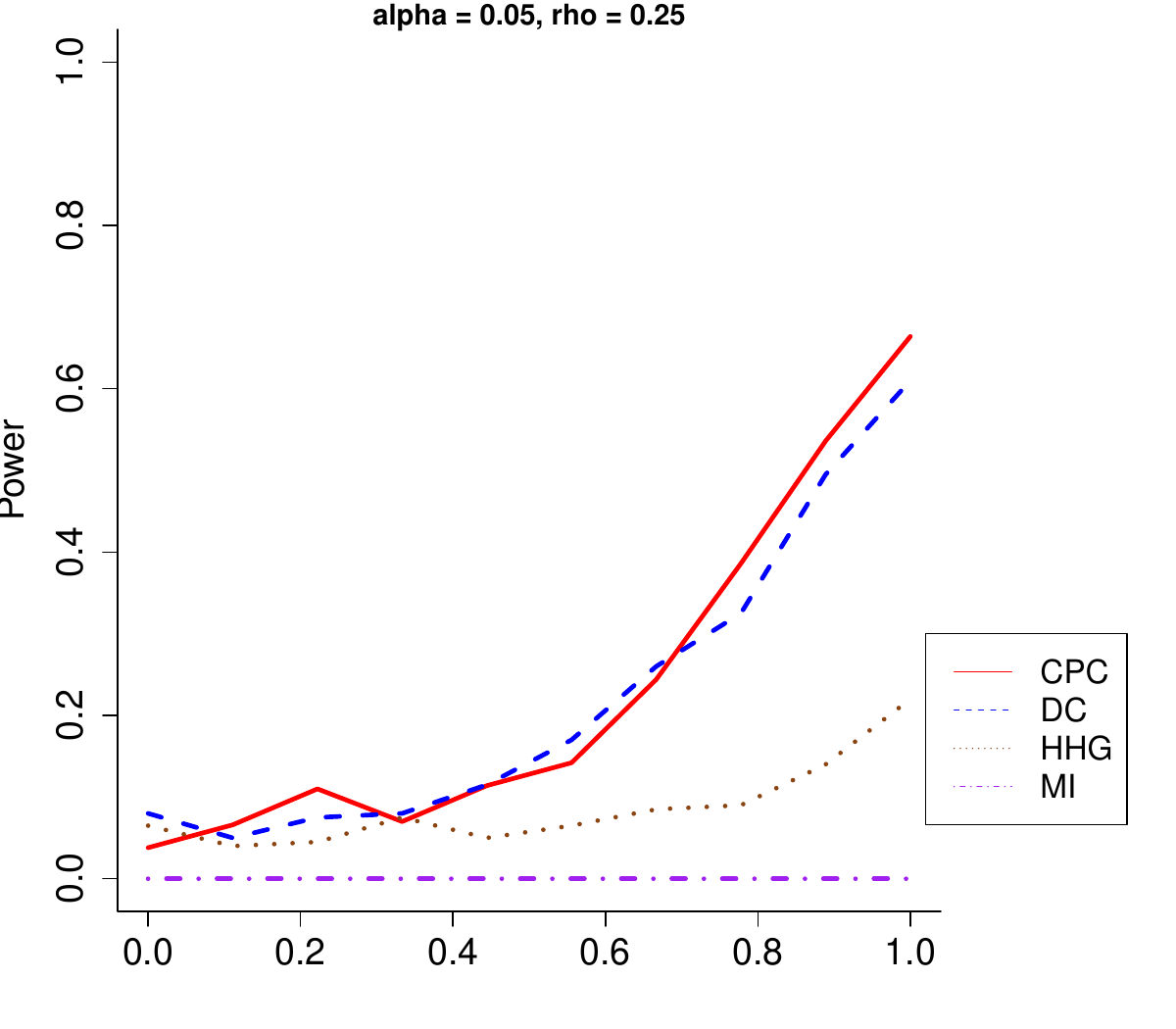}
	\end{subfigure}
	\hfill
	\begin{subfigure}[b]{0.45\textwidth}
		\centering
		\includegraphics[scale=0.37]{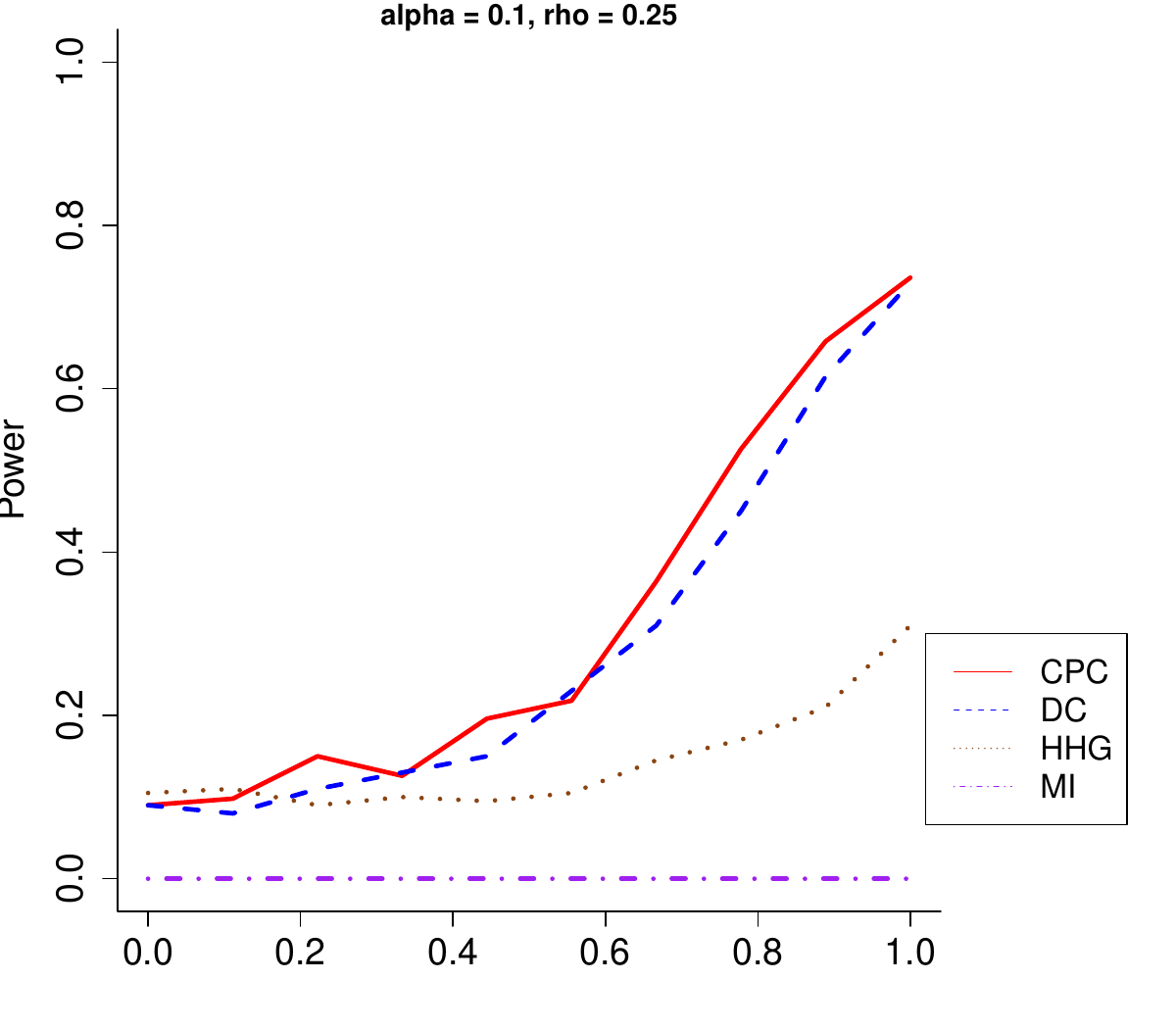}
	\end{subfigure}
	\centering
	\begin{subfigure}[b]{0.45\textwidth}
		\centering
		\includegraphics[scale=0.37]{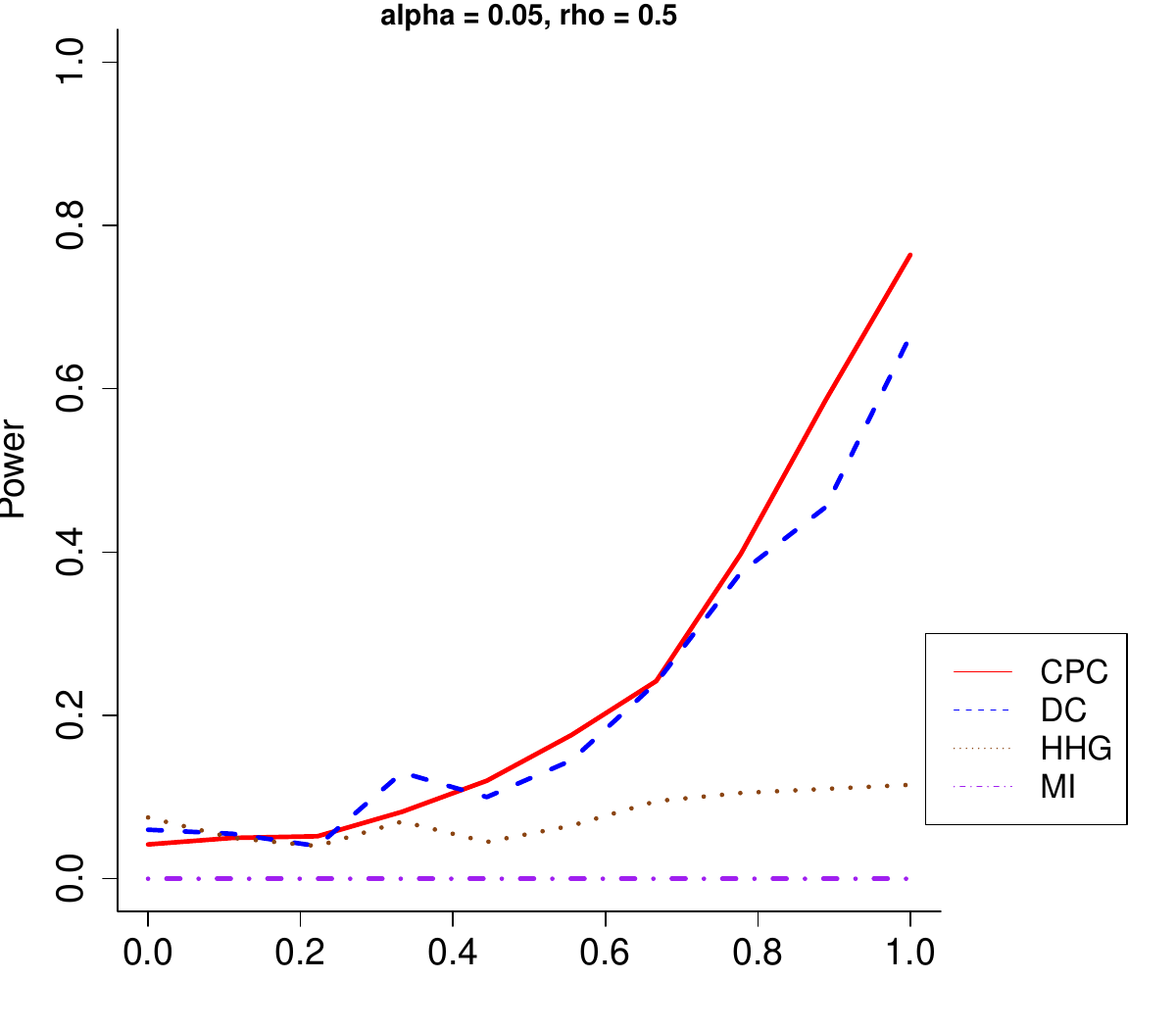}
	\end{subfigure}
	\hfill
	\begin{subfigure}[b]{0.45\textwidth}
		\centering
		\includegraphics[scale=0.37]{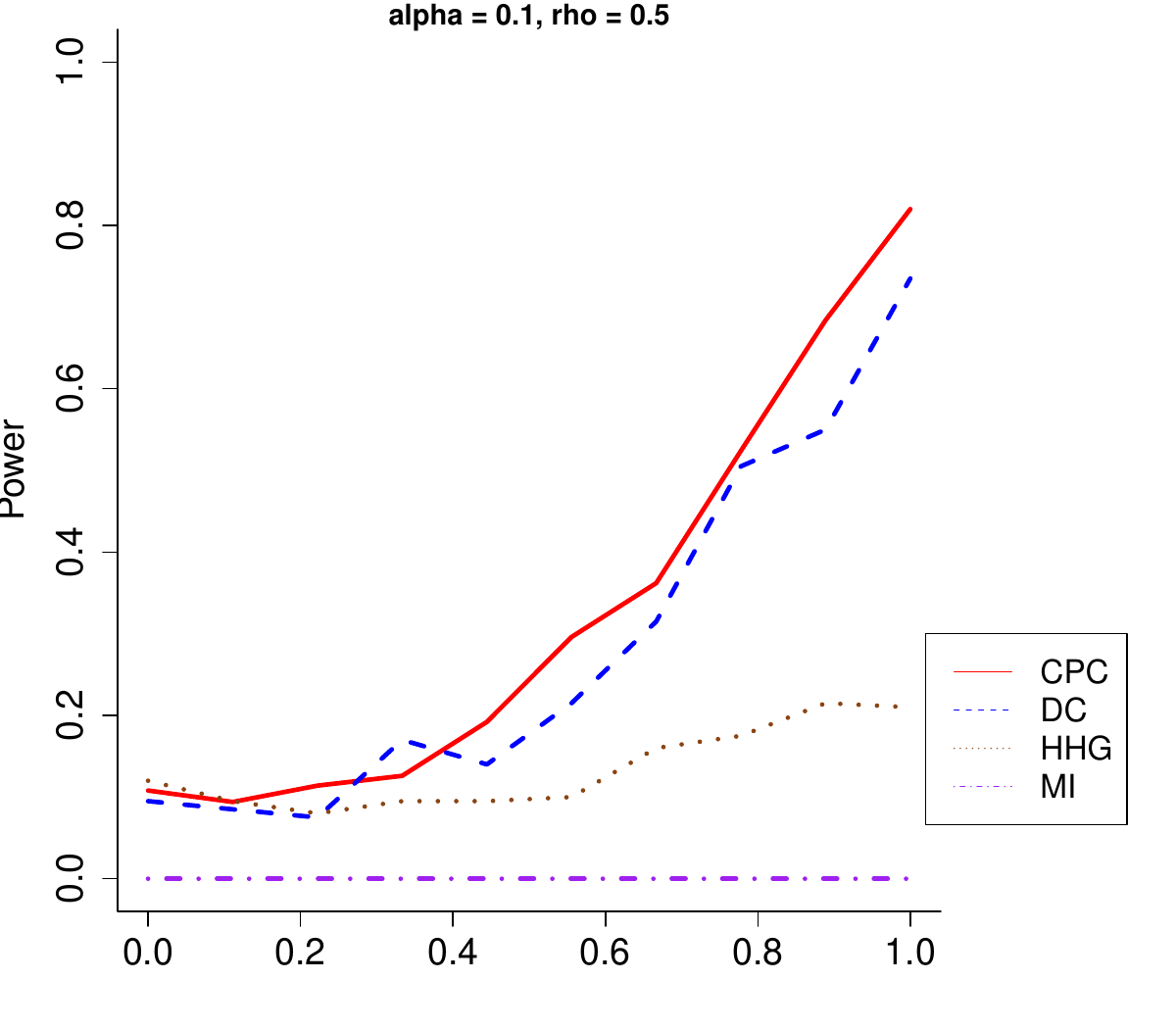}
	\end{subfigure}
	\begin{subfigure}[b]{0.45\textwidth}
		\centering
		\includegraphics[scale=0.37]{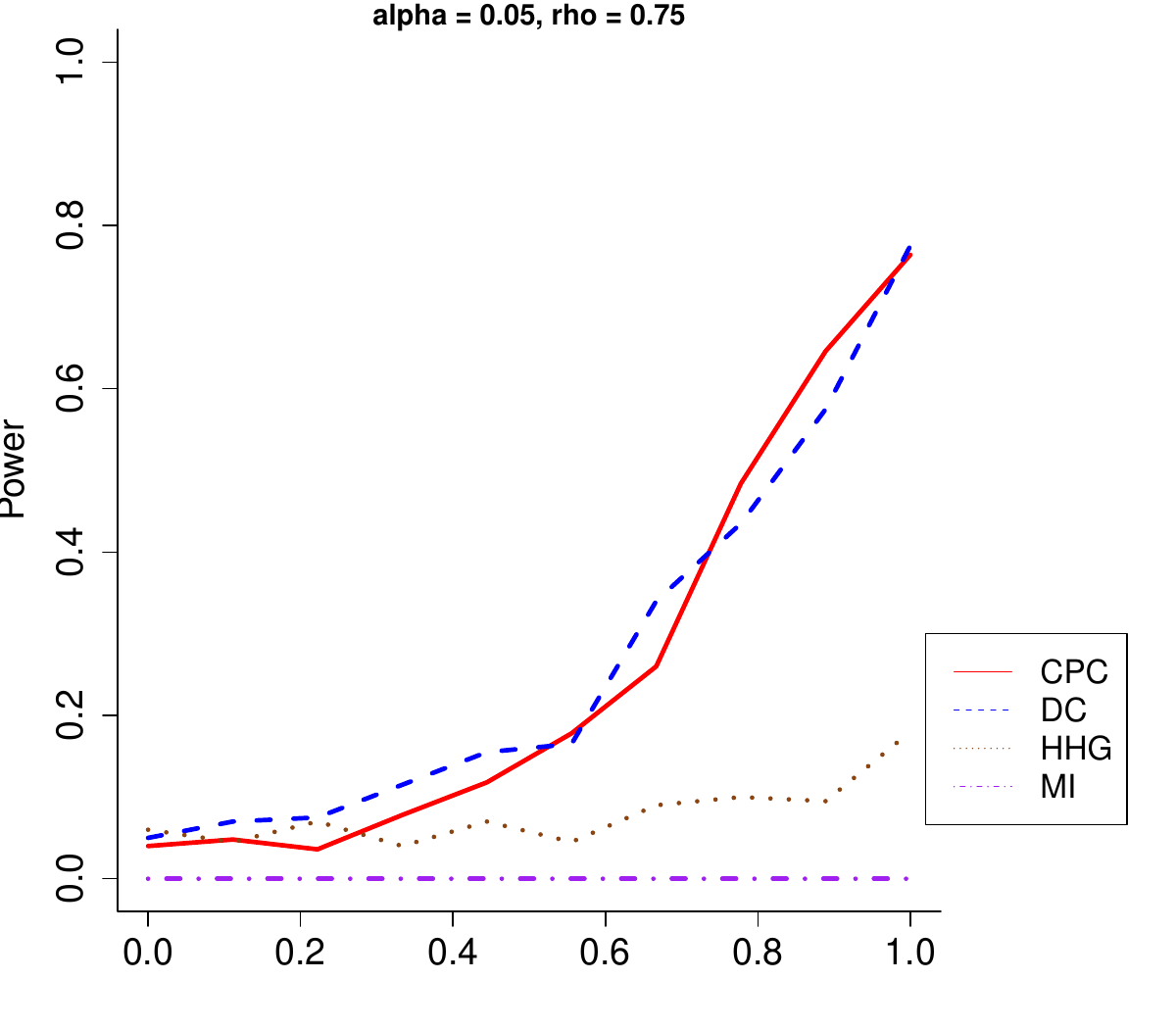}
	\end{subfigure}
	\hfill
	\begin{subfigure}[b]{0.45\textwidth}
		\centering
		\includegraphics[scale=0.37]{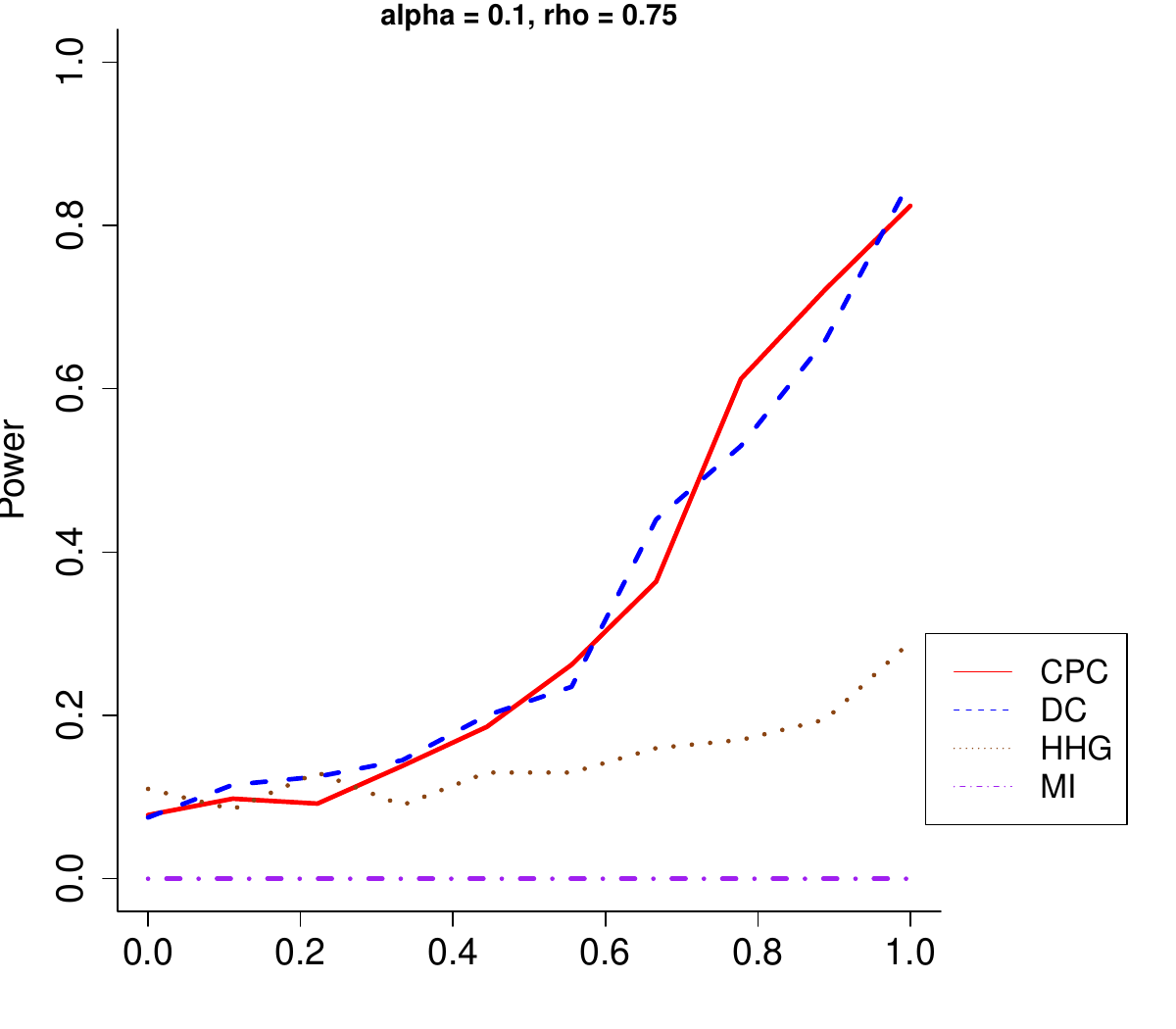}
	\end{subfigure}
	\caption{ The power versus the signal of the competing tests when the data has heavy-tailed distribution and is correlated.  $n=1000$, $d_1=d_2=200$.}\label{power_case_heavy}
\end{figure}

Lastly, we report the simulations where the data is correlated and has heavy tailed distribution in Figure \ref{power_case_heavy}. We assume that both $X$ and $Y$ follow multivariate $t$-distribution with 2 degrees of freedom. The multivariate $t$-distribution has location parameter $\mathbf{0}$ and scale matrix $\Sigma$, where $\Sigma_{i,j} = \rho^{|i-j|}$. Let $\rho\in\{0.25, 0.5, 0.75\}$. We report the power curves for all the methods when $\alpha = 0.05$ and $0.1$. We still work with the linear model, where $Y_1 = a X_1 + \epsilon$, and $\epsilon\sim N(0, 1)$.

\section{Other Choice of Test Statistics}\label{sec:other_choice}

In the paper, we have focused on using the rank sum test to distinguish $\wh\theta(X,Y)$ and $\wh\theta(X',Y')$. In fact one can use other two-sample tests under the same framework. For example, one may use a version of the two-sample $t$-statistic
\beqrs
\frac{1}{n_2}\sum_{i \in\calI_2}\left\{ \wh\theta (X_{i}, Y_{i}) - \wh\theta (X_{i}', Y_{i}') \right\}
\eeqrs
and reject for large values of the test statistic. One may also estimate the  the KL-divergence
\beqrs
\frac{1}{n_2}\sum_{i \in\calI_2}\left[\log\left\{ \frac{\wh\theta (X_{i}, Y_{i})}{1-\wh\theta (X_{i}, Y_{i})}\right\}-\log\left\{ \frac{\wh\theta (X_{i}', Y_{i}')}{1-\wh\theta (X_{i}', Y_{i}')}\right\}\right]
\eeqrs
since $\wh\theta (X_{i}, Y_{i})/ \{ 1 - \wh\theta (X_{i}, Y_{i})\}$ is an estimate of the likelihood ratio.
However, these test statistics would require additional assumptions on the distributions of $\wh\theta(X,Y)$, $\wh\theta(X',Y')$, and are more likely to be sensitive to outliers, which may not be very plausible in practice, especially when the quality of $\wh\theta(\cdot)$ is not fully guaranteed.

\section{Numerical verification of Conditions}\label{sec:verify_condition}

In this section, we numerically verifiy the condition $\mu_*-\mu = o_p(n_2^{-1/2})$ under the local alternative hypothesis. Consider the example where $(X, Y)\sim N(0, \Sigma)$, with
\beqrs
\Sigma = &\begin{bmatrix}
	1 & \frac{\rho}{\sqrt n}\\
	\frac{\rho}{\sqrt n}& 1 
\end{bmatrix}
\eeqrs
for some $\rho>0$ and $(X', Y')\sim N(0, I_2)$, where $I_2$ is the $2\times 2$ identity matrix. Then using the Quadratic Discriminant Analysis (QDA), we have
{\small 
	\beqrs
	\theta(X, Y) &=& \frac{\frac{1}{\sqrt{2\pi |\Sigma|}}\exp{ -\frac{1}{2}(X, Y)\Sigma^{-1}(X, Y)'} }{\frac{1}{\sqrt{2\pi |\Sigma|}}\exp{ -\frac{1}{2}(X, Y)\Sigma^{-1}(X, Y)'}+\frac{1}{\sqrt{2\pi}}\exp{ -\frac{1}{2}(X, Y)(X, Y)'}} \\
	&=&\frac{ D_1 }{ D_1+D_2}
	\eeqrs}
where
\beqrs
D_1 &=& (1-\rho^2/n)^{-1/2}\exp{-(1-\rho^2/n)^{-1}(X^2-2\rho XY/\sqrt n +Y^2)/2  }, \\
D_2 &=&  \exp{-(X^2+Y^2)/2}.
\eeqrs
$\wh\theta(X,Y)$ can be obtained by replacing $\rho$ with its sample estimator $\wh \rho$. For example, we may use the maximum likelihood estimator $\wh \rho = n_1^{-1/2}\sum_{i=1}^{n_1}X_i Y_i$. We numerically evaludate the functions $\wh\theta()$ and $\theta()$ with a sequence of sample sizes, and calculate $\mu_*-\mu$. The results are summarized in Table \ref{table: condition}. As we can see, as the sample increases, $n_2^{1/2}(\mu_*-\mu )$ goes to zero. This numerically verified that the condition is satisfied under large samples.

\begin{table}[ht]
	\centering
	\caption{Numerical validation of the condition $\mu_*-\mu = o_p(n_2^{-1/2})$ under the local alternative hypothesis. $\rho = 5$. }
	\begin{tabular}{cc}  \hline
		$n$ & $n_2^{1/2}(\mu_*-\mu )$ \\   \hline 
		100 & 0.06638 \\   
		500 & 0.00981 \\   
		1000 & 0.00116 \\   
		10000 & 0.00016 \\    \hline
	\end{tabular}\label{table: condition}
\end{table}

\section{Technical Lemmas}

The analysis of the asymptotic distribution relies on the following crucial lemma.

\begin{lemma}\label{marginal_proj_h1}
	The rank-sum test statistic satisfy
	\beqr\nonumber
	R  - \mu_*&=& \frac{1}{n_2}\sum_{i\in\calI_2}\left[1 - F_{2*}\{\wh\theta(X_i, Y_i)\} - \mu_*\right] + \frac{1}{n_2}\sum_{j\in\calI_2}\left[ F_{1*}\{\wh\theta(X_j', Y_j')\}  - \mu_*\right]+O_p(n_2^{-1}),\\\label{eq: Rp_projection}
	R'  - \mu&=& \frac{1}{n_2}\sum_{i\in\calI_2}\left[1 - F_2\{\theta(X_i, Y_i)\} - \mu\right] + \frac{1}{n_2}\sum_{j\in\calI_2}\left[ F_1\{\theta(X_j', Y_j')\}  - \mu\right]+O_p(n_2^{-1}).
	\eeqr
\end{lemma}

Recall the definition for likelihood ratio (1). 
Let $L_1 = L(X, Y)$, $L_2 = L(X', Y')$, where  $(X,Y)$, $(X',Y')$ are independent realizations from $\mathcal P_{X,Y}$ and $\mathcal P_X\times \mathcal P_Y$ respectively. Let $F_{1*}(x, y)$ be the density function of $\mathcal P_{X,Y}$ and $F_{2*}(x, y)$ be the density function of $\mathcal P_X\times \mathcal P_Y$. We further define the estimated version of $L$:
\begin{equation}\label{eq:L_hat}
	\wh L(x, y) = \frac{\wh\theta(x,y)}{1-\wh\theta(x,y)}, \ \wh L_1 = \wh L(X, Y), \ \wh L_2 = \wh L(X', Y').
\end{equation}

\begin{lemma}\label{theta_to_L}
	Let  $(X,Y)$ be independent realizations from $\mathcal P_{X,Y}$ and   $(X',Y'), $  $(X'', Y'')$  be another two independent realizations from $\mathcal P_X\times \mathcal P_Y$. $ L_2' =  L(X'', Y'')$ and $\wh L_2' = \wh L(X'', Y'')$. Then
	\beqrs
	\P_* \{\wh\theta(X, Y)<\wh\theta(X', Y')\} = \P_*(\wh L_1 <\wh L_2)= \E_*\{L_2'\indic{(\wh L_2'<\wh L_2)}\},
	\eeqrs
	\beqrs
	\P \{\theta(X, Y)<\theta(X', Y')\} = \P( L_1 < L_2)= \E\{L_2'\indic{( L_2'< L_2)}\}.
	\eeqrs
\end{lemma}

\begin{lemma}\label{bound_mu}
	Let  $(X,Y)$, $(X',Y')$ be independent realizations from $\mathcal P_{X,Y}$ and $\mathcal P_X\times \mathcal P_Y$. $\wh\theta$ is any estimated classifier of the true classifier $\theta$. Then
	\beqrs
	\left|\P_* \{\wh\theta(X, Y)<\wh\theta(X', Y')\} - \P _*\{\theta(X, Y)<\theta(X', Y')\}\right| \leq 2 \E_*|\wh L_2 - L_2|\,.
	\eeqrs
\end{lemma}

\begin{lemma}\label{lem: local alternative}
	Assume the distance between the null and local alternative as defined in (6) for a sequence $\delta=o(1)$, then $\sigma^2 - \sigma_0^2 =O(\delta)$.
\end{lemma}

\begin{lemma}\label{SLLN}(Strong Law of Large Numbers for Dependent Variables)
	Let $(Z_i: i \ge 1)$ be a sequence of random variables with mean $0$ and $\sup_{i\ge 1} \mathbb E (Z_{i}^4)<\infty$. Assume that $Z_i$ and $Z_j$ are independent whenever $|i-j|\notin\{0,1,n-1\}$. Let $S_n = \sum_{i=1}^n Z_i$. Then
	\beqrs
	\lim_{n\rightarrow\infty}\frac{S_n}{n} = 0 \quad \text{almost surely.}
	\eeqrs
\end{lemma}

We state two definitions for the readers' convenience.

\begin{definition}($\ep$-bracket)
	Let $U\in\mR^m$ be a random vector. 
	Given two functions $l(\cdot)$ and $u(\cdot)$, the bracket $[l, u]$ is the set of all functions $f\in\calF$ with $l(U)\leq f(U)\leq u(U)$, for all $U\sim \calP_U$. An $\ep$-bracket is a bracket $[l,u]$ with $\E_U|l(U)-u(U)|<\ep$.
\end{definition}
\begin{definition}(Bracketing number)
	When $U\sim\calP_U$, the bracketing number $N_{[\mkern5mu ]} (\ep, \calF, \calP_U)$ is the minimum number of $\ep$-brackets needed to cover $\calF$.
\end{definition}

\begin{lemma}\label{GC}
	$\calM = \{ M(X, Y, X', Y'; \beta) ,  \beta\in \mathcal{B}\}$ be a class of measurable functions such that $N_{[\mkern5mu ]} (\ep, \calM, \calP)<\infty$ for every $\ep>0$, and and $\E\left[\{  M(X, Y, X', Y'; \beta) \}^4\right]<\infty$. Then 
	\beqrs
	\sup_{M\in\calM} |\frac{1}{n_2}\sum_{i\in\calI_2}M(X_i, Y_i, X_{i}, Y_{i+1}; \beta) - \E\{M(X, Y, X', Y'; \beta) \}|\rightarrow 0, \quad \text{ almost surely.}
	\eeqrs
\end{lemma}

\section{Proofs }

\subsection{Proof of Proposition 1}

\begin{proof} 
	Let $r(\cdot)=(d\mathcal P/d\mathcal Q)(\cdot)$. Let $W',W\stackrel{iid}{\sim}\mathcal Q$. The claimed result follows by combining the equality
	\begin{equation}\label{eq:TV_bound_1}
		(1/2) - \mathbb E r(W')\ind\{r(W')<r(W)\} = \frac{1}{4}\mathbb E|r(W)-r(W')|
	\end{equation}
	and the inequality
	\begin{equation}\label{eq:TV_bound_2}
		d_{\rm tv}(\mathcal P,\mathcal Q)\le \mathbb E|r(W)-r(W')|\le 2 d_{\rm tv}(\mathcal P,\mathcal Q)\,.
	\end{equation}
	To prove \eqref{eq:TV_bound_1}, 
	\begin{align*}
		&\mathbb E|r(W)-r(W')|\\
		=& \mathbb E (r(W)-r(W'))\ind\{r(W)>r(W')\} +\mathbb E (r(W')-r(W))\ind\{r(W')>r(W)\}\\
		=& 2\mathbb E \{r(W)-r(W')\}\ind\{r(W)>r(W')\} \\
		=& 2\left[\mathbb E r(W)\ind\{r(W)>r(W')\} - 1/2 + 1/2-\mathbb E r(W')\ind\{r(W')<r(W)\}\right]\\
		=& 4 \left[1/2-\mathbb E r(W')\ind\{r(W')<r(W)\}\right]\,,
	\end{align*}
	where the last equality follows from
	$$
	\mathbb E r(W)\ind\{r(W)>r(W')\}=\mathbb E r(W')\ind\{r(W')>r(W)\}
	$$
	and
	$$
	\mathbb E r(W')\ind\{r(W')>r(W)\} + \mathbb E r(W')\ind\{r(W')<r(W)\} = \mathbb E r(W')=1
	$$
	by the construction of $r(W)$ and its continuity.
	
	To prove \eqref{eq:TV_bound_2}, observe that $d_{\rm tv}(\mathcal P,\mathcal Q)=\mathbb E|r(W)-1|$. For the lower bound we have by Jensen's inequality
	\begin{align*}
		\mathbb E|r(W)-r(W')|=\mathbb E\left\{\mathbb E|r(W)-r(W')|\big| W\right\}\ge \mathbb E |r(W)-1|=d_{\rm tv}(\mathcal P,\mathcal Q)\,.
	\end{align*}
	For the upper bound,
	\begin{align*}
		\mathbb E |r(W)-r(W')|\le \mathbb E |r(W)-1|+\mathbb E|r(W')-1| = 2d_{\rm tv}(\mathcal P,\mathcal Q)\,.
	\end{align*}
\end{proof}

\subsection{Proof of Lemma \ref{marginal_proj_h1}}
\begin{proof} 
	It suffices to prove the first equation, and the other equation follows similar reasons. 
	For notation simplicity, define $$h\{(X_i, Y_i), (X_j', Y_j') \}  = \indic\left\{\wh\theta(X_i, Y_i)<\wh\theta(X_j', Y_j') \right\}.$$
	Then we have
	\beqr\label{eq:h1_lhs}
	R - \mu_* = \frac{1}{n_2^2}\sum_{i, j\in\calI_2} \left[h\{(X_i, Y_i), (X_j', Y_j') \} - \mu_*\right] . 
	\eeqr
	The terms $j\in \{i-1,i\}$ only contributes to $2n_2$ terms in the sum, and is    $O(1/n_2)$ after dividing by $n_2^2$ because $h\{(X_i, Y_i), (X_j', Y_j') \} - \mu\in [-1, 1]$.  For the other terms, we consider the marginal projection of the two-sample kernel $h$.  Let $(X,Y)$, $(X',Y')$ be independent samples from $\mathcal P_{X,Y}$ and $\mathcal P_X\times\mathcal P_Y$, respectively. Then, by continuity of $F_{1*}$ and $F_{2*}$,
	\begin{align*}
		\mathbb E_*\left[ h\{(X,Y),(X',Y')\}-\mu\mid X,Y\right] = & 1 -F_{2*}\{\wh\theta(X,Y)\}-\mu_*\,,\\
		\mathbb E_*\left[ h\{(X,Y),(X',Y')\}-\mu\mid X',Y'\right] = & F_{1*}\{\wh\theta(X',Y')\}-\mu_*\,.
	\end{align*}
	Define
	\beqrs
	h^{\dagger}\{(X_i, Y_i), (X_j', Y_j') \} &\defby&  h\{(X_i, Y_i), (X_j', Y_j') \} -  \mu_* - \left[1 - F_{2*}\{\wh\theta(X_i, Y_i)\} - \mu_*\right]  \\ 
	&&- \left[ F_{1*}\{\wh\theta(X_j', Y_j')\}  - \mu_*\right]\,,
	\eeqrs 
	so that
	\begin{align}
		h\{(X_i, Y_i), (X_j', Y_j') \} - \mu_*= & h^{\dagger}\{(X_i, Y_i), (X_j', Y_j') \}+\left[1 - F_{2*}\{\wh\theta(X_i, Y_i)\} - \mu_*\right]\nonumber  \\
		&+\left[ F_{1*}\{\wh\theta(X_j', Y_j')\}  - \mu_*\right]. \label{eq:marginal_projection}
	\end{align}
	Plugging \eqref{eq:marginal_projection} into \eqref{eq:h1_lhs} for the pairs $j\notin\{i-1,i\}$,
	each $F_{2*}\{\wh\theta(X_i,Y_i)\}$ and $F_{1*}\{\wh\theta(X_j', Y_j')\}$ appear exactly $n_2-2$ times in the sum. 
	Thus \eqref{eq:h1_lhs} and \eqref{eq:marginal_projection} imply
	\beqrs
	R-\mu_*&=&O(n_2^{-1})+\frac{1}{n_2}\sum_{i\in\calI_2}\left[1 - F_{2*}\{\wh\theta(X_i, Y_i)\} - \mu_*\right] + \frac{1}{n_2}\sum_{j\in\calI_2}\left[ F_{1*}\{\wh\theta(X_j', Y_j')\}  - \mu_*\right]\\
	&&+\frac{1}{n_2^2}\sum_{j\notin\{i-1,i\}} h^{\dagger}\{(X_i, Y_i), (X_j', Y_j') \} .
	\eeqrs
	It suffices to show that
	\beqr\label{f_tilde_h1}
	\frac{1}{n_2^2}\sum_{j\notin\{i-1,i\}} h^{\dagger}\{(X_i, Y_i), (X_j', Y_j') \} = O_P(n_2^{-1})\,.
	\eeqr
	Consider 
	\begin{align}\nonumber
		& \E_*\left[\left(\frac{1}{n_2^2}\sum_{j\notin\{i-1,i\}}h^{\dagger}\{(X_i, Y_i), (X_j', Y_j') \} \right)^2\right]\\\label{h_sum_h1}
		= & \frac{1}{n_2^4}\sum_{(i,j),(i',j')} \E_*\left[h^{\dagger}\{(X_i, Y_i), (X_j', Y_j') \}h^{\dagger}\{(X_{i'}, Y_{i'}), (X_{j'}', Y_{j'}') \}\right]
	\end{align}
	where the sum is over all pairs $(i,j)$ and $(i',j')$ such that $j\notin \{i-1,i\}$ and $j'\notin\{i'-1,i'\}$. Consider the following two scenarios:
	\begin{enumerate}
		\item [(a)] $i'=i$ or $i'\in\{j,j+1\}$;
		\item [(b)] $j'\in\{i,i-1\}$ or $j'\in\{j-1, j, j+1 \}$.
	\end{enumerate}
	Then \eqref{f_tilde_h1} follows by combining the following two facts: (i) If at most one of (a), (b) holds, then 
	\beqrs
	\E_*\left[h^{\dagger}\{(X_i, Y_i), (X_j', Y_j') \} h^{\dagger}\{(X_{i'}, Y_{i'}), (X_{j'}', Y_{j'}') \}\right] =0
	\eeqrs
	because at least one of $(X_i,Y_i),(X_j', Y_j'),(X_{i'},Y_{i'}),(X_{j'},Y_{j'+1})$ is independent of the other three and the conditional expectation of $h^{\dagger}\{(X_i, Y_i), (X_j', Y_j') \} h^{\dagger}\{(X_{i'}, Y_{i'}), (X_{j'}', Y_{j'}') \}$ given the other three is zero, and (ii) If both (a) and (b) hold, then the number of terms in \eqref{h_sum_h1} is $O(n_2^2)$.
\end{proof}

\subsection{Proof of Lemma \ref{theta_to_L}}
\begin{proof} 
	The proof of the two results are identical and it suffices to show the first one.
	$\P_* \{\wh\theta(X, Y)<\wh\theta(X', Y')\} = \P_*(\wh L_1 <\wh L_2)$ follows trivially because $\wh L(x,y)$ is a monotone increasing transformation of $\wh\theta(x,y)$. Furthermore, by definition
	\beqrs
	\P_*(\wh L_1 <\wh L_2)&=&  \E_*\{\indic{(\wh L_1 <\wh L_2)}\} \\
	&=& \E_*\{\frac{F_{2*}(x, y)}{F_{1*}(x, y)}\frac{F_{1*}(x, y)}{F_{2*}(x, y)}\indic{(\wh L(x, y) <\wh L(x', y'))} \}\\
	&=&\int_{\indic{\{\wh L(x, y) <\wh L(x', y')\}}} F_{2*}(x, y) L(x, y) F_{2*}(x', y') dx dy dx' dy'\\
	&=&\E_*\{L_2'\indic{(\wh L_2'<\wh L_2)}\}.
	\eeqrs
	The second to the last equality follows because $L(x,y) = F_{1*}(x,y)/F_{2*}(x,y)$. 
	The last equality follows by replacing the notation $(x, y)$ with $(x'', y'')$.
\end{proof}

\subsection{Proof of Lemma \ref{bound_mu}}

\begin{proof} 
	By Lemma \ref{theta_to_L}, using the notation in \eqref{eq:L_hat}, we have
	\beqrs
	&&\left|\P _*\{\wh\theta(X, Y)<\wh\theta(X', Y')\} - \P \{\theta(X, Y)<\theta(X', Y')\}\right|  \\
	&=&\left|\P_* \{\wh L_1<\wh L_2\} - \P\{L_1 < L_2\}\right|\\
	&=& \left| \E_*\{L_2'\indic{(\wh L_2'<\wh L_2)}\} - \E\{L_2'\indic{( L_2'< L_2)}\}\right|\\
	&=& \left| \E_*\{L_2'\indic{(\wh L_2'<\wh L_2)}\} - \E_*\{L_2'\indic{( L_2'< L_2)}\}\right|.
	\eeqrs
	The last equality follows because $L_2'\indic{( L_2'< L_2)}$ is independent of the first subset of the data. 
	Now we study $\E_*\{L_2'\indic{(\wh L_2'<\wh L_2)}\}$. Let $\gamma = L_2 - \wh L_2 + \wh L_2' - L_2'$. Thus we have
	\beqrs
	&&\E_*\{L_2'\indic{(\wh L_2'<\wh L_2)}\} \\
	&=& \E_*\{L_2'\indic{( L_2'< L_2)}\}  + \E_*\{ L_2' \indic{(L_2'\leq L_2\leq L_2' +\gamma, \gamma>0)}\} \\
	&&-\E_*\{ L_2' \indic{(L_2' +\gamma\leq L_2\leq L_2', \gamma<0)}\}\\
	&=& \E_*\{L_2'\indic{( L_2'< L_2)}\}  + \E_*\{ L_2' \indic{(L_2'\leq L_2\leq L_2' +\gamma, \gamma>0)}\} \\
	&&-\E_*\{ L_2 \indic{(L_2'< L_2\leq L_2' +\gamma, \gamma>0)}\},\\
	\eeqrs
	where the last equality follows by changing the roles of $(X', Y')$ and $(X'', Y'')$ in the expectation.
	Applying this result we have
	\beqrs
	&&\left| \E_*\{L_2'\indic{(\wh L_2'<\wh L_2)}\} - \E_*\{L_2'\indic{( L_2'< L_2)}\}\right|\\
	&=& \left|\E_*\{ (L_2' - L_2) \indic{(L_2'\leq L_2\leq L_2' +\gamma, \gamma>0)}\}   \right|\\
	&\leq& \E _*\{|L_2' - L_2|\indic (|L_2 - L_2'|\leq|\gamma|)\}\\
	&\leq&\E_*|\gamma|\leq 2 \E_*|\wh L_2 - L_2|\,.
	\eeqrs
\end{proof}

\subsection{Proof of Lemma \ref{lem: local alternative}}

By Proposition 1, we know that $d_{tv}(F_1, F_2)\le d_{tv}(\mathcal P_{X,Y},\mathcal P_X\times\mathcal P_Y)\lesssim \delta$, where ``$\lesssim$'' means ``upper bounded up to a constant factor''. Thus 
\beqrs
\sup_t|F_1(t) - F_2(t)| \le C\delta\,,	
\eeqrs 
where $C$ is a constant.
Thus $\var(F_1\{\theta(X_2, Y_3)\}) = \var(F_2\{\theta(X_2, Y_3)\} + C\delta) = 1/12 + O(\delta)$. Similarly, $\var(F_2\{\theta(X_2, Y_2)\}) = 1/12 + O(\delta)$.

Furthermore, by definition of the total variation distance, we can construct $(\tilde X_2,\tilde Y_2)\sim \mathcal P_{X}\times \mathcal P_Y$ such that $\mathbb P((\tilde X_2,\tilde Y_2)\neq (X_2,Y_2))\le d_{tv}(\mathcal P_{X,Y},\mathcal P_X\times\mathcal P_Y)\le C\delta$. 
Hence 
$$\left|\cov(F_1\{\theta(X_1, Y_2)\}, F_1\{\theta(X_2, Y_3)\})\right| \le \left|\cov(F_1\{\theta(X_1, \tilde Y_2)\}, F_1\{\theta(\tilde X_2, Y_3)\})\right|+O(\delta)=O(\delta)\,.$$

With the preparations above, we are ready to calculate the $\cov(V_1+V_2+V_3,V_2)$.
Note that $V_i = F_{1}\{\theta(X_i', Y_i')\}  - F_{2}\{\theta(X_i, Y_i)\}$. We have
\beqrs
&&\cov(V_1+V_2+V_3,V_2) \\
&=& \cov(F_1\{\theta(X_1, Y_2)\}- F_2\{\theta(X_1, Y_1)\}, F_1\{\theta(X_2, Y_3)\}- F_2\{\theta(X_2, Y_2)\})  \\
&&+ \cov(F_1\{\theta(X_2, Y_3)\}- F_2\{\theta(X_2, Y_2)\}, F_1\{\theta(X_2, Y_3)\}- F_2\{\theta(X_2, Y_2)\})   \\
&&+\cov(F_1\{\theta(X_3, Y_4)\}- F_2\{\theta(X_3, Y_3)\}, F_1\{\theta(X_2, Y_3)\}- F_2\{\theta(X_2, Y_2)\})\\
&:=& B_1+B_2+B_3.
\eeqrs
We calculate each term separately.
\beqrs
B_1 &=& \cov(F_1\{\theta(X_1, Y_2)\}, F_1\{\theta(X_2, Y_3)\})  -   \cov(F_1\{\theta(X_1, Y_2)\}, F_2\{\theta(X_2, Y_2)\})\\
&=& O(\delta)- \cov(F_1\{\theta(X_1, Y_2)\}, F_2\{\theta(X_2, Y_2)\}) .
\eeqrs
The other two terms in $B_1$ are equal to 0 because $(X_1, Y_1)$ is independent of $(X_2, Y_3)$ and $(X_2, Y_2)$.
\beqrs
B_2 &=& \cov(F_1\{\theta(X_2, Y_3)\}, F_1\{\theta(X_2, Y_3)\})  + \cov(F_2\{\theta(X_2, Y_2)\}, F_2\{\theta(X_2, Y_2)\})  \\
&&- \cov(F_1\{\theta(X_2, Y_3)\}, F_2\{\theta(X_2, Y_2)\})  - \cov(F_2\{\theta(X_2, Y_2)\}, F_1\{\theta(X_2, Y_3)\}) \\
&=& 1/6 -  2\cov(F_2\{\theta(X_2, Y_2)\}, F_1\{\theta(X_2, Y_3)\}) + O(\delta).
\eeqrs
Now we deal with $B_3$.
\beqrs
B_3 &=& \cov(F_1\{\theta(X_3, Y_4)\}, F_1\{\theta(X_2, Y_3)\}) 
- \cov(F_2\{\theta(X_3, Y_3)\}, F_1\{\theta(X_2, Y_3)\})\\
&=&\cov(F_1\{\theta(X_1, Y_2)\}, F_1\{\theta(X_2, Y_3)\}) 
- \cov(F_2\{\theta(X_2, Y_2)\}, F_1\{\theta(X_1, Y_2)\})\\
&=&O(\delta) - \cov(F_2\{\theta(X_2, Y_2)\}, F_1\{\theta(X_1, Y_2)\}).
\eeqrs
The two terms in $B_3$ are also zero because $(X_3, Y_3)$ and $(X_3,Y_4)$ are independent of $(X_2, Y_2)$.
Combining $B_1$, $B_2$ and $B_3$, we get $\sigma^2 - \sigma_0^2 = O(\delta)$.

\subsection{Proof of Lemma \ref{SLLN}}

\begin{proof} 
	By the Chebyshev inequality, $\forall\ep>0$,
	\beqrs
	\P(|S_n|>n\ep) \leq \frac{1}{(n\ep)^4}\E(S_n^4).
	\eeqrs
	Now we study the upper bound for $\E(S_n^4)$. For simplicity, we call the pair of index $(i, j)$ dependent pair if $|i-j|\in\{0, 1, n-1\}$. Note that
	\beqrs
	\E(S_n^4) = \E(\sum_{i,j,k,l} Z_iZ_jZ_kZ_l) = \sum_{i,j,k,l} \E(Z_iZ_jZ_kZ_l).
	\eeqrs
	where the sum is over all $1\leq i,j,k,l\leq n$. Denote $A^* = \{(i,j), (i,k), (i,l), (j,k), (j,l), (k,l) \}$. Consider the following scenarios:
	\begin{itemize}
		\item[(a)] $A^*$ contains at most one dependent pairs. Then $\E(Z_iZ_jZ_kZ_l) = 0$.
		\item[(b)] $A^*$ contains at least two dependent pairs. $\E(Z_iZ_jZ_kZ_l)$ may not be 0. But the number of such terms $\E(Z_iZ_jZ_kZ_l)$ is of order $O(n^2)$.
	\end{itemize} 
	Thus there exists a constant $C>0$, such that $\E(S_n^4) \leq C n^2 $ for all positive integers $n$. It follows that
	\beqrs
	\sum_{n\geq 1} \P(|S_n|>n\ep)\leq \sum_{n\geq 1} \frac{C}{n^2\ep^4}< \infty.
	\eeqrs
	The claimed result follows from the Borel-Cantelli lemma.
\end{proof}

\subsection{Proof of Lemma \ref{GC}}

\begin{proof} 
	Let $\ep>0$ be a fixed number. We begin with choosing finitely many $\ep$-brackets $[l_i, u_i]$ whose union covers $\calM$. For simplicity, let $Z_j^*$ denote $(X_j, Y_j, X_{j}, Y_{j+1})$, and $Z^*$ denote $(X, Y, X', Y')$. Then for every $M\in\calM$, there exists a bracket such that
	\beqrs
	\frac{1}{n}\sum_{j=1}^nM(Z_j^*; \beta)- \E M(Z^*; \beta) &\leq& \{\frac{1}{n}\sum_{j=1}^nu_i(Z_j^*) - \E u_i(Z^*)\} + \E u_i(Z^*) - \E M(Z^*; \beta) \\
	&\leq& \{\frac{1}{n}\sum_{j=1}^nu_i(Z_j^*) - \E u_i(Z^*)\} + \ep.
	\eeqrs
	Thus we have
	\beqrs
	\sup_{M\in\calM}  \frac{1}{n}\sum_{j=1}^nM(Z_j^*; \beta) - \E M(Z^*; \beta) \leq \max_{i}\{\frac{1}{n}\sum_{j=1}^nu_i(Z_j^*) - \E u_i(Z^*)\} + \ep.
	\eeqrs
	By Lemma \ref{SLLN}, the right hand side converges almost surely to $\ep$. Similarly, we have
	\beqrs
	\frac{1}{n}\sum_{j=1}^nM(Z_j^*; \beta) - \E M(Z^*; \beta) &\geq& \{\frac{1}{n}\sum_{j=1}^nl_i(Z_j^*) - \E l_i(Z^*)\} + \E l_i(Z^*) - \E M(Z^*; \beta) \\
	&\geq& \{\frac{1}{n}\sum_{j=1}^nl_i(Z_j^*) - \E l_i(Z^*)\} - \ep.
	\eeqrs
	Thus we have
	\beqrs
	\inf_{M\in\calM}  \frac{1}{n}\sum_{j=1}^nM(Z_j^*; \beta) - \E M(Z^*; \beta) \geq \min_{i}\{\frac{1}{n}\sum_{j=1}^nl_i(Z_j^*) - \E l_i(Z^*)\} - \ep.
	\eeqrs
	Similarly, the right hand side converges to $-\ep$ almost surely. It follows that
	\beqrs
	&&\sup_{M\in\calM} | \frac{1}{n}\sum_{j=1}^nM(Z_j^*; \beta) - \E M(Z^*; \beta)| \\
	=&&\max \left\{\sup_{M\in\calM}  \frac{1}{n}\sum_{j=1}^nM(Z_j^*; \beta) - \E M(Z^*; \beta),  -\inf_{M\in\calM}  \frac{1}{n}\sum_{j=1}^nM(Z_j^*; \beta) - \E M(Z^*; \beta) \right\}.
	\eeqrs
	Thus $\lim\sup  | \frac{1}{n}\sum_{j=1}^nM(Z_j^*; \beta) - \E M(Z^*; \beta)|_{M\in\calM}\leq \ep$ almost surely for every $\ep>0$. Thus it holds almost surely that
	\beqrs
	\sup_{M\in\calM} |\frac{1}{n_2}\sum_{i\in\calI_2}M(X_i, Y_i, X_{i}, Y_{i+1}; \beta) - \E\{M(X, Y, X', Y'; \beta) \}|\rightarrow 0.
	\eeqrs
	
\end{proof}

\subsection{Proof of Theorem 3}

\begin{proof} 
	
	Under $H_0$.  We have $F_{1*}(\cdot) = F_{2*}(\cdot)$.  By Lemma \ref{marginal_proj_h1},
	\beqr\label{eq: R_projection_null}
	R  - \frac{1}{2} &=& \frac{1}{n_2}\sum_{i\in\calI_2}\left[\frac{1}{2} - F_{2*}\{\wh\theta(X_i, Y_i)\} \right] + \frac{1}{n_2}\sum_{i\in\calI_2}\left[ F_{2*}\{\wh\theta(X_i', Y_i')\}  - \frac{1}{2}\right]+O_p(n_2^{-1}).
	\eeqr
	Thus $R - 1/2 = \tilde{R} + O_p(n_2^{-1})$. 
	Let $g_1(X)=\E [F_{2*}\{\wh \theta(X, Y)\}|X]-1/2$ and $g_2(Y)=\E [F_{2*}\{\wh \theta(X, Y)\}|Y]-1/2$ and $\wt g (X, Y)= F_{2*}\{\wh \theta(X, Y)\} - g_1(X) - g_2(Y) - 1/2$.  Then
	\begin{align*}
		\tilde R = &  \frac{1}{n_2}\sum_{i\in\calI_2} \left[\wt g(X_i,Y_{i+1}) + g_1(X_i)+g_2(Y_{i+1}) -  \wt g(X_i,Y_{i})-g_1(X_i)-g_2(Y_{i})\right]\\
		=& \frac{1}{n_2} \sum_{i\in\calI_2} \wt g(X_i,Y_{i+1}) - \frac{1}{n_2}\sum_{j\in\calI_2}\wt g(X_j,Y_{j})\,,
	\end{align*}
	Note that $\E _*\{\wt g(X_j,Y_j)\wt g(X_i,Y_{i+1})\}=0$ for all $i,j$.  This is because when $i\neq j$ and $i\neq j+1$, the two terms are independent and $\tilde g(X,Y)$ has mean $0$.  When $i=j$, it reduces to 
	\beqrs
	\E _*\{\wt g(X,Y)\wt g(X,Y')\}=\E_*\left[\E_*\left(\{\wt g(X,Y)\wt g(X,Y')|X\right\}\right]=\E_*\left[\E_*\left\{\wt g(X,Y)|X\right\}\E_*\left\{\wt g(X,Y')|X\right\}\right]=0.
	\eeqrs
	The case of $i=j+1$ is similar. Therefore we have
	$$
	\var_*(\tilde R) = \frac{2}{n_2}\var_*\{\tilde g(X_1,Y_1)\}\,.
	$$
	By assumption, $\tilde g(X,Y)=F_{2*}\{\wh \theta(X, Y)\} - g_1(X)-g_2(Y)-1/2$ is non-degenerate and  $\var_*\{\tilde g(X_1,Y_1)\}:=\sigma_*^2/2>0$. 
	Moreover, $F_{1*}(\cdot)$ and $F_{2*}(\cdot)$ both follow uniform distribution and has variance $1/12$. By Lemma \ref{marginal_proj_h1}, we calculate that
	\beqrs
	\var_*(\wt R) = \frac{1}{6n_2} - \frac{2}{n_2}\mathrm{Cov}_*\left[F_{2*}\{\wh\theta(X_2, Y_2)\},F_{1*}\{\wh\theta(X_{1}, Y_{2})\} \right]-
	\frac{2}{n_2}\mathrm{Cov}_*\left[F_{2*}\{\wh\theta(X_2, Y_2)\},F_{1*}\{\wh\theta(X_{2}, Y_{3})\} \right],
	\eeqrs
	which shows that $\var_*(\tilde R)= n_2^{-1}\sigma_*^2$. 
	By construction and the null hypothesis, the $2n_2$ random variables $\tilde g(X_i,Y_{i})$ and $\tilde g(X_i',Y_i')$ have a 3-regular dependence graph, therefore by Theorem 2.2 of \cite{baldi1989normal} we have
	
	\begin{equation}\label{eq:proof_br}
		\sup_{s\in\mR}\left| \P_*\left(\frac{\sqrt{n_2}\tilde R  }{\sigma_* }\leq s\right) - \Phi(s) \right| \leq c(\sqrt{A_3} +\sqrt{A_4})
	\end{equation}


	Now we proceed under the assumption that $n_2^{1/6}\sigma_*\stackrel{p}{\rightarrow}\infty$ (note that in this case we are assuming that $(n_1,n_2)$ changes simultaneously).  Fix an $\epsilon>0$, the assumption $n_2^{1/6}\sigma_*\stackrel{p}{\rightarrow}\infty$ guarantees there exists $(n_{1,0},n_{2,0})$ such that $\mathbb P(n_2^{1/6}\sigma_* \le \epsilon^{-2/3})\le \epsilon$ whenever $n_1\geq n_{1,0}$ and $n_2\geq n_{2,0}$.
	
	Let $T=\sqrt{n_2}\tilde R/\sigma_*$.   Now the $\sqrt A_3$ term dominates the right hand side of \eqref{eq:proof_br}, which can be bounded by $C/(n_2^{1/4}{\sigma_*}^{3/2})$ for some universal constant $C$. Then
	\begin{align*}
		\mathbb P(T\le s) \le & \mathbb P(T\le s |n^{1/6}\sigma_* \ge \epsilon^{-2/3})\mathbb P(n_2^{1/6}\sigma_*\ge \epsilon^{-2/3})+\mathbb P(n_2^{1/6}\sigma_*< \epsilon^{-2/3})\\
		\le & (\Phi(s)+C\epsilon)+\epsilon =\Phi(s) +(1+C)\epsilon\,.
	\end{align*}
	On the other hand
	\begin{align*}
		\mathbb P(T\le s) \ge &  \mathbb P(T\le s |n^{1/6}\sigma_* \ge \epsilon^{-2/3})\mathbb P(n_2^{1/6}\sigma_*\ge \epsilon^{-2/3})\\
		\ge & (\Phi(s)-C \epsilon)(1-\epsilon) \ge \Phi(s)-(1+C)\epsilon\,.
	\end{align*}
	This establishes that $T$ converges in distribution to $N(0,1)$ unconditionally.
	
	Now we analyze $\hat\sigma^2$. Using the Dvoretzky-Kiefer-Wolfowitz inequality we have $\|\hat F-F\|_\infty=O_P(1/\sqrt{n_2})$, so
	\beqrs
	&&\left|\wh F_{1*}\{\wh\theta(X_i, Y_i)  \}\wh F_{1*}\{\wh\theta(X_i, Y_{i+1}) \} - F_{1*}\{\wh\theta(X_i, Y_i)  \} F_{1*}\{\wh\theta(X_i, Y_{i+1}) \}\right| \\
	= && \left|\wh F_{1*}\{\wh\theta(X_i, Y_i)  \} \left[ \wh F_{1*}\{\wh\theta(X_i, Y_{i+1}) \} -  F_{1*}\{\wh\theta(X_i, Y_{i+1}) \} \right] \right|  \\
	&&+\left|F_{1*}\{\wh\theta(X_i, Y_{i+1}) \}\left[ \wh F_{1*}\{\wh\theta(X_i, Y_i)  \} -   F_{1*}\{\wh\theta(X_i, Y_i)  \} \right]  \right|\\
	=&& O_p(n_2^{-1/2}).
	\eeqrs
	Combining this with the fact that the difference between $\sigma_*^2$ and the empirical version using the true $F$ function is just the difference between sample mean and the population mean for a random varialbe uniformly bounded by $1$, we have
	\begin{align*}
		\hat\sigma^2-\sigma_*^2 = O_P(1/\sqrt{n_2})
	\end{align*}
	So that
	\begin{align*}
		\frac{\hat\sigma^2}{\sigma_*^2}-1 = O_{P}(1)\frac{1}{\sqrt{n_2}\sigma_*^2} = o_P(1)
	\end{align*}
	because by assumption $n_2^{1/2}\sigma_*^2\ge n_2^{1/3}\sigma_*^2\stackrel{p}{\rightarrow}\infty$.
	
	Finally,
	\begin{align*}
		\frac{\sqrt{n_2}(R-1/2)}{\hat\sigma} = &\frac{\sigma_*}{\hat\sigma} \frac{\sqrt{n_2}(R-1/2)}{\sigma_*}\\
		=&\frac{\sigma_*}{\hat\sigma} \left( \frac{\sqrt{n_2}\tilde R}{\sigma_*}+ \frac{\sqrt{n_2}(R-1/2-\tilde R)}{\sigma_*}\right)\\
		\rightsquigarrow N(0,1)
	\end{align*}
	because $\hat\sigma/\sigma_*=1+o_P(1)$ and $\sqrt{n_2}(R-1/2-\tilde R)/\sigma_*=O_P(1/(n_2^{1/2}\sigma_*))=o_P(1)$.
\end{proof}

\subsection{Proof of Theorem 6}

\begin{proof} 
	We first show that $R = \mu_* + O_p(n_2^{-1/2})$.  Let $$R_{\mu} = \frac{1}{n_2}\sum_{i\in\calI_2}\left[1 - F_{2*}\{\wh\theta(X_i, Y_i)\} - \mu_*\right] + \frac{1}{n_2}\sum_{j\in\calI_2}\left[ F_{1*}\{\wh\theta(X_j', Y_j')\}  - \mu_*\right].$$ 
	By Lemma \ref{marginal_proj_h1},  $R - \mu_*= R_{\mu} +O_p(n_2^{-1})$.  Note that $\mathbb E_*R_\mu =0$  and $\var_*(R_{\mu})=\sigma_*^2/n_2 = \cov_*(\wh V_1+\wh V_2+\wh V_3,\wh V_2)/n_2$ similarly as in the proof of Theorem 3. $\wh V_i = F_{1*}\{\wh\theta(X_i', Y_i')\} - F_{2*}\{\wh\theta(X_i, Y_i)\}$ Thus $R_{\mu} = O_p(n_2^{-1/2})$ and $R = \mu_* + O_p(n_2^{-1/2})$.

	Note that the rank sum comparison is invariant with respect to any monotone transformation on $\wh\theta$. Thus one can easily replace $\wh\theta$ with $g(\wh\theta)$, where $g(\cdot)$ is a strictly monotone function. By Lemma \ref{bound_mu} and condition (5), we have
	\beqrs
	\left|\mu_* - \P \{\theta(X, Y)<\theta(X', Y')\}\right| \leq \frac{1}{2} - \mu - 2c.
	\eeqrs
	Because $\P \{\theta(X, Y)<\theta(X', Y')\}<1/2$ under $H_1$, we have $\mu_*<1/2 - 2c$ holds with probability tending to 1. Thus as $n_1, n_2\rightarrow\infty$,
	$$
	\sqrt{n_2}(R - 1/2) = \sqrt{n_2}(\mu_* - 1/2) + O_p(1) \rightarrow -\infty
	$$
	holds in probability. The result follows because $\wh\sigma^2$ is upper bounded by constant $7/6$.
\end{proof}

\subsection{Proof of Theorem 7}

\begin{proof} 
	
	Because
	\beqrs
	\frac{\sqrt{n_2}(R - 1/2)}{\wh \sigma} &=&    \frac{\sqrt{n_2}(R'- \mu)}{\wh \sigma} + \frac{\sqrt{n_2}(R - R')}{\wh \sigma} + \frac{\sqrt{n_2}(\mu- 1/2)}{\wh \sigma} \\
	&=&  \frac{\sqrt{n_2}(R'- \mu)}{ \sigma_0}\frac{\sigma_0}{\wh\sigma}  + \frac{\sqrt{n_2}(R - R')}{ \sigma}\frac{\sigma}{\wh\sigma} + \frac{\sqrt{n_2}(\mu- 1/2)}{\wh \sigma} 
	\eeqrs
	To deal with the three terms, we can divide our proof into four steps.
	
	{\bf Step 1: } 	We begin by showing that ratio $\sigma/\wh\sigma  = 1 + o_p(1)$, and $\sigma_0/\wh\sigma = 1+o_p(1)$. Following similar proof as in Theorem 3, we can show that $\sigma_*^2/\wh\sigma^2-1=o_p(1)$. By Lemma \ref{lem: local alternative},  $\sigma_0^2 /\sigma^2 -1=  o(1)$.
	Note that
	\beqrs
	\frac{\sigma}{\wh\sigma} = \frac{\sigma}{\sigma_*}\times \frac{\sigma_*}{\wh\sigma},\quad \mbox{and}\quad \frac{ \sigma_0}{\wh\sigma} = \frac{\sigma_0}{\sigma}\times\frac{\sigma}{\sigma_*}\times\frac{\sigma_*}{\wh\sigma}.
	\eeqrs
	Thus it suffices to show that $\sigma^2/\sigma_*^2 - 1 = o_p(1)$.

	We first have
	\beqrs
	&&\E_*\left[ F_{2*}\{\wh\theta(X_2, Y_2)\}F_{1*}\{\wh\theta(X_{1}, Y_{2})\} \right]- \E\left[ F_2\{\theta(X_2, Y_2)\}F_1\{\theta(X_{1}, Y_{2})\}\right] \\
	=&&\E_*\left[ F_{2*}\{\wh\theta(X_2, Y_2)\}F_{1*}\{\wh\theta(X_{1}, Y_{2})\} - F_2\{\theta(X_2, Y_2)\}F_1\{\theta(X_{1}, Y_{2})\}\right] \\
	=&&\E_*\left[ F_{2*}\{\wh\theta(X_2, Y_2)\}F_{1*}\{\wh\theta(X_{1}, Y_{2})\} -  F_{2*}\{\wh\theta(X_2, Y_2)\}F_1\{\theta(X_{1}, Y_{2})\}  + \right. \\
	&&\left. F_{2*}\{\wh\theta(X_2, Y_2)\}F_1\{\theta(X_{1}, Y_{2})\} -
	F_2\{\theta(X_2, Y_2)\}F_1\{\theta(X_{1}, Y_{2})\}\right] \\
	\leq&&\E_*\left[ \left|F_{1*}\{\wh\theta(X_{1}, Y_{2})\} -  F_1\{\theta(X_{1}, Y_{2})\}\right| \right]  +\E_* \left[\left|F_{2*}\{\wh\theta(X_2, Y_2)\}-
	F_2\{\theta(X_2, Y_2)\}\right|\right] = o_p(1),
	\eeqrs
	where the last equation holds by condition (8). 
	By the assumption of $\mu_* - \mu = o_p(n_2^{-1/2})$, we have $\E\left[F_1\{\theta(X_i', Y_i')\}\right] =\E_*\left[F_{1*}\{\wh\theta(X_i', Y_i')\}\right] +o_p(1)$. Similarly,  $\E\left[F_2\{\theta(X_i, Y_i)\}\right] = \E_*\left[F_{2*}\{\wh\theta(X_i, Y_i)\}\right] +o_p(1)$.  
	It follows that
	\beqrs
	&&\cov_*\left[F_{2*}\{\wh\theta(X_2, Y_2)\},F_{1*}\{\wh\theta(X_{1}, Y_{2})\} \right] - \cov\left[F_2\{\theta(X_2, Y_2)\},F_1\{\theta(X_{1}, Y_{2})\} \right]\\
	&=&\E_*\left[ F_{2*}\{\wh\theta(X_2, Y_2)\}F_{1*}\{\wh\theta(X_{1}, Y_{2})\} \right] - \E\left[ F_2\{\theta(X_2, Y_2)\}F_1\{\theta(X_{1}, Y_{2})\}\right] - \\
	&&\E_*\left[ F_{2*}\{\wh\theta(X_2, Y_2)\}\right]  \E_*\left[F_{1*}\{\wh\theta(X_{1}, Y_{2})\} \right] + \E\left[ F_2\{\theta(X_2, Y_2)\}\right]  \E\left[F_1\{\theta(X_{1}, Y_{2})\}\right] = o_p(1).
	\eeqrs
	Thus we have shown that $\sigma_*^2 = \sigma^2 + o_p(1)$. Under the condition $\sigma^2\geq C>0$, it follows that
	\beqrs
	\frac{\sigma_*^2}{\sigma^2} -1 =  \frac{\sigma_*^2 - \sigma^2}{\sigma^2}  = o_p(1)
	\eeqrs
	
	{\bf Step 2:} We then deal with  $\frac{\sqrt{n_2}(R'- \mu)}{ \sigma_0}\frac{\sigma_0}{\wh\sigma}$. By Lemma \ref{marginal_proj_h1}, $R'  - \mu = R'_{\mu} + O_p(n_2^{-1})$, where
	\beqrs
	R'_{\mu} = \frac{1}{n_2}\sum_{i\in\calI_2}\left[1 - F_2\{\theta(X_i, Y_i)\} - \mu\right] + \frac{1}{n_2}\sum_{j\in\calI_2}\left[ F_1\{\theta(X_j', Y_j')\}  - \mu\right].
	\eeqrs
	The dependence graph of the $2n_2$ random variables 
	\beqrs
	\left\{ 1 - F_{2}\{\wh\theta(X_i, Y_i)\}- \mu, i\in\calI_2 \right\} \cup \left\{  F_{1}\{\wh\theta(X_j', Y_j')\}  - \mu, j\in\calI_2\right\}
	\eeqrs
	is 3-regular. Note that  $\sigma_0^2>c\geq0$,  thus we have $n_2^{1/3}\sigma_0^2\rightarrow\infty$. Similar as the proof in Theorem 3, we have
	\beqrs
	\frac{\sqrt{n_2}(R'- \mu)}{ \sigma_0} \overset{d}{\rightarrow} N(0, 1).
	\eeqrs
	It follows that
	\beqrs
	\frac{\sqrt{n_2}(R' - \mu)}{\wh \sigma} = \frac{\sqrt{n_2}(R' - \mu)}{ \sigma_0}\frac{\sigma_0}{\wh \sigma} = Z(1+o_p(1)), 
	\eeqrs
	where $Z$ converges to a standard normal distribution as $n_1$ and $n_2$ goes to infinity.

	{\bf Step 3:} We now deal with $\sqrt{n_2}(R - R')$. 
	By Lemma \ref{marginal_proj_h1}, we know that
	\beqrs
	R  - \mu_* &=& \frac{1}{n_2}\sum_{i\in\calI_2}\left[1 - F_{2*}\{\wh\theta(X_i, Y_i)\} - \mu_*\right] + \frac{1}{n_2}\sum_{j\in\calI_2}\left[ F_{1*}\{\wh\theta(X_j', Y_j')\}  - \mu_*\right]+O_p(n_2^{-1}),\\
	R'  - \mu&=& \frac{1}{n_2}\sum_{i\in\calI_2}\left[1 - F_2\{\theta(X_i, Y_i)\} - \mu\right] + \frac{1}{n_2}\sum_{j\in\calI_2}\left[ F_1\{\theta(X_j', Y_j')\}  - \mu\right]+O_p(n_2^{-1}).
	\eeqrs
	
	Thus $\sqrt{n_2}(R - R') $ is equal to
	\beqrs
	\sqrt{n_2}(R - R') &=&\frac{\sqrt{n_2}}{n_2}\sum_{i\in\calI_2}\left[F_2\{\theta(X_i, Y_i)\} -F_{2*}\{\wh\theta(X_i, Y_i)\} \right] + \\
	&&\frac{\sqrt{n_2}}{n_2}\sum_{i\in\calI_2}\left[F_{1*}\{\wh\theta(X_i', Y_i')\} -F_1\{\theta(X_i', Y_i')\} \right]  + \sqrt n_2(\mu - \mu_*) + O_p(n_2^{-1/2})\\
	&:=& A_1+A_2+A_3+ O_p(n_2^{-1/2}).
	\eeqrs
	First of all, we know that $A_3 = o_p(1)$ by assumption. 
	To deal with $A_1$, note that the conditional expectation of each term in $A_1$ is
	\beqrs
	&&\E_*\left[F_2\{\theta(X_i, Y_i)\} -F_{2*}\{\wh\theta(X_i, Y_i)\} \right] \\
	=&& \P_* \{\theta(X', Y')<\theta(X, Y)\} - \P _*\{\wh\theta(X', Y')<\wh\theta(X, Y)\} \\
	=&& 1-\mu - (1-\mu_*)  = \mu_* - \mu  = o_p(n_2^{-1/2}),
	\eeqrs
	where the inequality follows the assumption. Thus $\E(A_1) = o_p(1)$. Now consider the conditional variance of $F_2\{\theta(X_i, Y_i)\} -F_{2*}\{\wh\theta(X_i, Y_i)\}$:
	\beqrs
	\var_*\Big(F_2\{\theta(X_i, Y_i)\} -F_{2*}\{\wh\theta(X_i, Y_i)\}\Big) &\leq &\E_*\Big(F_2\{\theta(X_i, Y_i)\} -F_{2*}\{\wh\theta(X_i, Y_i)\}\Big)^2\\
	&\leq&  \E_*\Big|F_2\{\theta(X_i, Y_i)\} -F_{2*}\{\wh\theta(X_i, Y_i)\}\Big| = o_p(1),
	\eeqrs
	where the equation holds by condition (8).  Thus we have $\var_*(A_1) = o_p(1)$. It follows that $A_1 = o_p(1)$.
	
	To deal with $A_2$, we need to consider the dependence between samples because $(X_i', Y_i')$ are no longer independent. First,  it follows similarly that 
	\beqrs
	&&\E_*\left[F_{1*}\{\wh\theta(X_i', Y_i')\} -F_1\{\theta(X_i', Y_i')\} \right]  =  o_p(n^{-1/2})\\
	&&\var_*\Big(F_{1*}\{\wh\theta(X_i', Y_i')\} -F_1\{\theta(X_i', Y_i')\} \Big) =  o_p(1)
	\eeqrs
	And when $i$ and $j$ are dependent pairs (as defined in Lemma \ref{SLLN}),
	\beqrs
	\cov_*\Big(F_{1*}\{\wh\theta(X_i', Y_i')\} -F_1\{\theta(X_i', Y_i')\}, F_{1*}\{\wh\theta(X_{j}', Y_{j}')\} -F_1\{\theta(X_{j}', Y_{j}')\} \Big) =  o_p(1)
	\eeqrs
	When  $i$ and $j$ are not dependent pairs, 
	\beqrs
	\cov_*\Big(F_{1*}\{\wh\theta(X_i', Y_i')\} -F_1\{\theta(X_i', Y_i')\}, F_{1*}\{\wh\theta(X_{j}', Y_{j}')\} -F_1\{\theta(X_{j}', Y_{j}')\} \Big) =0.
	\eeqrs
	It follows that $\var(A_2) = o_p(1)$. Because $\E(A_2) = o_p(1)$ by the assumption of $\mu_* - \mu = o_p(n_2^{-1/2})$, we have $A_2 = o_p(1)$.  Thus we have shown that $\sqrt{n_2}(R - R') = o_p(1)$.
	It follows that
	\beqrs
	\frac{\sqrt{n_2}(R - R')}{\wh \sigma} = \frac{\sqrt{n_2}(R - R')}{ \sigma}\frac{\sigma}{\wh \sigma} = o_p(1)(1+o_p(1)) = o_p(1).
	\eeqrs

	{\bf Step 4:}
	By Lemma \ref{theta_to_L} and the continuous assumption of $\theta$, we know that
	\beqrs
	\mu  &=& \E\{L_2'\indic{( L_2'< L_2)}\} \\
	&=& \frac{1}{2}\Big(\E\{L_2'\indic{( L_2'< L_2)}\} +\E\{L_2\indic{( L_2< L_2')}\}  \Big)\\
	&=& \frac{1}{2}\Big(1 - \E\{L_2'\indic{( L_2< L_2')}\} +\E\{L_2\indic{( L_2< L_2')}\}  \Big)\\
	&=& \frac{1}{2}\Big(1 - \E\{(L_2' - L_2)\indic{( L_2< L_2')}\}  \Big)\\
	&=& \frac{1}{2}\Big(1 - \frac{1}{2}\E\{|L_2' - L_2| \}  \Big)\\	
	&=& \frac{1}{2} - \frac{1}{4}\E\{|L_2' - L_2| \} \\	
	\eeqrs

	Finally, 
	\beqrs
	\frac{\sqrt{n_2}(R - 1/2)}{\wh \sigma} &=&   \frac{\sqrt{n_2}(R'- \mu)}{ \sigma}\frac{\sigma}{\wh\sigma} -\frac{\sqrt n_2 \delta}{4\sigma}(1+o_p(1)) + o_p(1)\\
	&=& Z  -\frac{\sqrt n_2 \delta}{4\sigma} + o_p(1).
	\eeqrs
	where $Z$ converges to a standard normal distribution as the sample size goes to infinity.
\end{proof}

\subsection{Proof of Theorem 9}\label{proof:m_est}

\begin{proof} 
	Because $\wh\beta_n$ is the maximizer of $M_n(\beta)$, we know that
	\beqrs
	M_n(\wh\beta_n) \geq M_n(\beta_0) - o_p(1).
	\eeqrs
	By the uniform consistency in Lemma \ref{GC}, we have that $M_n(\beta_0) = \E\{M(X, Y, X', Y'; \beta_0) \} + o_p(1)$.
	Thus $M_n(\wh\beta_n)\geq \E\{M(X, Y, X', Y'; \beta_0) \} - o_p(1)$. It follows that
	\beqrs
	\E\{M(X, Y, X', Y'; \beta_0) \} - \E\{M(X, Y, X', Y'; \wh\beta_n) \}&\leq& M_n(\wh\beta_n) -  \E\{M(X, Y, X', Y'; \wh\beta_n) \}+o_p(1)\\
	&\leq& o_p(1)\overset{p}{\rightarrow}0,
	\eeqrs
	where the last inequality follows by Lemma \ref{GC}.
	By the identifiability condition, for every $\ep>0$, $d(\beta,\beta_0)\geq\ep$, there exist an $\eta>0$ such that $\E\{M(X, Y, X', Y'; \beta) \} <\E\{M(X, Y, X', Y'; \beta_0) \} -\eta$. Thus $\P\{d(\beta,\beta_0)\geq\ep\}\leq \P\{\E\{M(X, Y, X', Y'; \beta) \} <\E\{M(X, Y, X', Y'; \beta_0) \}  -\eta\} \rightarrow0$. This completes the proof.
\end{proof}

\subsection{Proof of Theorem 10}\label{proof:hd_theory}

\begin{proof}
	Empirically, we optimize the regularized optimization problem (10). Because $\wh\beta$ is optimal, we have 
	\beqrs
	\wh\beta^T\wh\Gamma\wh\beta - 2\wh\gamma^T\wh\beta +\lambda\|\wh\beta \|_1  \leq \beta^{*T}\wh\Gamma\beta^* - 2\wh\gamma^T\beta^*+\lambda\|\beta^* \|_1
	\eeqrs
	Let $\wh\Delta = \wh\beta-\beta^*$.	After some basic algebra, we obtain that
	\beqrs
	\frac{1}{n}\|\Xi\wh\Delta\|_2^2\leq \frac{1}{n}2 w^T\Xi\wh\Delta + \lambda\|\beta^* \|_1- \lambda\|\wh\beta \|_1.
	\eeqrs
	Because  $\beta^*$ is supported on a subset $S\subset\{1,2,\dots, m\}$ with $|S| = s_2$, we can write
	\beqrs
	\|\beta^* \|_1- \|\wh\beta \|_1 =\|\beta^*_S \|_1- \|\beta^*_S+ \wh\Delta_S \|_1 - \|\wh\Delta_{S^c} \|_1 .
	\eeqrs
	Thus we have
	\beqrs
	0&\leq& \frac{1}{n}\|\Xi\wh\Delta\|_2^2\leq \frac{1}{n}2 w^T\Xi\wh\Delta + \lambda\left(\|\beta^*_S \|_1- \|\beta^*_S+ \wh\Delta_S \|_1 - \|\wh\Delta_{S^c} \|_1 \right) \\
	&\leq& 2\|\frac{ \Xi^T w }{n}\|_{\infty} \|\wh\Delta\|_{1} + \lambda\left( \|\wh\Delta_S \|_1 - \|\wh\Delta_{S^c} \|_1 \right) \\
	&\leq&\frac{\lambda}{2}\left(3 \|\wh\Delta_S \|_1 - \|\wh\Delta_{S^c} \|_1 \right).
	\eeqrs
	The second inequality follows from Holder's inequality and triangle inequality. The last inequality follows from the assumption on $\lambda$. Thus we have  $\wh\Delta\in\mathcal{C}_3(S)$. By the restricted eigenvalue condition,
	\beqrs
	\kappa \|\wh\Delta\|_2^2 \leq \frac{1}{n}\|\Xi\Delta\|_2^2 \leq  \frac{3\lambda}{2}  \|\wh\Delta_S \|_1 \leq   \frac{3\lambda}{2} \sqrt{s_2} \|\wh\Delta_S \|_2
	\eeqrs
	Thus we have
	\beqrs
	\|\wh\Delta\|_2 \leq  \frac{3}{2\kappa} \sqrt{s_2}\lambda.
	\eeqrs
	Now it suffices to derive the upper bound for $\| \Xi^T w /n\|_{\infty}$. For notation simplicity, we let $p_1(Z)$ be the density of $Z$, and let $p_0(Z')$ be the density of $Z'$. Then by assumption, we have $$ \xi(Z)^T\beta^* = g(Z) = \frac{p_1(Z)}{p_1(Z) + p_0(Z)}.$$
	Thus for any function $q(z)\in\mR$, we have 
	\beqr\label{eq: exp_w}
	&&\E[w_1(Z) q(Z) + w_0(Z')q(Z') ] \\\nonumber
	=&&\int \left(1 - \frac{p_1(Z)}{p_1(Z) + p_0(Z)}\right)q(Z)p_1(Z) dZ + \int \left(- \frac{p_1(Z)}{p_1(Z) + p_0(Z)}\right)q(Z)p_0(Z)	dZ = 0.
	\eeqr
	Let $\Xi^T w \defby (\zeta_1,\dots, \zeta_m)^T$. Thus by (\ref{eq: exp_w}), we have $\E\zeta_j = 0$. Moreover, $\zeta_j$ is only a function of $Z = (Z_1,\dots, Z_n)$, written as $\zeta_j(Z)$. Let $\wt Z^i = (Z_1,\dots, Z_{i-1}, \wt Z_i, Z_{i+1},\dots, Z_n)$, where $\wt Z_i$ is an independent copy of $Z_i$. Following similar reasoning as in the proof of Theorem 3, the dependence graph of $(Z_1,\dots, Z_n, Z_1',\dots, Z_n')$ is 3-regular. Thus $|\zeta_j(Z) - \zeta_j(\wt Z^i)|\leq 6B$. By the McDiarmid’s inequality, we have  
	\beqrs
	\P(|\zeta_j|\geq nt)\leq 2\exp\left(-\frac{2nt^2}{36B^2} \right)
	\eeqrs
	Thus,
	\beqrs
	\P\left( \|\frac{ \Xi^T w }{n}\|_{\infty} \geq t\right)\leq \sum_{j=1}^{m}\P(|\zeta_j|\geq nt)\leq 2m\exp\left(-\frac{2nt^2}{36B^2} \right)
	\eeqrs
	Replacing $t$ with $C_1\sqrt{\log m/n}$, we obtain that
	\beqrs
	\P\left( \|\frac{ \Xi^T w }{n}\|_{\infty} \geq C_1\sqrt{\frac{\log m}{n}} \right)\leq m^{-1}.
	\eeqrs
	$C_1$ is a constant related to $B$. This completes the proof.
\end{proof}

\bibliography{cairui}		
\bibliographystyle{refstyle}

\end{document}